\documentclass[aps]{revtex4}
\usepackage{amsfonts}
\usepackage{amsmath}
\usepackage{amssymb,epsf}

\begin{document}

\title{Thermodynamical Structure of AdS Black Holes in Massive Gravity\\
with Stringy Gauge-Gravity Corrections}
\author{S. H. Hendi$^{1,2}$\footnote{email address: hendi@shirazu.ac.ir}, B. Eslam Panah$^{1}$\footnote{
email address: behzad.eslampanah@gmail.com}, and S.
Panahiyan$^{1,3}$ \footnote{email address:
sh.panahiyan@gmail.com}} \affiliation{$^1$ Physics Department and
Biruni Observatory, College of Sciences, Shiraz
University, Shiraz 71454, Iran\\
$^2$ Research Institute for Astronomy and Astrophysics of Maragha (RIAAM),
P. O. Box 55134-441, Maragha, Iran\\
$^3$ Physics Department, Shahid Beheshti University, Tehran 19839, Iran}

\begin{abstract}
Motivated by gauge/gravity group in the low energy effective theory of the
heterotic string theory and novel aspects of massive gravity in the context
of lattice physics, the minimal coupling of Gauss-Bonnet-massive gravity
with Born-Infeld electrodynamics is considered. At first, the metric
function is calculated and then the geometrical properties of the solutions
are investigated. It is found that there is an essential singularity at the
origin and the intrinsic curvature is regular elsewhere. In addition, the
effects of massive parameters are studied and black hole solutions with
multi horizons are found in this gravity. Also the conserved and
thermodynamic quantities are calculated, and it is shown that the solutions
satisfy the first law of thermodynamics. Furthermore, using heat capacity of
these black holes, thermal stability and phase transitions are investigated.
The variation of different parameters and related modifications on the
(number of) phase transition are examined. Next, the critical behavior of
the Gauss-Bonnet-Born-Infeld-massive black holes in the context of extended
phase space is studied. It is shown that how the variation of the different
parameters affects the existence and absence of phase transition. Also, it
is found that for specific values of different parameters, these black holes
may enjoy the existence of new type of phase transition which to our
knowledge was not observed in black hole physics before.
\end{abstract}

\maketitle

\section{Introduction}

This fact that the Universe expands with acceleration follows directly from
the observation of high red-shift supernova \cite{Perlmutter} and indirectly
from the measurement of angular fluctuations of cosmic microwave background
fluctuations \cite{Lee}. Since Einstein's theory can not explain current
acceleration of the Universe, without fine tuning, various modified
gravities have been proposed. For example one can refer to brane world
cosmology \cite{Hamed,Ida}, scalar-tensor theories \cite{Brans} and $F(R)$
gravity theories \cite{Odintsov2,Capozziello,Hendi2012}.

One of the interesting modified gravity theories is Lovelock gravity \cite%
{Lovelock}. It is the most generalization that satisfies properties of
Einstein's tensor in higher dimensions \cite{Lovelock}. In addition, by
taking the Gaussian null coordinates into account, one can find that the
near null surface behavior of gravitational field equations is equivalent
with the conventional first law of thermodynamics (see \cite%
{Sumanta1,Sumanta2} for more details). On the other hand,
evolution of the spacetime has been investigated by using degrees
of freedom of bulk and surface in Lovelock gravity. It was also
pointed out that considering Lovelock gravity provides a natural
backdrop to test different conceptual and mathematical aspects of
Einstein's theory \cite{Padmanabhan}. Lovelock gravity also enjoys
only first and the second-order derivatives of metric function in
its field equations and it was shown that it is a ghost free
theory of gravity. Furthermore, the effects of higher curvature
terms of this generalization present themselves in higher
dimensions. Indeed Gauss-Bonnet (GB) gravity is a topological
invariant term in $4$-dimensions, and therefore, it has no
contribution to the equations of motion.

Generalization of Einstein gravity to higher curvature Lovelock
theory enables us to obtain a better insight into the phenomena in
the context of gravity. The first three terms of Lovelock gravity
called GB gravity in the presence of cosmological constant (GB
gravity). The first term is the cosmological term and the second
and third terms are the Einstein and second order Lovelock (GB)
terms, respectively. The GB gravity has interesting properties
which are listed below.

In order to have a ghost-free action, the quadratic curvature corrections to
the Einstein-Hilbert action should be proportional to the GB term \cite%
{Zwiebach}. In addition, the natural next-to-leading order term of the
heterotic string effective action which plays a fundamental role in
Chern-Simons gravitational theories is GB term \cite{Chamseddine}. In Ref.
\cite{Pastras}, it was pointed out that considering GB gravity will lead to
obtain the modified Renyi entropies. These entropies violate specific
inequality which must be hold for Renyi entropy. In addition, in case of
AdS/CFT correspondence, it was shown that considering GB gravity will modify
shear viscosity, entropy, thermal conductivity and electrical conductivity
\cite{conductivity}.

The existence of singularity at the origin for a point-like
charge, represents a shortcoming of Maxwell theory. In order to
overcome this problem, Born and Infeld \cite{BI} generalized
linear Maxwell theory to a nonlinear one. The nonlinear form of
the Born-Infeld theory results into a bounded electric field
associated to a point-like charge everywhere. This leads into a
finite self-energy which resolves the shortcoming of the Maxwell
theory in this regard. The spherically symmetric solutions of
Einstein gravity in the presence of Born-Infeld (BI) theory was
obtained in Ref. \cite{Hoffmann}. Various applications of BI
theory in the context of black hole physics have been investigated
by many authors \cite{Clement}. On the other hand, it was shown
that BI theory could be derived in open super strings and D-branes
in the context of this theory are free from physical singularities
\cite{Fradkin}. For getting a better picture of properties of BI
theory in string theory see Ref. \cite{Gibbons}

In Einstein theory of gravity, gravitons are massless particles. In order to
make a massive theory, one can simply add massive terms to Einstein gravity.
This will lead to introduction of a massive theory with a massive spin $2$
particle propagation in which for $m\rightarrow 0$, massless Einstein theory
of gravity is recovered.

Manipulation of anti-de Sitter/conformal field theory (AdS/CFT)
correspondence may lead to obtain a massive gravity theory. In usual AdS/CFT
correspondence, the graviton in AdS side is corresponding to the
energy-momentum tensor of CFT. In this usual regard, diffeomorphism
invariance in AdS gravity translates to energy-momentum conservation in the
CFT side. Motivated by generalized energy-momentum dispersion relation in UV
regime \cite{Rainbow} and multi-trace deformation \cite{multi}, one may
deform the CFT in a way that violates the energy-momentum conservation.
Accordingly, a gravitational Higgs effect appears in AdS gravity and
diffeomorphism invariance gets broken, and therefore, the graviton acquires
a mass (for more details see \cite{Niarchos}). In addition, a model of
cosmological topologically massive gravity in the context of AdS/CFT
relation has been presented in Ref. \cite{CTMG}.

In addition, using an equivalent of the Brout-Englert-Higgs mechanism \cite%
{BEH} for gravity, gravitons (spin 2 particles) can attain mass by
spontaneous local symmetry breaking \cite{tHooft}. Also, two models of
exhibiting Higgs mechanism for gravitons and the criteria for having massive
graviton based on the process of spontaneous symmetry breaking of
diffeomorphisms have been constructed in \cite{Higgs}. Moreover, a modified
gravity based on supersymmetry (where the graviton has acquired a mass) has
been used by Sj\"{o}rs \cite{thesis}. He has taken into account the
Vainshtein mechanism to obtain a constraint on the graviton mass, which is
proportional to $10^{-2} H_{0}$. Regarding the mass of the graviton
proportional to the inverse Hubble scale today, the cosmological constant
problem may be resolved and the value of $\Lambda$ is in agreement with
todays' observations \cite{LambdaProb}. Besides, based on SO(3) symmetry and
time reparameterization invariance, a class of massive gravity theory was
proposed \cite{SO3}.

On the other side, the one-loop mass shifts for the graviton, has been
calculated in the context of bosonic string theory \cite{loop}. Based on
Kawai-Lewellen-Tye formulae \cite{KLT}, the soft behavior of gravitons to
sub-leading order for superstring amplitudes with massive insertions to open
and closed strings has been investigated in Refs. \cite{soft1,soft2}. In
addition, the consistent equations of motion for the massive spin-$2$ field
interacting with gravity in the context of both field theory and string
theory have been obtained \cite{gravstring}.

Following the massive gravity model of Visser \cite{Visser}, an interesting
cosmological model was obtained which has acceptable value for the age of
the Universe. This consistent model can be fitted by the present
cosmological supernovae type Ia data without considering dark energy \cite%
{SNIa}. Finally, we should note that although some models of general
relativity put a limit on the graviton mass, such models do not rule out
interesting cosmological and gravitational effects of massive gravity. In
this direction, inability of LIGO's observation \cite{LIGO} for
distinguishing general relativity and its massive extension has been
recently reported by Deser \cite{inability}.

Several types of massive gravity with their specific characteristics have
been introduced and their properties have been investigated \cite%
{Fierz,Boulware,Hassan,Minjoon,Rham,Saridakis}. In Ref.
\cite{Babichev} massive gravity is explored in more details.
Another class of massive theory was introduced by Vegh in Ref.
\cite{Vegh}. The black holes solutions of this theory in AdS
spacetime has been investigated \cite{Hassan2011}. One of the
interesting aspects of this theory is the lattice like behavior of
the graviton in holographic conductor model. It was shown that
limit of massless gravity leads to a Drude peak which approaches
to delta function. This is the behavior of the lattice in theory
of the field. In addition, this theory has been employed to study
stability conditions and metal-insulator transition in AdS/CFT
context \cite{MassiveADSCFT,Davison}. Several studies regarding
thermodynamical aspects of the Vegh's massive gravity have been
conducted \cite{Cai2015,Xu2015,HendiPEM}. In the context of
holography, the conductivity and phase transitions of this
theories of massive gravity has been investigated \cite{Davison}.
Charged BTZ black hole solutions and their thermodynamics with
Maxwell and BI fields in massive gravity have been studied in
\cite{HendiEPBTZmassive}. In addition, the generalization to GB
gravity and BI nonlinear electromagnetic fields were done. Their
thermodynamical behavior have been investigated in details \cite%
{HendiPEGBmass,HendiEPEN-BImass}. In this paper, we consider these two
generalization for gravity and matter field of the action and black holes in
GB-BI-massive gravity.

Thermodynamical aspects of black holes plays a crucial role toward theory of
quantum gravity \cite{Myung}. Of the greatest interest is thermal stability
of the black holes. In order to black holes be stable, thermal stability
conditions indicate that the heat capacity must be positive valued. This is
known as canonical ensemble. Investigation of the heat capacity also enables
one to study the phase transitions of the black holes. It was pointed out
that roots of the denominator and numerator of the heat capacity are denoted
as two different types of the phase transition.

Recently, there has been a growing interest in considering cosmological
constant as a thermodynamical variable. It was pointed out that this
consideration will enrich thermodynamical structure \cite{cosmological} of
the black holes and leads to removing ensemble dependency \cite{ensemble}.
On the other hand, it was shown that interpretation of the negative
cosmological constant as thermodynamical pressure may lead to a van der
Waals like behavior of liquid/gas for black holes \cite{van}. Consideration
of the cosmological constant as a thermodynamical variable could be
justified through AdS/CFT. In the first case, the Yang-Mills theory residing
on the boundary of the AdS spacetime enjoys the existence of variation of
color which is related to the variation of cosmological constant in the bulk
spacetime \cite{yang}. In the second case, an RG-flow in the context of
field theory corresponds to extended phase space which is obtained by
consideration of cosmological constant as thermodynamical variable \cite{RG}%
. In the space of the field theory, the way that number of degrees of
freedom runs with the energy scale is codified by isotherm curves.

The main motivation of this article comes from string theory. As
mentioned before, BI theory and GB-gravity arise from the
low-energy limit of the open and heterotic string theories,
respectively. So, we add massive spin-$2$ particle to these
theories and obtain the black hole solutions and investigate phase
transition and their other properties.

Regarding the motivations for considering GB-BI-massive set up,
one can point out following ones; The Einstein theory of gravity
has specific problems which could be solved by consideration of
generalization to GB gravity. Then again, the GB gravity has
special properties that are mentioned before which make its
studying crucial. But this set up of gravity (EN-GB) provides
massless gravitons. In order to have massive gravitons, it is
necessary to include mass terms such as those introduced by Vegh's
theory of massive gravity. In addition, it is well-known that BI
theory enjoys vast number of properties that are absent in the
Maxwell theory which among them
one can name: the absence of shock waves and birefringence phenomena \cite%
{absence}, existence of electric-magnetic duality \cite{duality} and etc (
see also \cite{Dulaization}). Therefore, for enriching our solutions to
include these properties, we consider BI generalization of Maxwell theory as
well.

Adding these terms to action of our gravitational system will have specific
contributions into geometric and thermodynamics of black holes. These
contributions and their corresponding modifications will introduce new
phenomenologies to physics of black holes which are of our interest in this
paper. We are especially interested in thermal structure, stability
condition and phase transitions of black holes which are solutions to this
specific set up of gravitational system. We are interested to see how the
effects of these generalization will modify the thermodynamical picture of
black holes.

Another important property of such set up is the fact that some of the
symmetries of the system are broken (due to presence of massive gravity).
This symmetry breaking could be used to conduct specific studies in the
context of gauge/gravity duality. For example, the massive theory which is
employed here has interpretation of lattice. As it was stated before, the
graviton in this theory in specific limits presents a behavior which leads
to interpretation of lattice. Now, the generalization to GB and BI will
affect this lattice like behavior to some degrees which could introduce new
limits for having lattice like behavior and new physics for it.

On the other hand, properties such as shear viscosity, Renyi entropy and
conductivity of the usual GB-BI black holes will be modified in the presence
of massive gravity. These modifications motivate one to consider this
specific set up for our gravitational system. Here, we will not address all
the issues, but rather providing required information to some level, so one
is able to conduct mentioned studies.

Different theories that are introduced regarding physical systems
may have specific shortcomings. In order to remove these
shortcomings, introducing new phenomena for physical systems and
obtaining more accurate and reliable solutions, one should apply a
number of generalization. Having this in mind, we would like to
impose three generalizations into Reissner-Nordstr\"{o}m black
holes: I) Gauss-Bonnet gravity which is considered to solve some
of the shortcomings of Einstein gravity such as renormalization
problem. II) Born-Infeld gravity which is taken into account to
remove problems of Maxwell theory such singularity at the origin.
III) Finally, massive gravity which solves the absence of massive
gravitons in the Einstein and Gauss-Bonnet gravity models. It is
worthwhile to mention that existence of massive graviton was
proposed by studies in the brane-world gravity theories. These
theories are also motivated from different aspects of physics
which were pointed out before. But the main idea is to see how the
combination of these different generalizations and corrections,
motivated from different branches of physics, would modify the
geometry and thermodynamics of the black holes in different
contexts. In this paper, we try to provide a picture regarding
this matter. In order to provide a situation for considering a
combination of all GB, BI and massive gravity, the best candidate
is a black hole in high energy and high curvature regimes. This
situation is more subtle because the geometrical behavior and
thermodynamical aspect of the solutions are not, in general, as
trivial as in usual general relativity.

The paper is organized as follow. In Sec. II, GB-BI-massive gravity action
and corresponding field equations will be introduced. Then, in Sec. III a
class of black holes with this configuration is obtained and geometrical
properties are investigated. Next, validation of the first law of
thermodynamics is investigated through obtained thermodynamical and
conserved quantities. Sec. V will be devoted to study thermal stability in
canonical ensemble for GB-BI-massive black hole solutions. Next, we employ
the analogy between cosmological constant and thermodynamical pressure in
extended phase space and study critical behavior of the obtained solutions.
We finish our paper with some concluding remarks.

\section{Basic Equations}

The $d$-dimensional action of GB-massive gravity with a nonlinear
electrodynamics is
\begin{equation}
\mathcal{I}=-\frac{1}{16\pi }\int d^{d}x\sqrt{-g}\left( \mathcal{R}-2\Lambda
+\alpha L_{GB}+L(\mathcal{F})+m^{2}\sum_{i}^{4}c_{i}\mathcal{U}%
_{i}(g,f)\right),  \label{Action}
\end{equation}
where $\mathcal{R}$, $\Lambda$, $m$ and $\alpha $ are the scalar curvature,
the cosmological constant, the massive parameter and the coefficient of GB
gravity, respectively, and $L_{GB}$ is the Lagrangian of GB gravity
\begin{equation}
L_{GB}=R_{\mu \nu \gamma \delta }R^{\mu \nu \gamma \delta }-4R_{\mu \nu
}R^{\mu \nu }+R^{2},
\end{equation}%
where $R_{\mu \nu }$ and $R_{\mu \nu \gamma \delta }$ are, respectively, the
Ricci and the Riemann tensors. $f$ is a fixed symmetric tensor, $c_{i}$'s
are constants and $\mathcal{U}_{i}$ are symmetric polynomials of the
eigenvalues of the $d\times d$ matrix $\mathcal{K}_{\nu }^{\mu }=\sqrt{%
g^{\mu \alpha }f_{\alpha \nu }}$, which can be written as
\begin{eqnarray}
\mathcal{U}_{1} &=&\left[ \mathcal{K}\right] , \\
\mathcal{U}_{2} &=&\left[ \mathcal{K}\right] ^{2}-\left[ \mathcal{K}^{2}%
\right] , \\
\mathcal{U}_{3} &=&\left[ \mathcal{K}\right] ^{3}-3\left[ \mathcal{K}\right] %
\left[ \mathcal{K}^{2}\right] +2\left[ \mathcal{K}^{3}\right] , \\
\mathcal{U}_{4} &=&\left[ \mathcal{K}\right] ^{4}-6\left[ \mathcal{K}^{2}%
\right] \left[ \mathcal{K}\right] ^{2}+8\left[ \mathcal{K}^{3}\right] \left[
\mathcal{K}\right] +3\left[ \mathcal{K}^{2}\right] ^{2}-6\left[ \mathcal{K}%
^{4}\right] .
\end{eqnarray}

One of the primitive motivations of studying the BI theory is due to its
relation to string effective actions. The functional form of BI Lagrangian $%
L(\mathcal{F})$ is presented by

\begin{equation}
L(\mathcal{F})=4\beta ^{2}\left( 1-\sqrt{1+\frac{\mathcal{F}}{2\beta ^{2}}}%
\right) ,  \label{L(F)}
\end{equation}%
where $\beta$ and $\mathcal{F}=F_{\mu \nu }F^{\mu \nu }$ are the BI
parameter and the Maxwell invariant, respectively, in which $F_{\mu \nu
}=\partial_{\mu}A_{\nu }-\partial _{\nu }A_{\mu }$ is the electromagnetic
field tensor and $A_{\mu }$ is the gauge potential. It is notable that, in
the limit $\beta \rightarrow \infty $, Eq. (\ref{L(F)}) reduces to the
standard Maxwell Lagrangian, as it should be.

Using the action (\ref{Action}) and variation of this action with respect to
the metric tensor ($g_{\mu \nu }$) and the Faraday tensor ($F_{\mu \nu }$),
one can obtain the following field equations
\begin{equation}
G_{\mu \nu }+\Lambda g_{\mu \nu }+H_{\mu \nu }-\frac{1}{2}g_{\mu \nu }L(%
\mathcal{F})-\frac{2F_{\mu \lambda }F_{\nu }^{\lambda }}{\sqrt{1+\frac{%
\mathcal{F}}{2\beta ^{2}}}}+m^{2}\chi _{\mu \nu }=0,  \label{Field equation}
\end{equation}%
\begin{equation}
\partial _{\mu }\left( \frac{\sqrt{-g}F^{\mu \nu }}{\sqrt{1+\frac{\mathcal{F}%
}{2\beta ^{2}}}}\right) =0,  \label{Maxwell equation}
\end{equation}%
in which $G_{\mu \nu }$ is the Einstein tensor, $H_{\mu \nu }$ and $\chi
_{\mu \nu }$\ are in the following forms
\begin{eqnarray}
H_{\mu \nu } &=&-\frac{\alpha }{2}\left( 8R^{\rho \sigma }R_{\mu \rho \nu
\sigma }-4R_{\mu }^{\rho \sigma \lambda }R_{\nu \rho \sigma \lambda
}-4RR_{\mu \nu }+8R_{\mu \lambda }R_{\nu }^{\lambda }+g_{\mu \nu
}L_{GB}\right) , \\
\chi _{\mu \nu } &=&-\frac{c_{1}}{2}\left( \mathcal{U}_{1}g_{\mu \nu }-%
\mathcal{K}_{\mu \nu }\right) -\frac{c_{2}}{2}\left( \mathcal{U}_{2}g_{\mu
\nu }-2\mathcal{U}_{1}\mathcal{K}_{\mu \nu }+2\mathcal{K}_{\mu \nu
}^{2}\right) -\frac{c_{3}}{2}(\mathcal{U}_{3}g_{\mu \nu }-3\mathcal{U}_{2}%
\mathcal{K}_{\mu \nu }  \notag \\
&&+6\mathcal{U}_{1}\mathcal{K}_{\mu \nu }^{2}-6\mathcal{K}_{\mu \nu }^{3})-%
\frac{c_{4}}{2}(\mathcal{U}_{4}g_{\mu \nu }-4\mathcal{U}_{3}\mathcal{K}_{\mu
\nu }+12\mathcal{U}_{2}\mathcal{K}_{\mu \nu }^{2}-24\mathcal{U}_{1}\mathcal{K%
}_{\mu \nu }^{3}+24\mathcal{K}_{\mu \nu }^{4}).
\end{eqnarray}

\section{GB-BI-massive black hole solutions}

Here, we are looking for obtaining the topological static black holes. For
this purpose, we consider the metric of $d$-dimensional spacetime as
\begin{equation}
ds^{2}=-f(r)dt^{2}+f^{-1}(r)dr^{2}+r^{2}h_{ij}dx_{i}dx_{j},\;\;\;\;
i,j=1,2,3,...,d-2,  \label{Metric}
\end{equation}%
in which $h_{ij}dx_{i}dx_{j}$ is the line element with constant curvature $%
\left( d-2\right) (d-3)\kappa $ and volume $V_{d-2}$. The constant $\kappa$
is related to the boundary curvature and may have a positive (elliptic),
negative (hyperbolic) or zero (flat) constant curvature.

Following \cite{Vegh}, we consider the ansatz metric in the following form
\begin{equation}
f_{\mu \nu }=diag(0,0,c^{2}h_{ij}),  \label{f11}
\end{equation}%
where $c$ is a positive constant. Considering Eq. (\ref{f11}), $\mathcal{U}%
_{i} $'s are \cite{Vegh}
\begin{equation}
\mathcal{U}_{1}=\frac{d_{2}c}{r},\;\;\;\;\mathcal{U}_{2}=\frac{%
d_{2}d_{3}c^{2}}{r^{2}},\;\;\;\;\mathcal{U}_{3}=\frac{d_{2}d_{3}d_{4}c^{3}}{%
r^{3}},\;\;\;\;\mathcal{U}_{4}=\frac{d_{2}d_{3}d_{4}d_{5}c^{4}}{r^{4}},
\end{equation}%
where $d_{i}=d-i$. We use the gauge potential ansatz $A_{\mu }=h(r)\delta
_{\mu }^{0}$ in Maxwell equation (\ref{Maxwell equation}) to obtain electric
field. Considering the metric (\ref{Metric}), one can obtain
\begin{equation}
h(r)=-\sqrt{\frac{d_{2}}{2d_{3}}}\frac{q}{r^{d_{3}}}\ \mathcal{H},
\end{equation}%
where $\mathcal{H}$ and $\Gamma $ are
\begin{eqnarray}
\mathcal{H} &=&{}_{2}F_{1}\left( \left[ \frac{1}{2},\frac{d_{3}}{2d_{2}}%
\right] ,\left[ \frac{3d_{7/3}}{2d_{2}}\right], -\eta \right) , \\
\eta &=&\frac{d_{2}d_{3}q^{2}}{2\beta ^{2}r^{2d_{2}}},
\end{eqnarray}%
in which $q$ is an integration constant related to the electric charge. In
addition, the Maxwell equation implies that the nonzero component of the
electromagnetic field tensor in $d$-dimensions is given by
\begin{equation}
F_{tr}=\frac{\sqrt{ \frac{d_{2}d_{3}}{2} }q}{r^{d_{2}}\sqrt{1+\eta }}.
\end{equation}

Now, we obtain the topological static GB-BI black hole solutions in massive
gravity. To do so, one may use any component of Eq. (\ref{Field equation})
and obtain metric function $f(r)$. It is a matter of calculation to show
that $rr$ and $tt$ are the same, whereas, the $x_{i}x_{i}$ components of Eq.
(\ref{Field equation}) are identical. Therefore, we write
\begin{eqnarray}
e_{tt} &=&2\beta ^{2}f\left( 1-\frac{1}{\Upsilon }\right) +\frac{f}{2r^{4}}%
\left\{ d_{2}d_{3}d_{4}\alpha f\left[ 2rf^{\prime }+d_{5}f\right]
-d_{2}d_{3}\kappa r^{2}\left[ \frac{\alpha d_{4}d_{5}\left( 2f-\kappa
\right) }{r^{2}}+\frac{2\alpha d_{4}f^{\prime }}{r}-1\right] \right.  \notag
\\
&&\left. -2\Lambda r^{4}-d_{2}\left[ d_{3}f+rf^{\prime }\right] r^{2}+m^{2}%
\left[ r^{2}d_{2}\left( cc_{1}r+d_{3}c^{2}c_{2}\right)
+d_{2}d_{3}d_{4}\left( c^{3}c_{3}r+d_{5}c^{4}c_{4}\right) \right] \right\} ,
\label{tteq} \\
&&  \notag \\
e_{x_{1}x_{1}} &=&2r^{2}\beta ^{2}\left[ \Upsilon -1\right] -\frac{1}{2r^{2}}%
\left\{ d_{3}d_{4}d_{5}\alpha f\left[ 4rf^{\prime }+d_{6}f\right]
-r^{4}\left( 2\Lambda +f^{\prime \prime }\right) +d_{3}d_{4}\left[ 2\alpha
\left( f^{\prime 2}+ff^{\prime \prime }\right) -f-\frac{2rf^{\prime }}{d_{4}}%
\right] r^{2}\right.  \notag \\
&&\left. -d_{3}d_{4}\kappa r^{2}\left[ \frac{\alpha d_{5}d_{6}\left(
2f-\kappa \right) }{r^{2}}+\frac{4\alpha d_{5}f^{\prime }}{r}+2\alpha
f^{\prime \prime }-1\right] +d_{3}m^{2}c\left[ \left(
c_{1}r+d_{4}cc_{2}\right) r^{2}+d_{4}d_{5}c^{2}\left(
c_{3}r+d_{6}cc_{4}\right) \right] \right\} ,  \label{x1x1eq}
\end{eqnarray}%
where $f$ and $h$ are functions of $r$ and also $\Upsilon =\sqrt{1-\left(
\frac{h^{\prime }}{\beta }\right) ^{2}}$. Using Eqs. (\ref{tteq}) and (\ref%
{x1x1eq}), we can obtain the metric function as
\begin{eqnarray}
f\left( r\right) &=&\kappa +\frac{r^{2}}{2\alpha d_{3}d_{4}}\left\{ 1-\sqrt{%
1+\frac{8\alpha d_{3}d_{4}}{d_{1}d_{2}}\left[ \Lambda +\frac{d_{1}d_{2}m_{0}%
}{2r^{d_{1}}}+\mathcal{A}+\mathcal{B}\right] }\right\} ,  \label{f(r)} \\
\mathcal{A} &=&-2\beta ^{2}\left( 1-\sqrt{1+\eta }\right) -\frac{d_{2}^{2}
q^{2}}{r^{2d_{2}}}\mathcal{H},  \notag \\
\mathcal{B} &=&-m^{2}d_{1} d_{2} \left[ \frac{d_{3}d_{4}c^{4}c_{4}}{2r^{4}}+%
\frac{d_{3}c^{3}c_{3}}{2r^{3}}+\frac{c^{2}c_{2}}{2r^{2}}+\frac{cc_{1}}{%
2d_{2}r}\right] ,  \notag
\end{eqnarray}%
where $m_{0}$ is an integration constant which is related to the total mass
of the black hole. It is notable that, the obtained metric function (\ref%
{f(r)}), satisfies all the components of Eq. (\ref{Field equation}).

In order to study the geometrical structure of this solution (\ref{f(r)}),
we look for the essential singularity(ies). For this purpose, we calculate
the Ricci and Kretschmann scalars and obtain following results
\begin{eqnarray}
\lim_{r\longrightarrow 0}R &\longrightarrow &\infty , \\
\lim_{r\longrightarrow 0}R_{\alpha \beta \gamma \delta }R^{\alpha \beta
\gamma \delta } &\longrightarrow &\infty ,
\end{eqnarray}%
the above results confirm that there is a curvature singularity at $r=0$. On
the other hand, in order to investigate the asymptotical behavior of the
solutions, we find the curvature scalars at $r\longrightarrow \infty $. So,
we have
\begin{eqnarray}
\lim_{r\longrightarrow \infty }R &\longrightarrow &\frac{dd_{1}\left(\sqrt{%
d_{2}^{2}+\frac{8d_{2}d_{3}d_{4}}{d_{1}}\alpha \Lambda }-1\right)}{2\alpha
d_{2}d_{3}d_{4}}, \\
\lim_{r\longrightarrow \infty }R_{\alpha \beta \gamma \delta }R^{\alpha
\beta \gamma \delta } &\longrightarrow &\frac{1}{d_{3}d_{4}\alpha }\left[
\frac{4\Lambda}{d_{2}}+\frac{d_{1}}{d_{3}d_{4}\alpha } \left( 1-d \sqrt{1+%
\frac{8 d_{3}d_{4}}{d_{1}d_{2}}\alpha \Lambda}\right) \right].
\end{eqnarray}

These results confirm that, the asymptotical behavior of the solutions are
(a)dS with an effective cosmological constant ($\Lambda_{eff}$). This
effective cosmological constant reduces to ordinary $\Lambda$ for vanishing $%
\alpha$. In other words, neither massive nor BI parts affect the
asymptotical behavior of the solutions.

It is worthwhile to mention that, in the absence of massive parameter ($m=0$%
), the solution (\ref{f(r)}) reduces to $d$-dimensional asymptotically adS
topological black hole solution which was found in Ref. \cite{GBBornInfeld}.
In addition, for $\beta \longrightarrow \infty$, obtained solution reduces
to GB-massive solution with Maxwell field \cite{HendiPEGBmass}. Moreover,
for vanishing $\alpha$ the solution of Einstein-BI-massive gravity may be
recovered \cite{HendiEPEN-BImass}.

\begin{figure}[tbp]
\epsfxsize=7cm \centerline{\epsffile{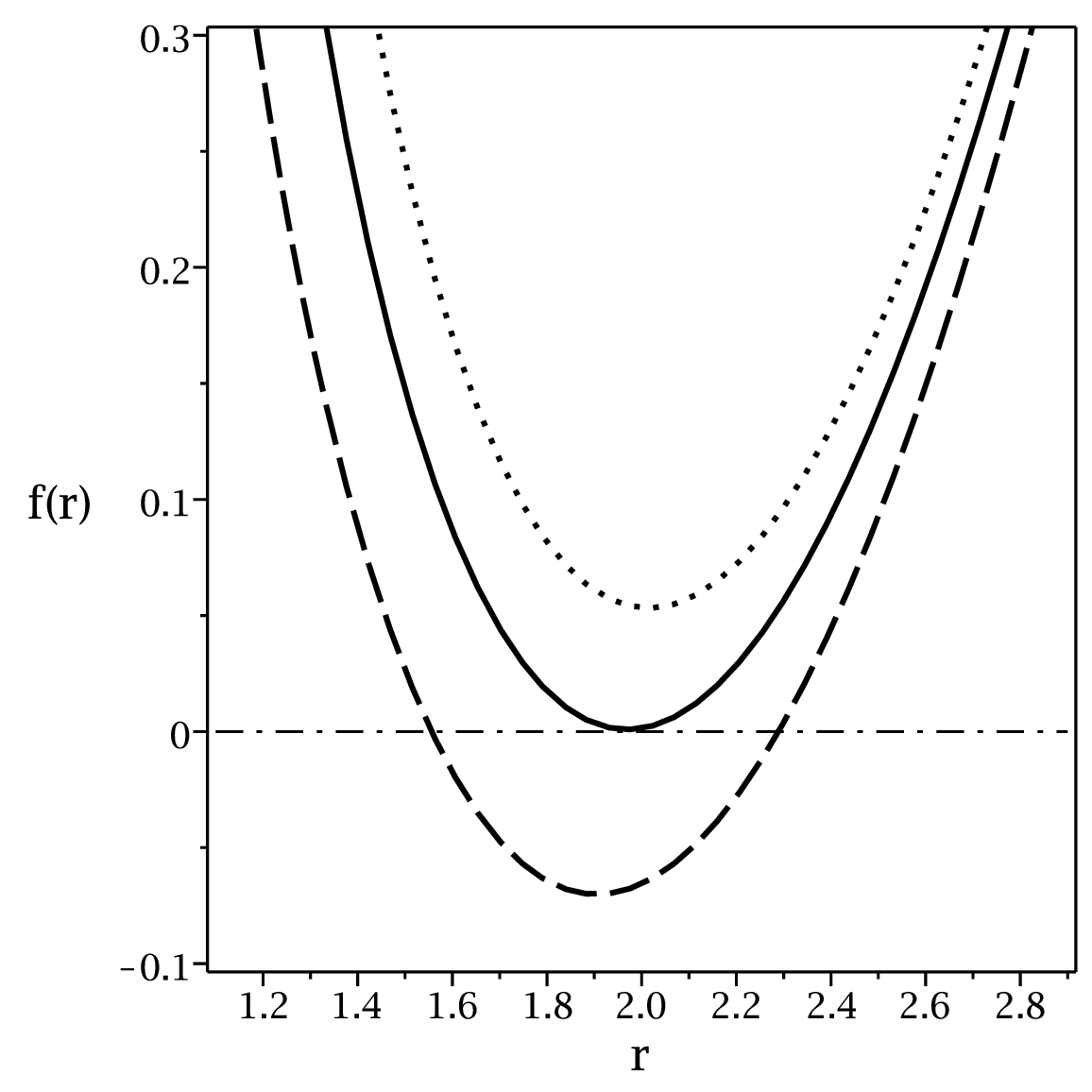}}
\caption{$f(r)$ versus $r$ for $\Lambda=-1$, $q=1$, $\protect\beta=0.9$, $%
\protect\alpha=0.4$, $c=1 $, $c_{1}=0.3$, $c_{2}=1$, $c_{3}=-4$, $c_{4}=2$, $%
m=1$, $\protect\kappa=1$ and $d=5$; $m_{0}=2.5$ (dashed line), $m_{0}=1.3$
(continues line) and $m_{0}=0.4$ (dotted line).}
\label{Figfr1}
\end{figure}

\begin{figure}[tbp]
\epsfxsize=7cm \centerline{\epsffile{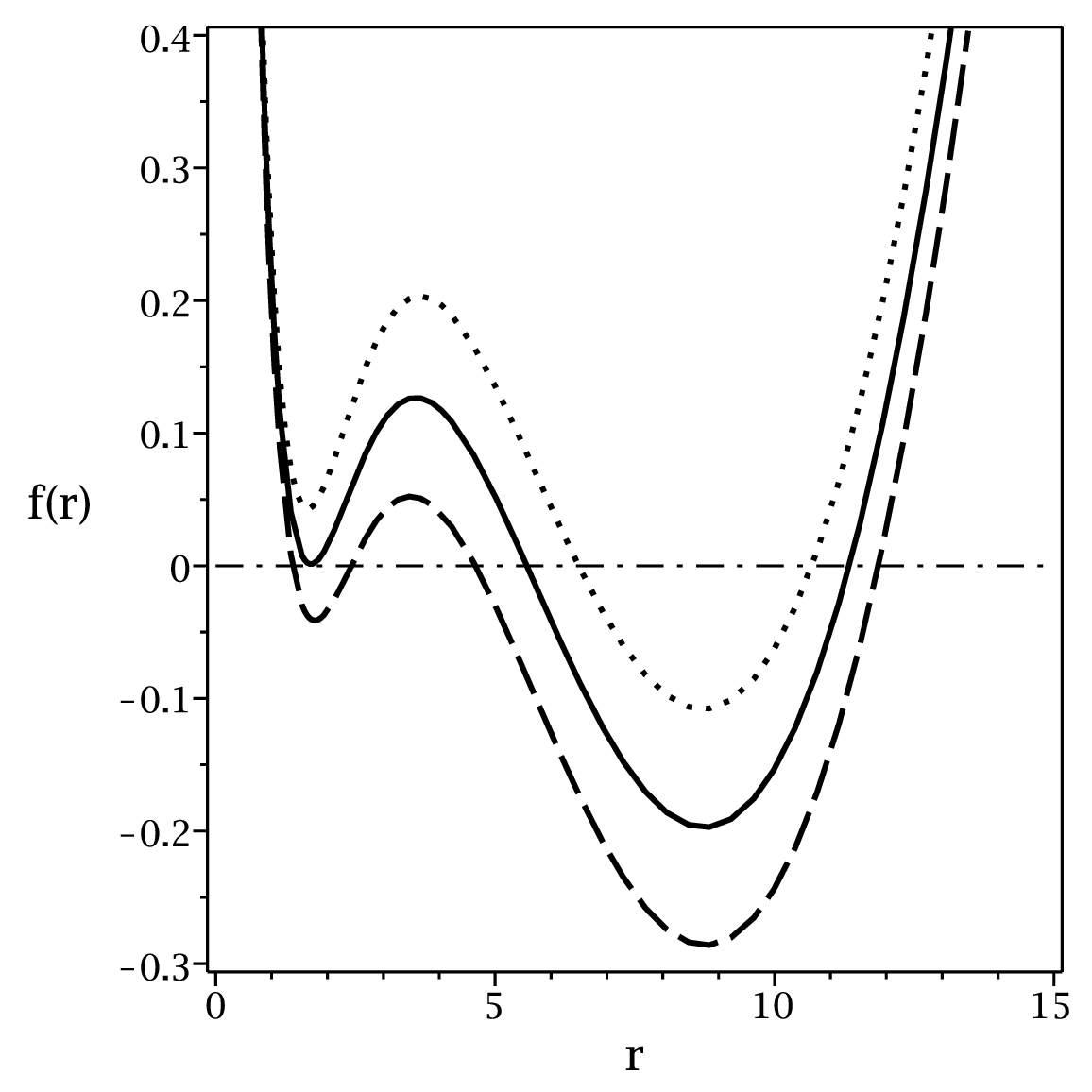}}
\caption{$f(r)$ versus $r$ for $\Lambda=-0.4$, $q=1$, $\protect\beta=0.6$, $%
\protect\alpha=0.3$, $m_{0}=3.1$, $c=0.8$, $c_{1}=-2.1$, $c_{3}=-4$, $%
c_{4}=1.9$, $m=1.4$, $\protect\kappa=1$ and $d=5$; $c_{2}=3.250$ (dashed
line), $c_{2}=3.325$ (continues line) and $c_{2}=3.4$ (dotted line).}
\label{Figfr2}
\end{figure}

In order to study the effects of the GB-BI-massive gravity on metric
function, we have plotted the diagrams related to this solution in Figs. \ref%
{Figfr1} and \ref{Figfr2}. The GB-BI-massive black holes may behave like
Reissner-Nordstr\"{o}m black holes. In other words, these black holes may
contain two horizons (inner and outer horizons), one extreme horizon or
without horizon (naked singularity) (see Fig. \ref{Figfr1} for more
details). On the other hand, by adjusting some of the parameters, we may
encounter with interesting behavior in which more than two horizons are
observed (Figs. \ref{Figfr2}). The existence of three and four horizons for
black holes is due to the presence of massive part of gravity \cite%
{HendiPEM,HendiPEGBmass,HendiEPEN-BImass}. Although multiple
horizon solutions may be applied to find some algebraic
mathematical relations for the horizons (see multiple horizon
relations in \cite{Multiple}), all thermodynamical analysis of the
black holes should be calculated at $r=r_{+}$ which is the
outermost horizon of the solutions. Speaking more precisely, the
outermost horizon of the black holes can be event horizon
($r_{+}$) or cosmological horizon ($r_{c}$, which c stands for
cosmological). It is notable that event horizon satisfies $\left.
\mathbf{\partial }_{r}g_{tt}\right\vert _{r=r_{+}}>0$
($g_{tt}=g^{rr}$) condition and for $r>r_{+}$, the metric function
is real and positive valued. On the contrary, for cosmological
horizon, $\left. \mathbf{\partial }_{r}g_{tt}\right\vert
_{r=rc}<0$ is satisfied and in case of $r>r_{c}$, the metric
function is negative. Regardless of the number of the horizons,
the outermost horizon (event horizon) of the black holes with
mentioned conditions should be employed to study thermodynamical
properties of the black holes. The existence of multiple horizon
solutions has been reported for other black holes in massive
gravity and $F(R)$ gravity as well. It was shown that the
existence of the multiple horizon solutions provides the
possibility of the anti-evaporation property for the black holes
\cite{anti}. Meaning that these specific configurations for the
number of horizons introduce a new phenomena which acts in
opposite of the evaporation of the black holes (see Ref.
\cite{anti} for more details).


\section{Thermodynamics}

In this section, we study thermodynamic properties of GB-BI-massive black
holes. In order to examine the first law, we should calculate the conserved
and thermodynamics quantities of the solutions in $d$-dimensions.

The Hawking temperature of these black holes can be obtained through the
definition of surface gravity
\begin{eqnarray}
&&T =\frac{\vartheta }{2\pi }=\frac{\sqrt{-\frac{1}{2}\left( \nabla _{\mu
}\chi _{\nu }\right) \left( \nabla ^{\mu }\chi ^{\nu }\right) }}{2\pi }=%
\frac{f^{\prime }\left( r_{+}\right) }{4\pi }=  \notag \\
&&\frac{1}{4\pi \mathcal{N}}\left[ \frac{m^{2}\left[ d_{3}d_{4}\left(
c^{3}c_{3}r_{+}+d_{5}c^{4}c_{4}\right) +r_{+}^{2}\left(
cc_{1}r_{+}+d_{3}c^{2}c_{2}\right) \right] }{r_{+}}+\frac{2r_{+}^{3}\left(
2\beta ^{2}-\Lambda \right) }{d_{2}}-\frac{4\beta ^{2}r_{+}^{3}}{%
d_{2}\Upsilon _{+}}+\frac{\kappa d_{3}\left( r_{+}^{2}+\alpha \kappa
d_{4}d_{5}\right) }{r_{+}}\right],  \label{TotalTT}
\end{eqnarray}%
where $\vartheta$ is the surface gravity, $\chi=\partial_{t}$ denotes the
(timelike) Killing vector, $\mathcal{N}=2\alpha \kappa d_{3}d_{4}+r_{+}^{2}$
and $\Upsilon _{+}=\Upsilon \left\vert _{r=r_{+}}\right. $.

Besides, using the Gauss's law, one can find the total electric charge of
the black holes
\begin{equation}
Q=\frac{V_{d_{2}}\ \sqrt{d_{2}d_{3}}}{4\pi }q.  \label{TotalQ}
\end{equation}

We obtain the electric potential at the horizon with respect to spacial
infinity as reference
\begin{equation}
U=A_{\mu }\chi ^{\mu }\left\vert _{r\rightarrow \infty }\right. -A_{\mu
}\chi ^{\mu }\left\vert _{r\rightarrow r_{+}}\right. =\sqrt{\frac{d_{2}}{%
2d_{3}}}\frac{q}{r_{+}^{d_{3}}}\ \mathcal{H}_{+},  \label{TotalU}
\end{equation}%
where $\mathcal{H}_{+}=\mathcal{H}\left\vert _{r=r_{+}}\right. $. The
calculation of the entropy of black holes depends on the gravity under
consideration. Regarding Einstein gravity, it was shown that the entropy of
black holes satisfies the so-called area law which states that the black
hole entropy is equal to one-quarter of horizon area \cite{AREA}. However,
it is not possible to use the area law for higher derivative gravity models
\cite{AreaViolate,Wald}. Depending on the asymptotical behavior of the
solutions, the entropy of higher derivative gravity theories can be obtained
from Wald formula or Gibbs-Duhem relation \cite%
{GBBornInfeld,Wald,blackholesGB}. It is straightforward to find that
\begin{equation}
S=\frac{V_{d_{2}}}{4}r_{+}^{d_{2}}\left( 1+\frac{2d_{2}d_{3}}{r_{+}^{2}}%
\kappa \alpha \right) ,  \label{TotalS}
\end{equation}%
which shows that area law is violated for GB-BI-massive black holes with
non-flat horizons.

In addition, for obtaining the total finite mass of the black holes, one can
use Hamiltonian approach which leads into following result
\begin{equation}
M=\frac{\ d_{2}\ V_{d_{2}}}{16\pi }m_{0}.  \label{TotalM}
\end{equation}

Having conserved and thermodynamic quantities, we are in a position to check
the first law of thermodynamics. For this purpose, we obtain the mass ($%
m_{0} $) of Eqs. (\ref{f(r)}) and (\ref{TotalM}), as a function of other
parameters
\begin{eqnarray}
M(r_{+},q) &=&\frac{\ d_{2}\ V_{d_{2}}}{16\pi }\left\{ d_{3}d_{4}\alpha
\kappa ^{2}r_{+}^{d_{5}}+\kappa r_{+}^{d_{3}}+\frac{2r_{+}^{d_{1}}}{%
d_{1}d_{2}}\left[ 2\beta ^{2}\left( 1-\sqrt{1+\eta _{+}}\right) -\Lambda %
\right] +\frac{2d_{2}q^{2}}{d_{1}r_{+}^{d_{3}}}\mathcal{H}_{+}\right.  \notag
\\
&&\left. +\frac{m^{2}r_{+}^{d_{5}}}{d_{2}}\left[ d_{2}d_{3}\left(
c^{3}c_{3}r_{+}+d_{4}c^{4}c_{4}\right) +r_{+}^{2}\left(
cc_{1}r_{+}+d_{2}c^{2}c_{2}\right) \right] \right\} ,  \label{Mm}
\end{eqnarray}%
where $\eta _{+}=\frac{d_{2}d_{3}q^{2}}{2\beta ^{2}r_{+}^{2d_{2}}}$. Using
Eq. (\ref{Mm}), and by defining the temperature and the electric potential
as
\begin{eqnarray}
T&=&\left( \frac{\partial M}{\partial r_{+}}\right) _{Q}\ \left( \frac{%
\partial r_{+}}{\partial S}\right) _{Q}  \label{TU1} \\
U&=&\left( \frac{\partial M}{\partial Q}\right) _{S},  \label{TU2}
\end{eqnarray}%
one finds that Eqs. (\ref{TU1}) and (\ref{TU2}) coincide with Eqs. (\ref%
{TotalTT}) and (\ref{TotalU}), and therefore, we find that these
thermodynamics quantities satisfy classical form of the first law of black
hole thermodynamics
\begin{equation}
dM=TdS+UdQ.
\end{equation}

\section{Heat Capacity and Stability in Canonical Ensemble}

Considering obtained conserved and thermodynamic quantities, we are in a
position to study thermal stability of the solutions. There are different
approaches for studying thermal stability which are in the context of
canonical and grand canonical ensembles. In grand canonical ensembles, by
employing the mass of the black holes as a thermodynamical potential and its
corresponding extensive parameters, one can build up the Hessian matrix. The
stability is investigated by studying the behavior of the determinant of
this Hessian matrix. The canonical ensemble approach is based on behavior of
the heat capacity. In this paper, we will investigate the stability
conditions in canonical ensemble. One can use following relation for
calculating the heat capacity
\begin{equation}
C_{Q}=T \left( \frac{\partial T}{\partial S}\right)^{-1} _{Q}.  \label{CQ}
\end{equation}

Using Eq. (\ref{TotalTT}), we obtain $\left( \frac{\partial T}{\partial S}%
\right) _{Q}$ in the following form
\begin{eqnarray}
\left( \frac{\partial T}{\partial S}\right) _{Q} &=&\frac{\kappa d_{3}\left(
\mathcal{N}-2r_{+}^{2}\right) }{\pi d_{2}\mathcal{N}^{3}r_{+}^{d_{5}}}+\frac{%
\left( 3\mathcal{N}-2r_{+}^{2}\right) }{\pi d_{2}^{2}\mathcal{N}%
^{3}r_{+}^{d_{7}}}\left( \frac{4\beta ^{2}\left( \Upsilon _{+}-1\right) }{%
\Upsilon _{+}}-2\Lambda \right) -\frac{4h^{\prime }h^{\prime \prime }}{\pi
d_{2}^{2}\mathcal{N}^{2}r_{+}^{d_{8}}\Upsilon _{+}^{3}} \\
&&-\frac{\alpha \kappa ^{2}d_{3}d_{4}d_{5}\left( \mathcal{N}%
+2r_{+}^{2}\right) }{\pi d_{2}\mathcal{N}^{3}r_{+}^{d_{3}}}-\frac{2m^{2}%
\mathcal{E}}{\pi d_{2}\mathcal{N}^{3}r_{+}^{d_{3}}}.
\end{eqnarray}%
where $\mathcal{E}$ is
\begin{equation}
\mathcal{E}=d_{3}d_{4}\left[ d_{5}c^{4}c_{4}\left( r_{+}^{2}+\frac{\mathcal{N%
}}{2}\right) +c^{3}c_{3}r_{+}\right] +r_{+}^{2}\left[ d_{3}c^{2}c_{2}\left(
r_{+}^{2}-\frac{\mathcal{N}}{2}\right) +cc_{1}r_{+}\left( r_{+}^{2}-\mathcal{%
N}\right) \right] ,
\end{equation}

\begin{figure}[tbp]
$%
\begin{array}{ccc}
\epsfxsize=5.5cm \epsffile{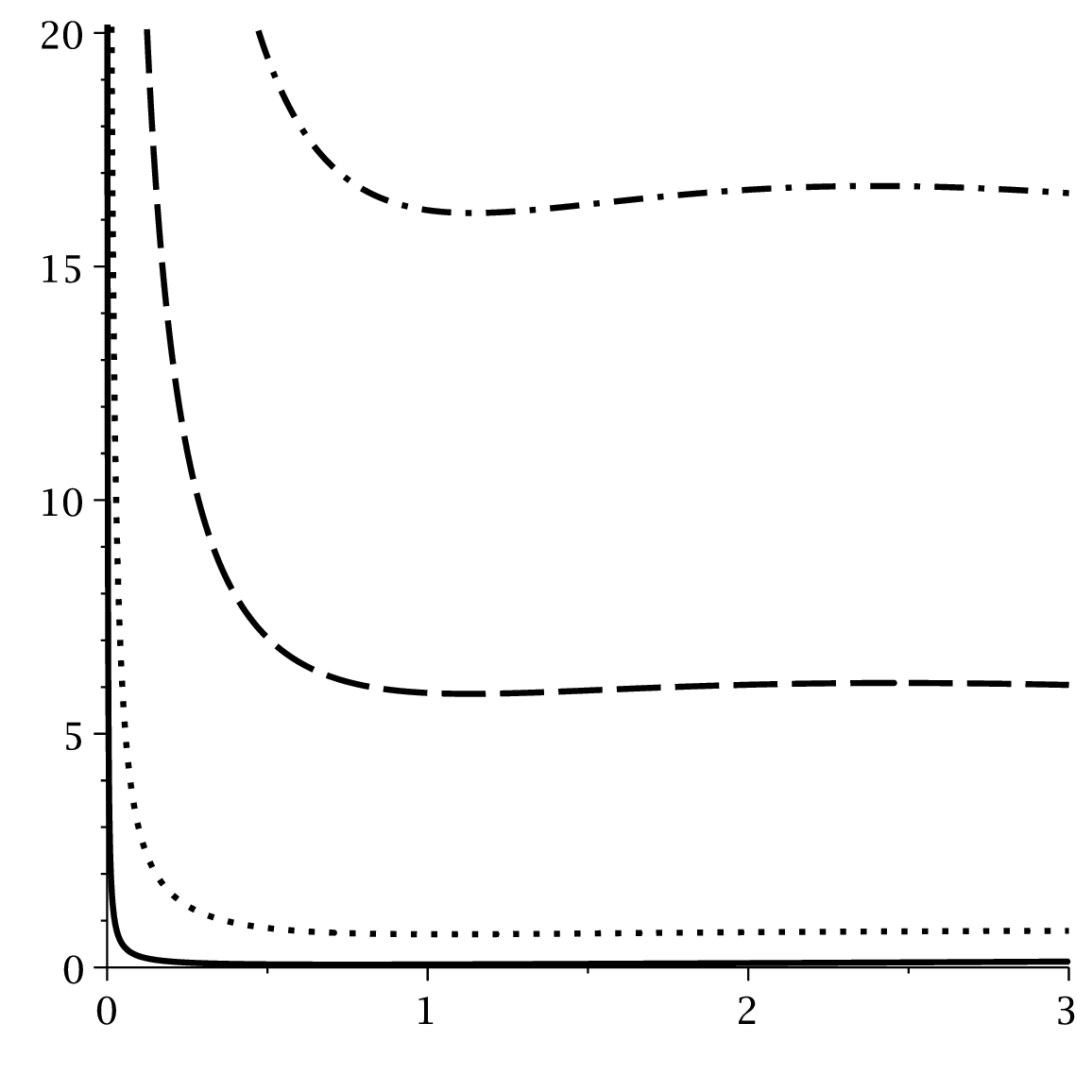} & \epsfxsize=5.5cm %
\epsffile{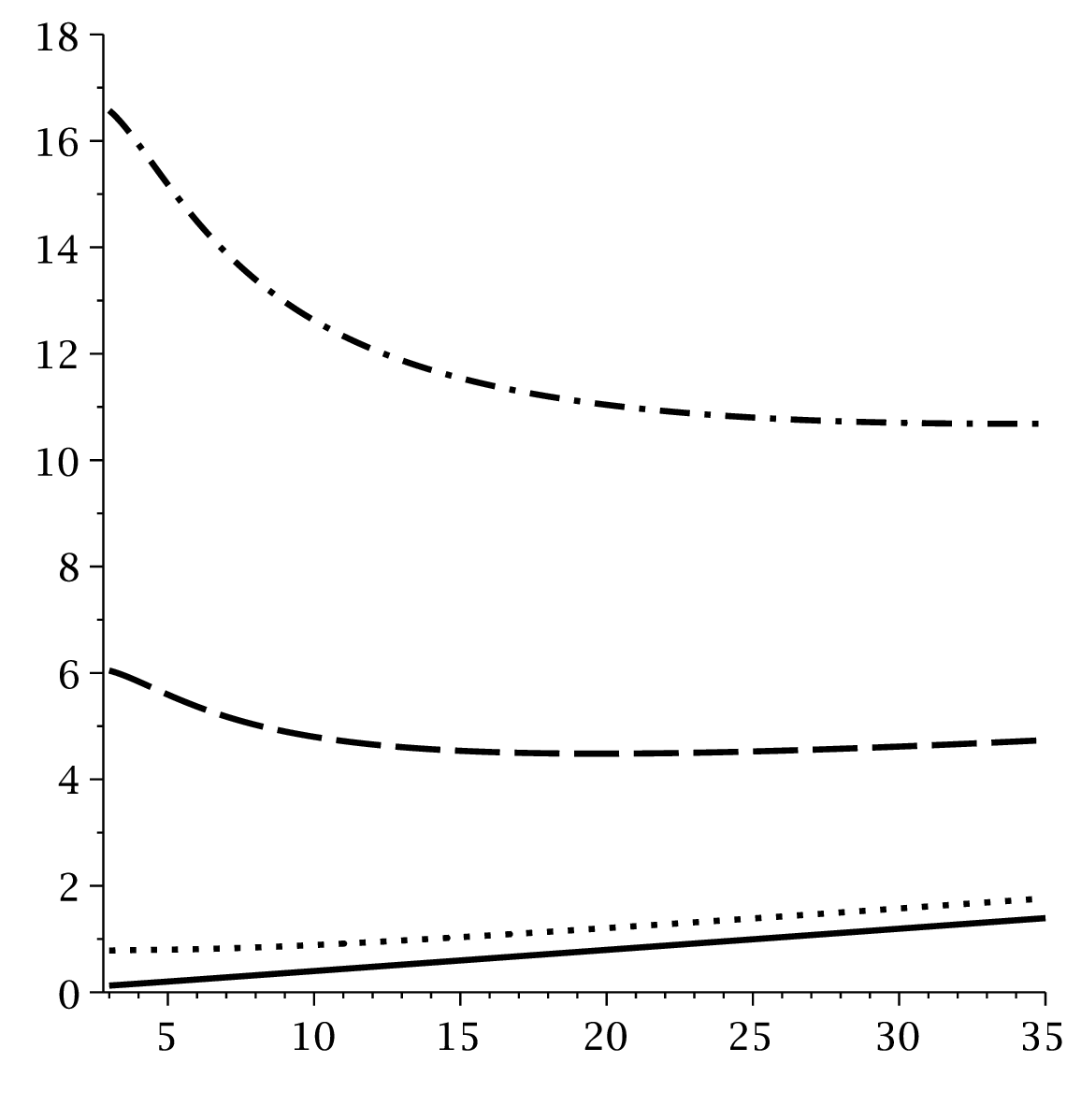} &  \\
\epsfxsize=5.5cm \epsffile{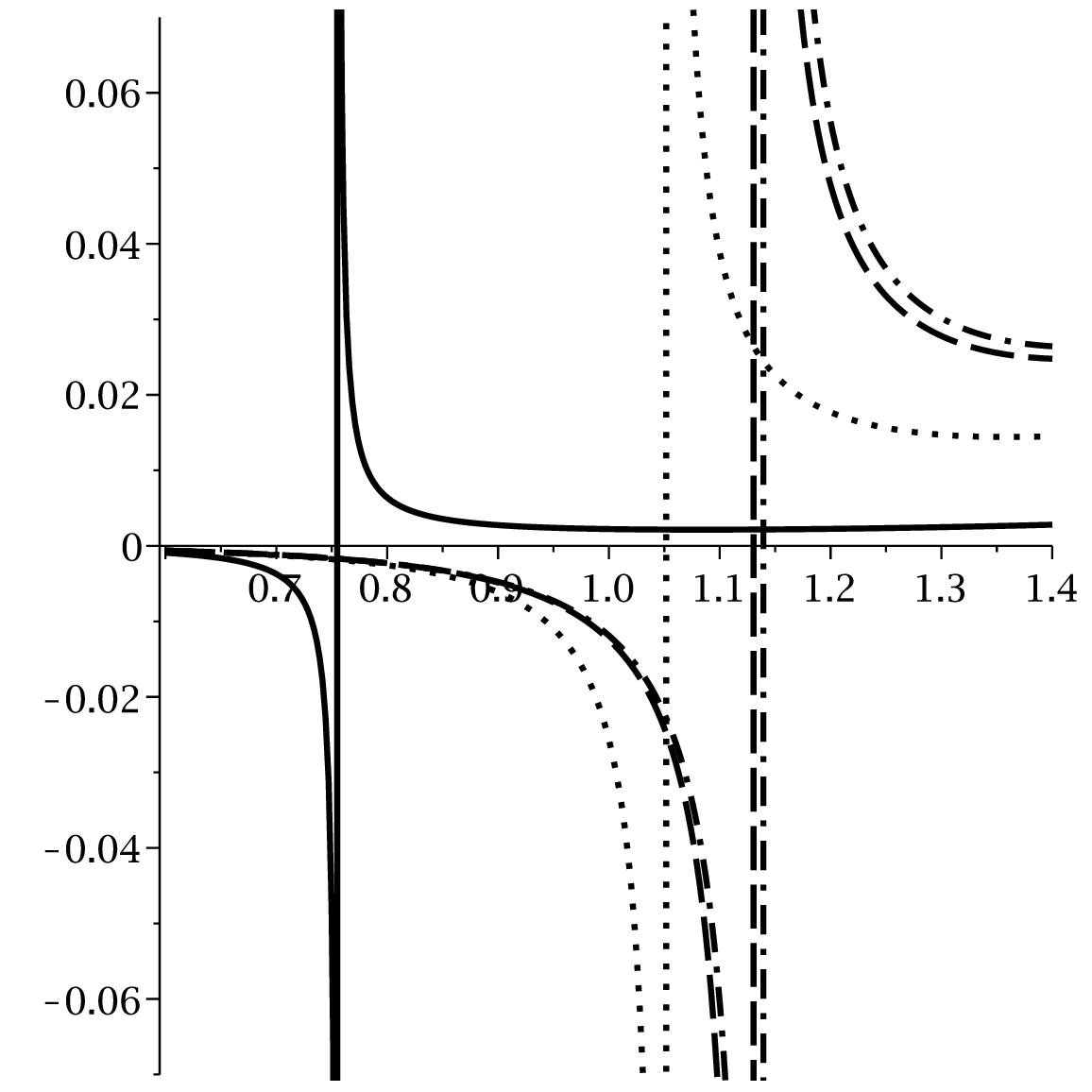} & \epsfxsize=5.5cm %
\epsffile{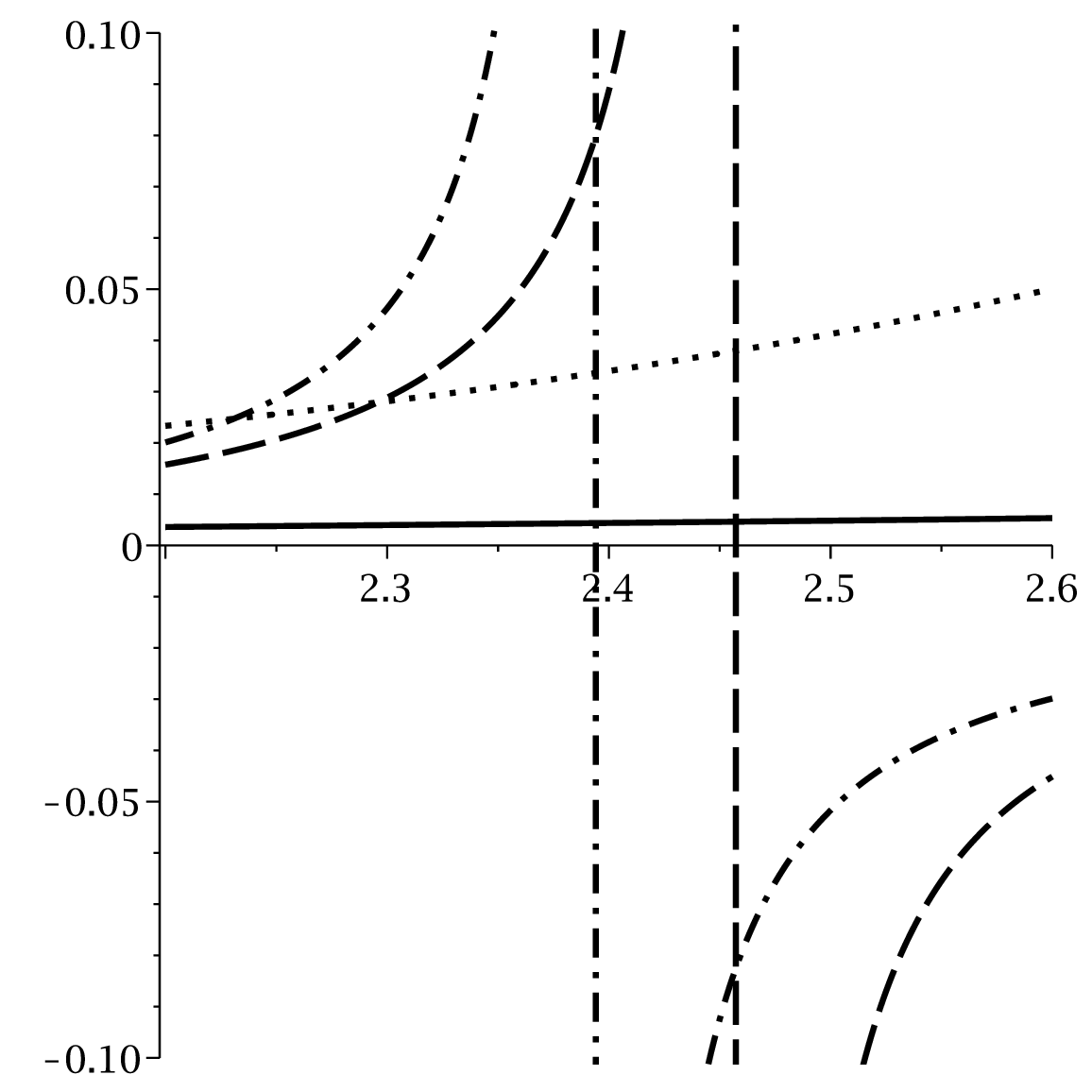} & \epsfxsize=5.5cm \epsffile{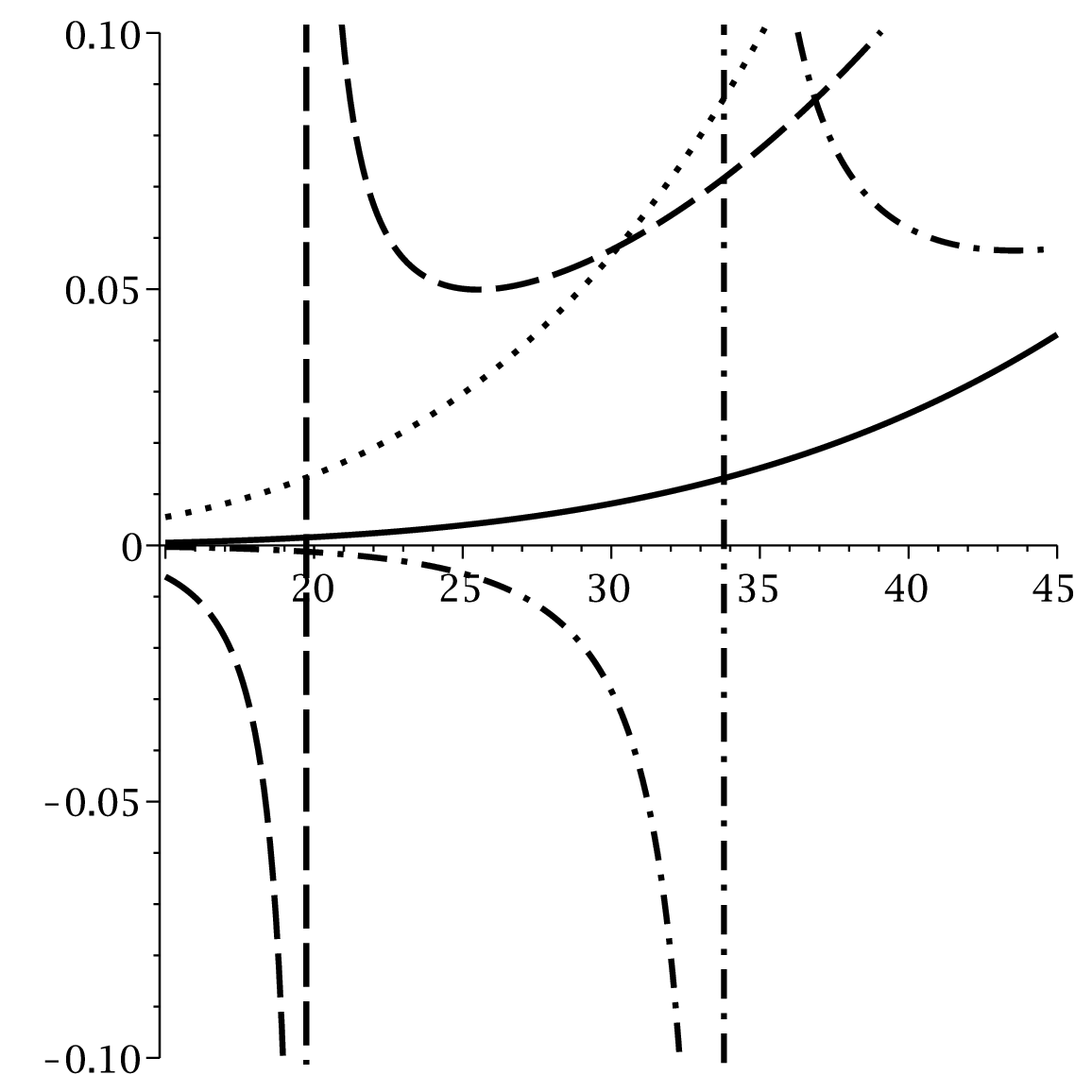}%
\end{array}
$%
\caption{For different scales: $C_{Q}$ (down panels) and $T$ (up panels)
versus $r_{+} $ for $q=1$, $\Lambda =-1$, $c=c_{1}=c_{2}=2$, $%
c_{3}=c_{4}=0.2 $, $\protect\beta=0.5$, $\protect\alpha=0.5$, $d=6$ and $%
\protect\kappa=1$; $m=0$ (continues line), $m=1$ (dotted line), $m=3$
(dashed line) and $m=5$ (dashes-dotted line).}
\label{Fig1}
\end{figure}

\begin{figure}[tbp]
$%
\begin{array}{ccc}
\epsfxsize=5.5cm \epsffile{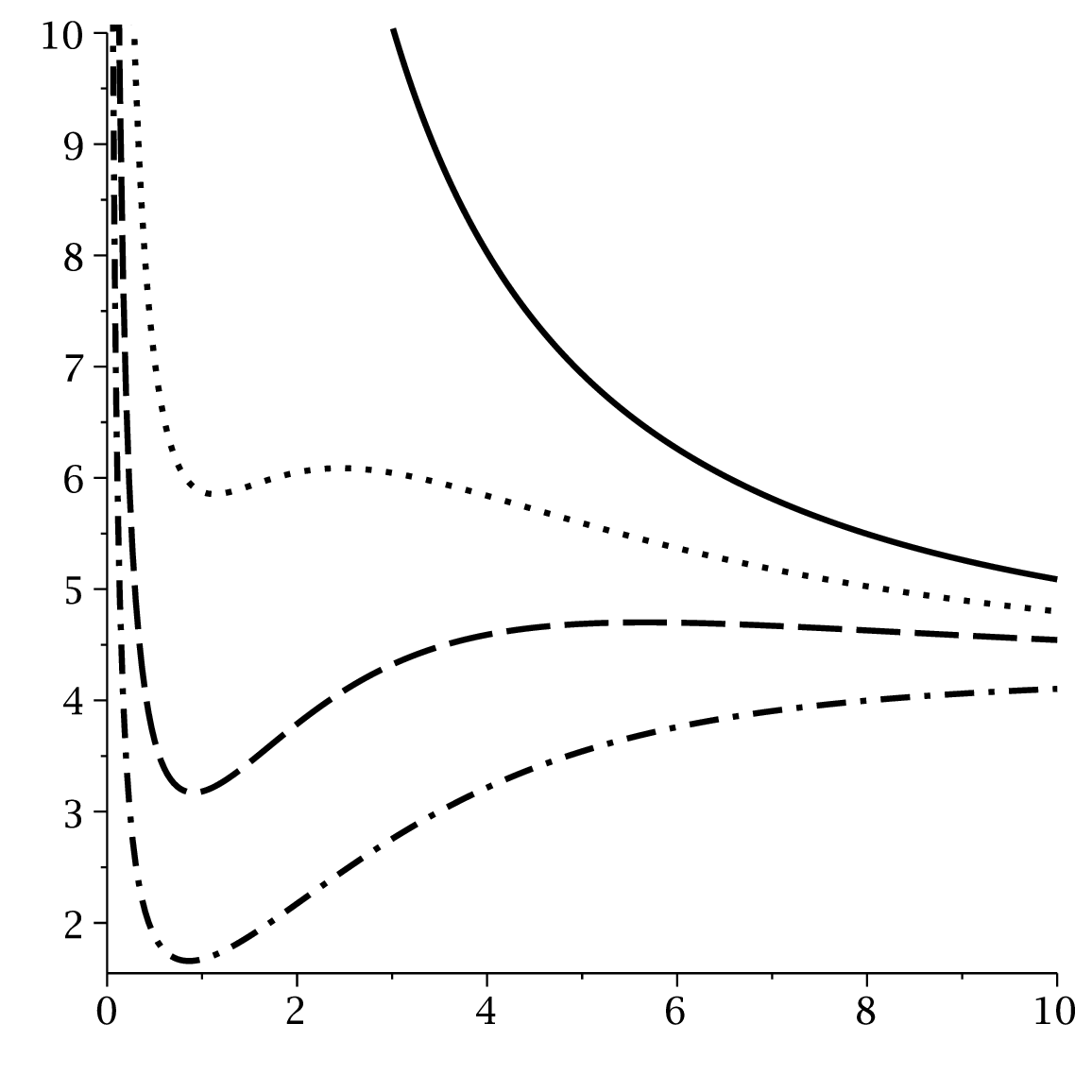} & \epsfxsize=5.5cm %
\epsffile{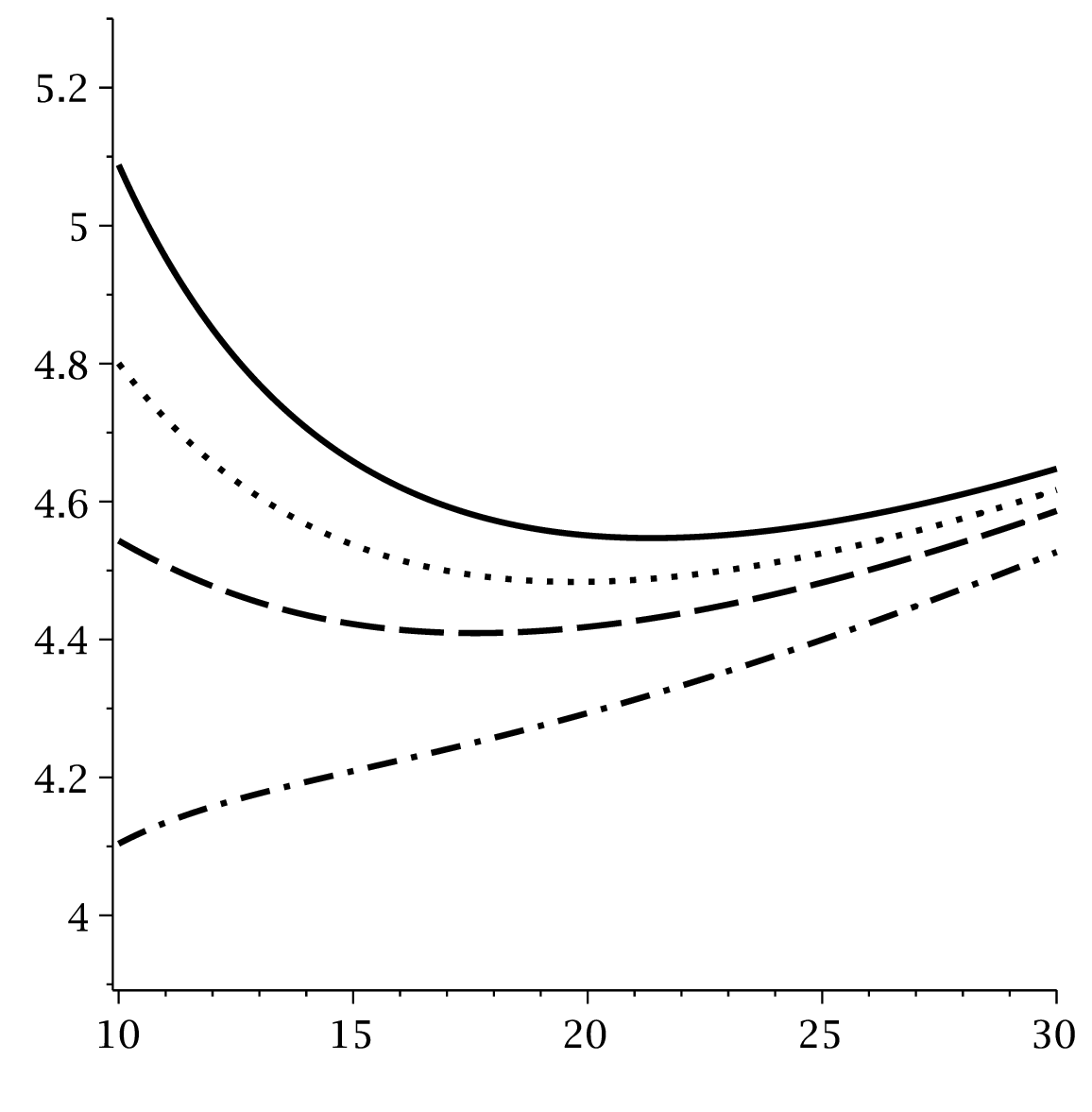} &  \\
\epsfxsize=5.5cm \epsffile{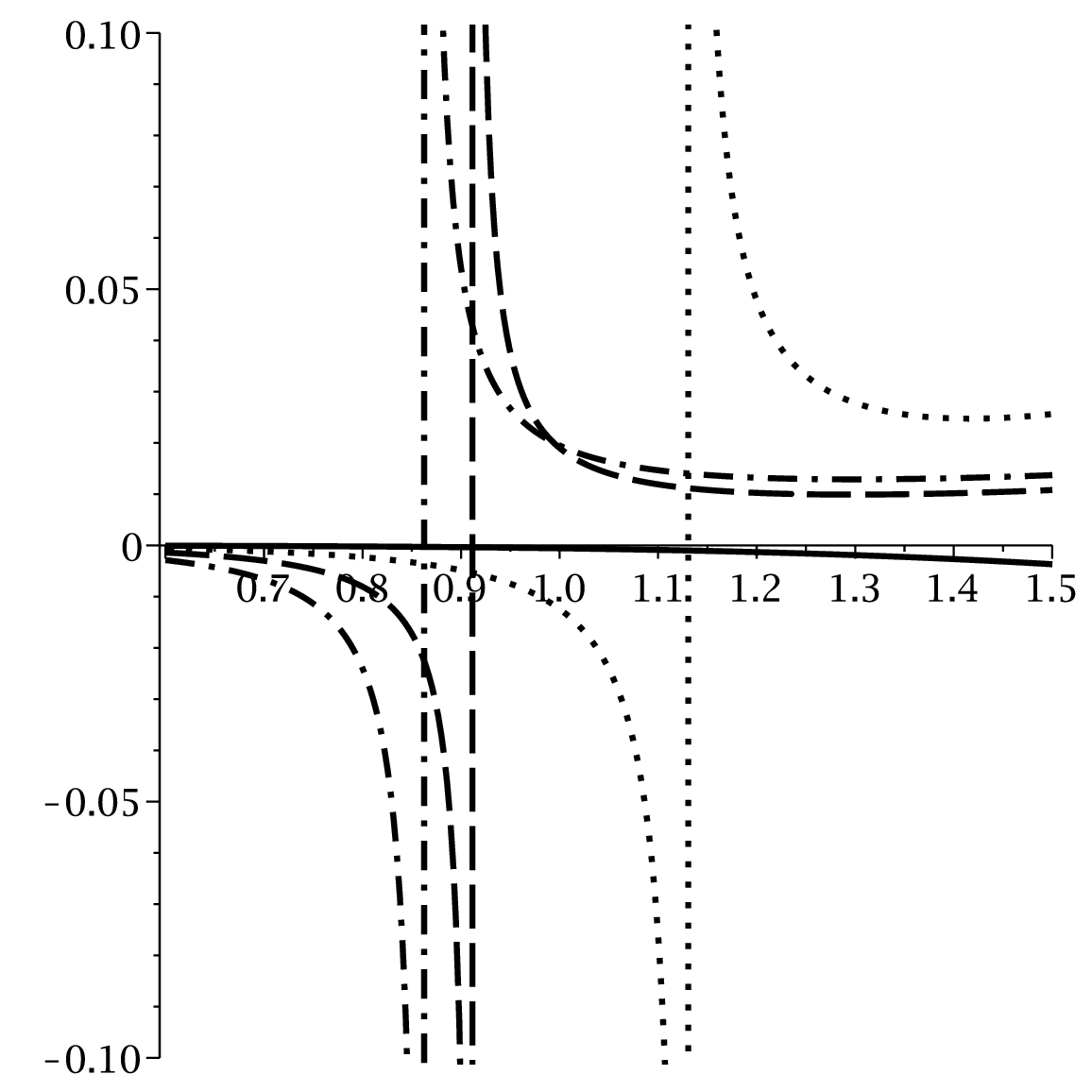} & \epsfxsize=5.5cm %
\epsffile{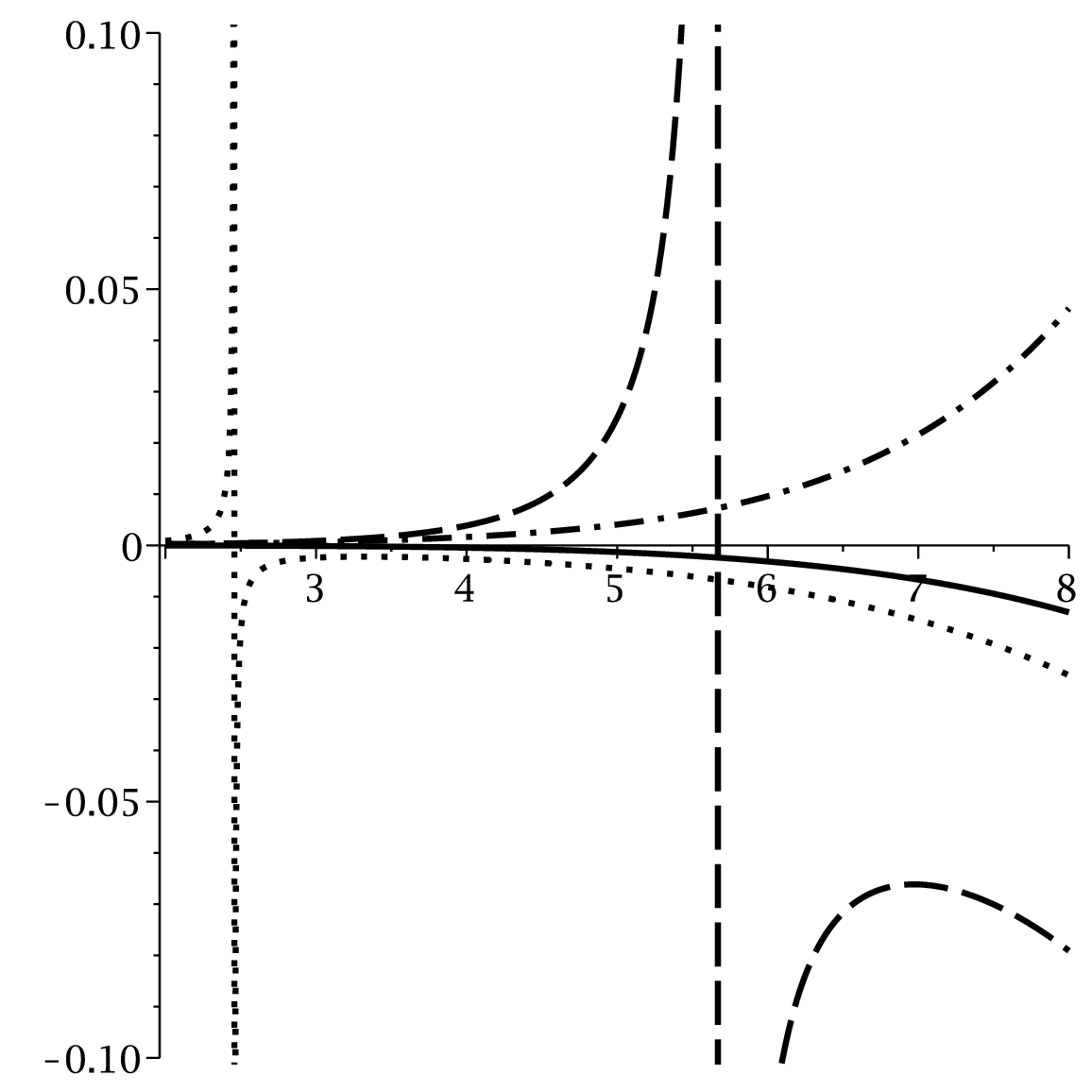} & \epsfxsize=5.5cm %
\epsffile{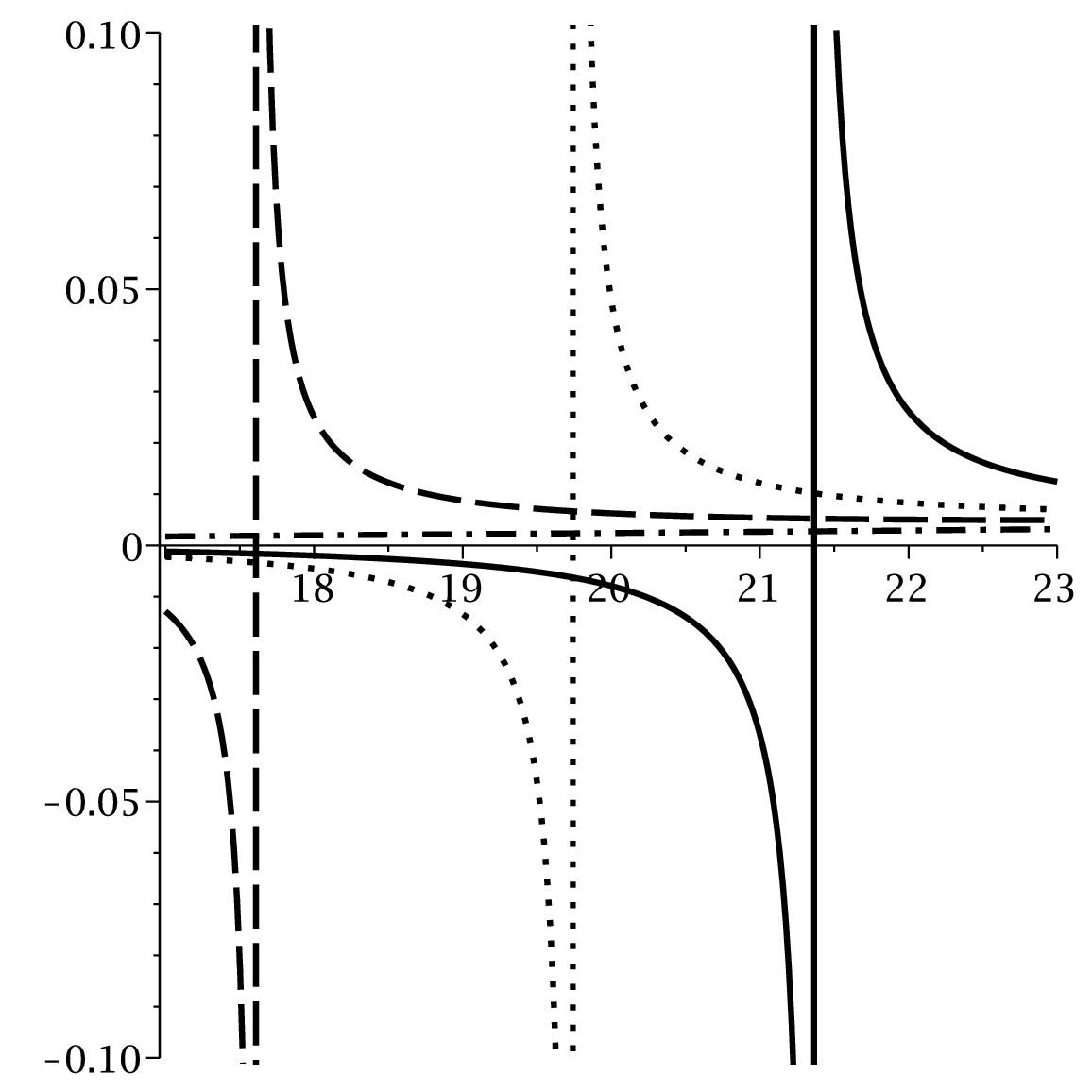}%
\end{array}
$%
\caption{For different scales: $C_{Q}$ (down panels) and $T$ (up panels)
versus $r_{+} $ for $q=1$, $\Lambda =-1$, $c=c_{1}=c_{2}=2$, $%
c_{3}=c_{4}=0.2 $, $m=3$, $\protect\beta=0.5$, $d=6$ and $\protect\kappa=1$;
$\protect\alpha=0$ (continues line), $\protect\alpha=0.5$ (dotted line), $%
\protect\alpha=1$ (dashed line) and $\protect\alpha=2$ (dashes-dotted line).}
\label{Fig2}
\end{figure}

\begin{figure}[tbp]
$%
\begin{array}{ccc}
\epsfxsize=5.5cm \epsffile{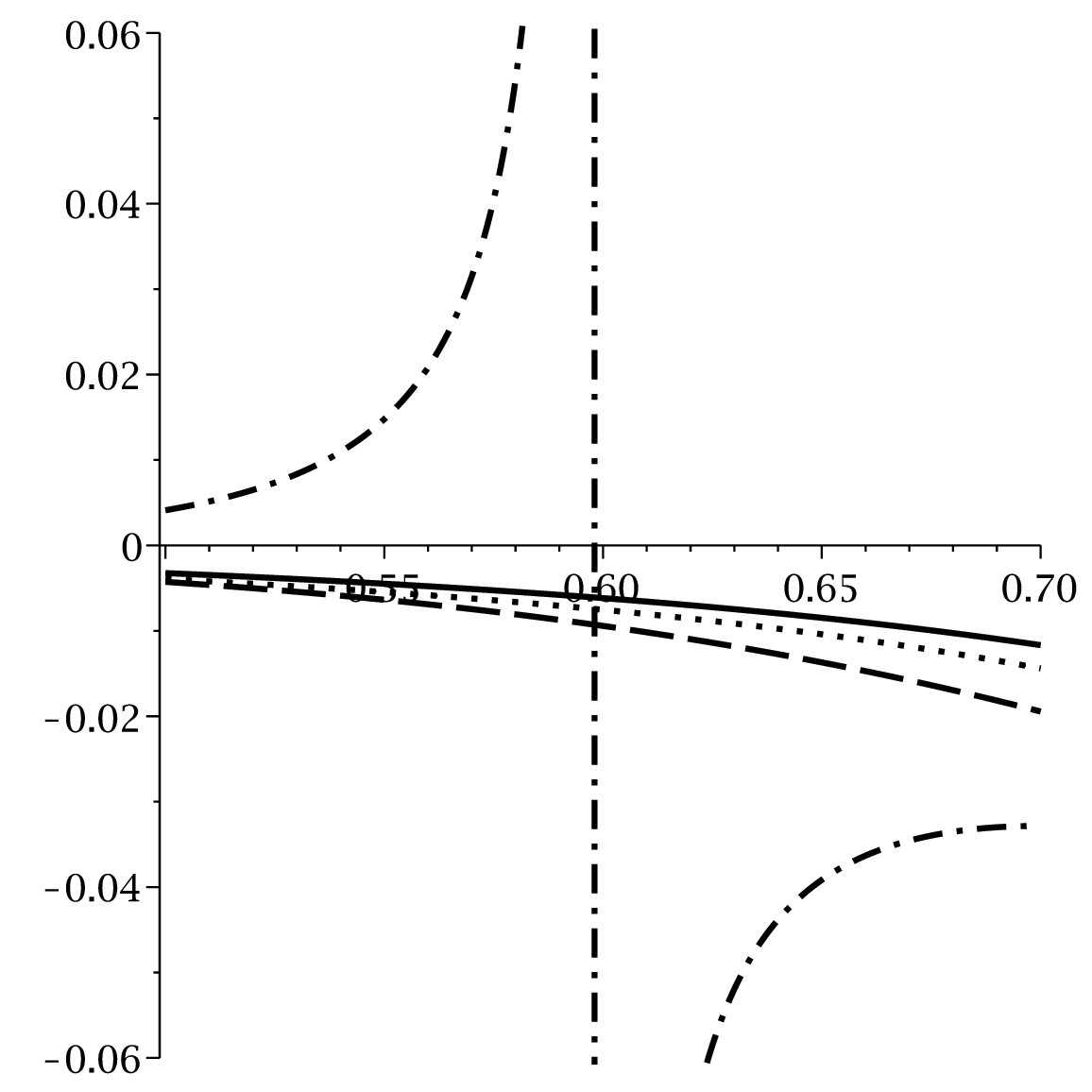} & \epsfxsize=5.5cm %
\epsffile{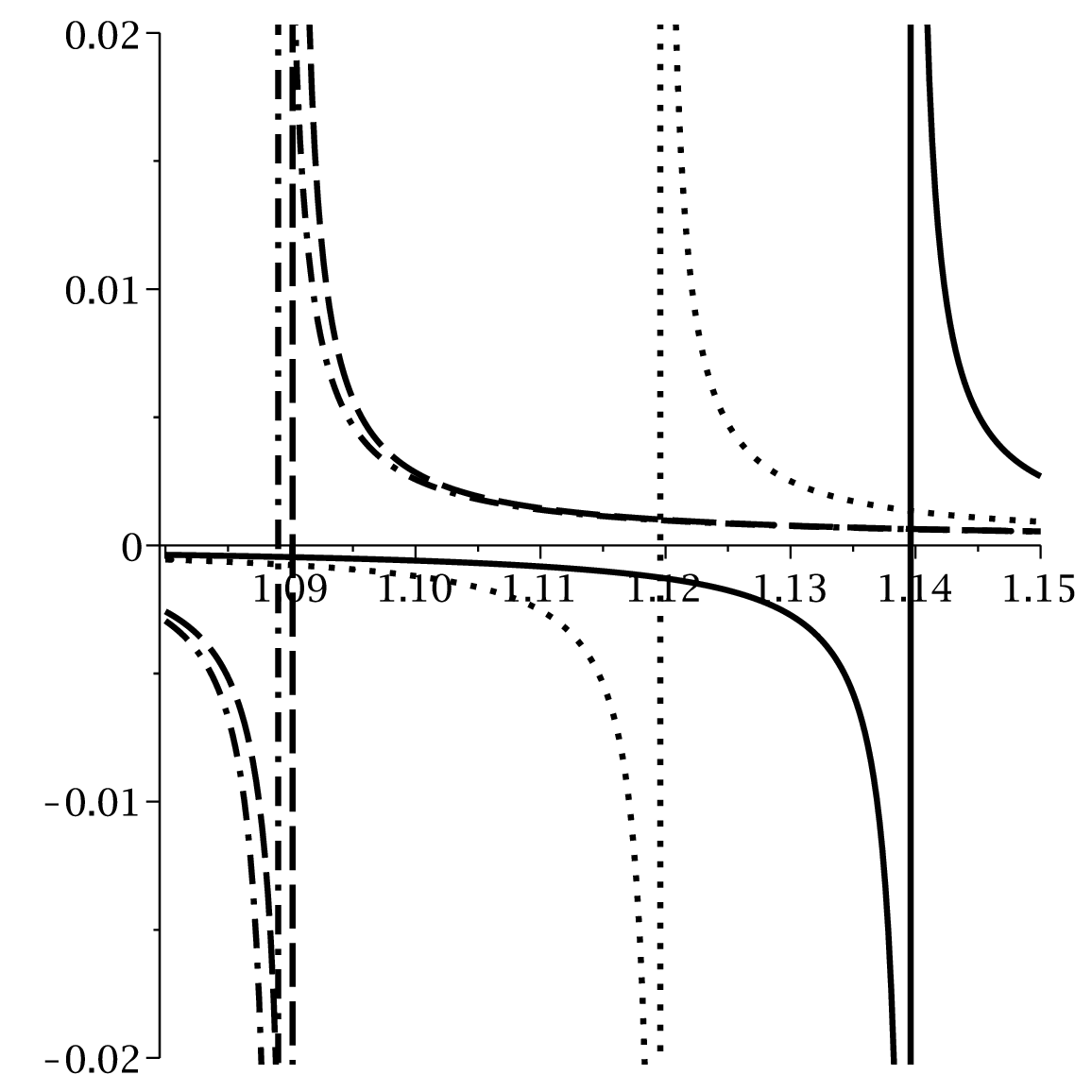} & \epsfxsize=5.5cm \epsffile{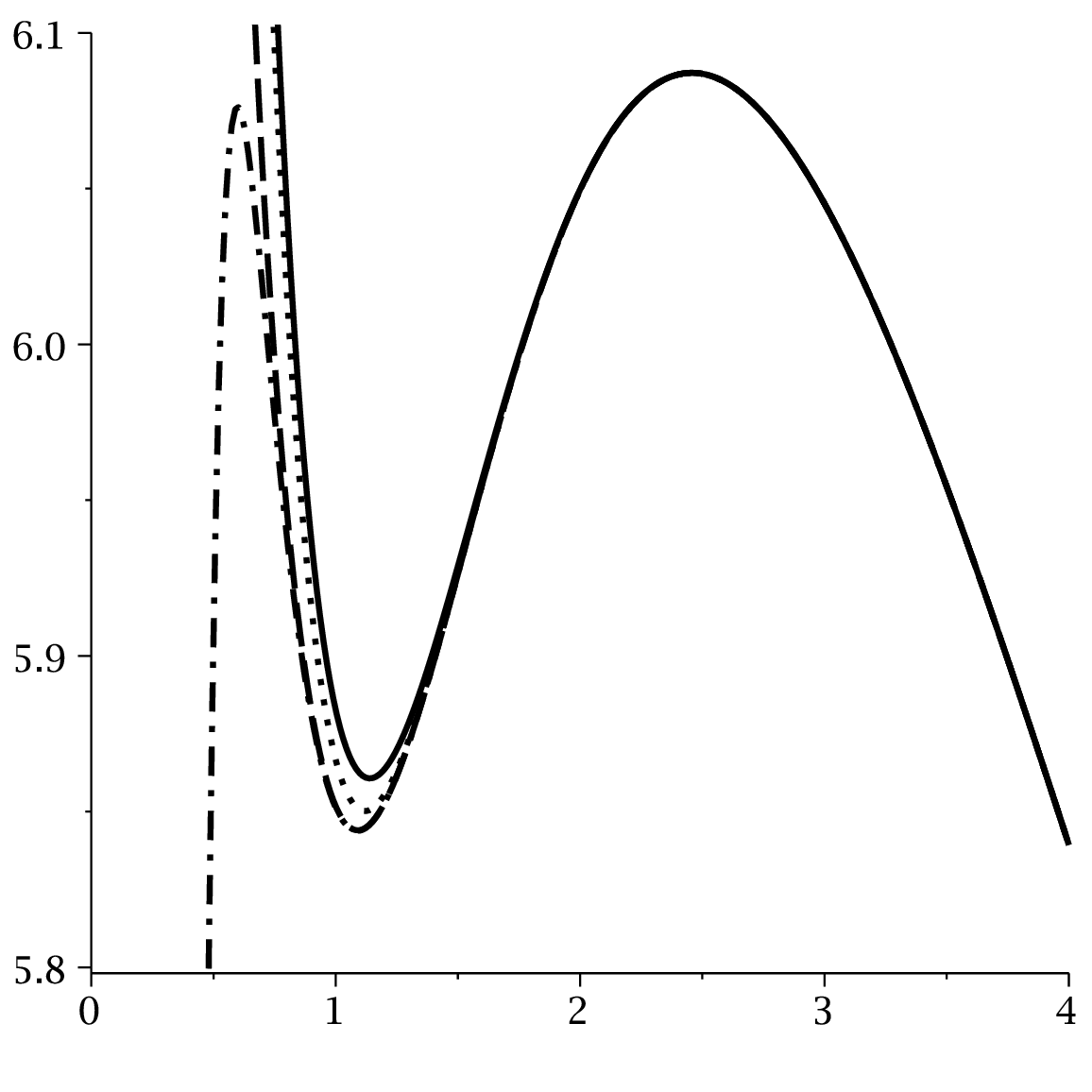}
\\
\epsfxsize=5.5cm \epsffile{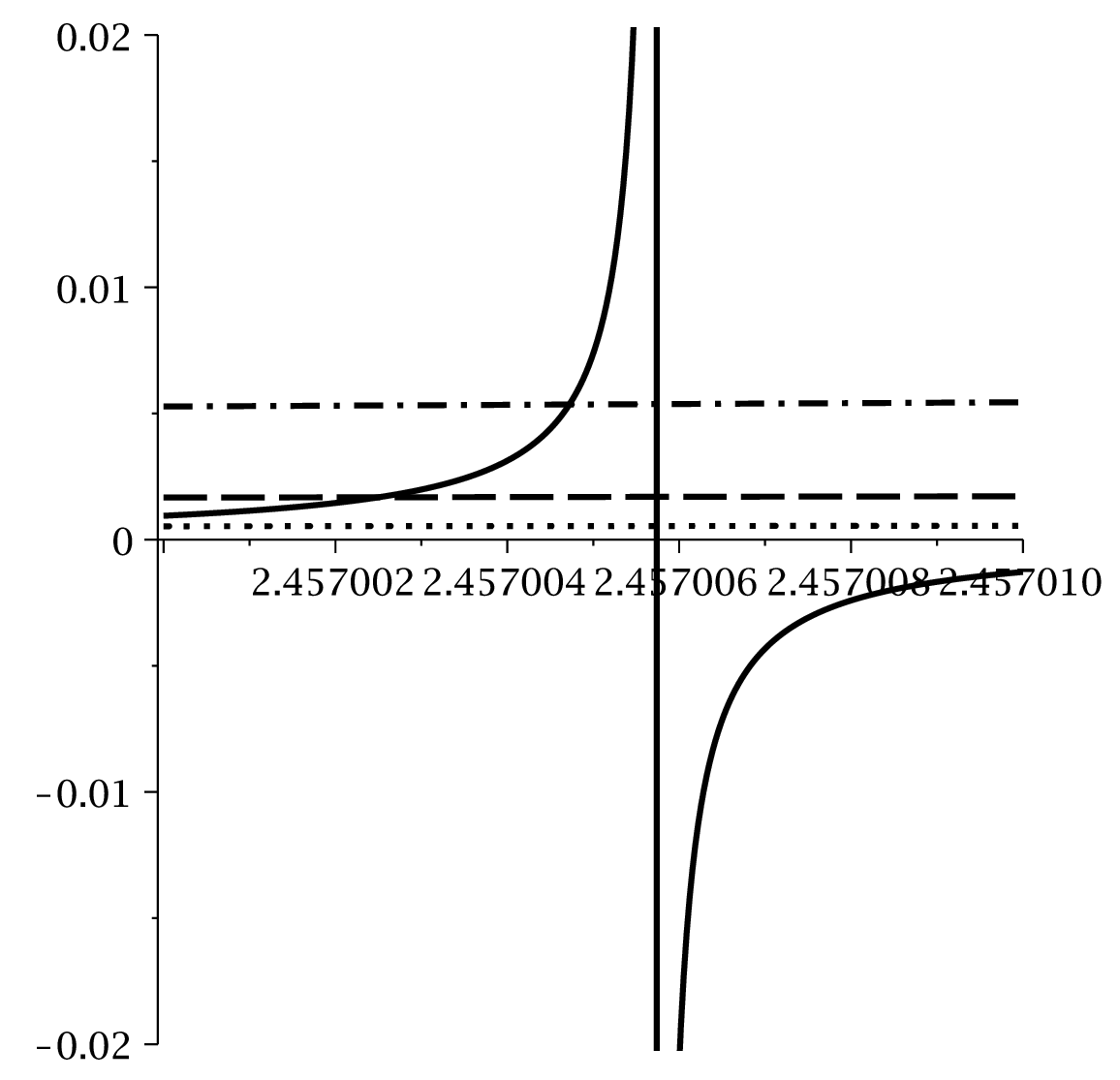} & \epsfxsize=5.5cm %
\epsffile{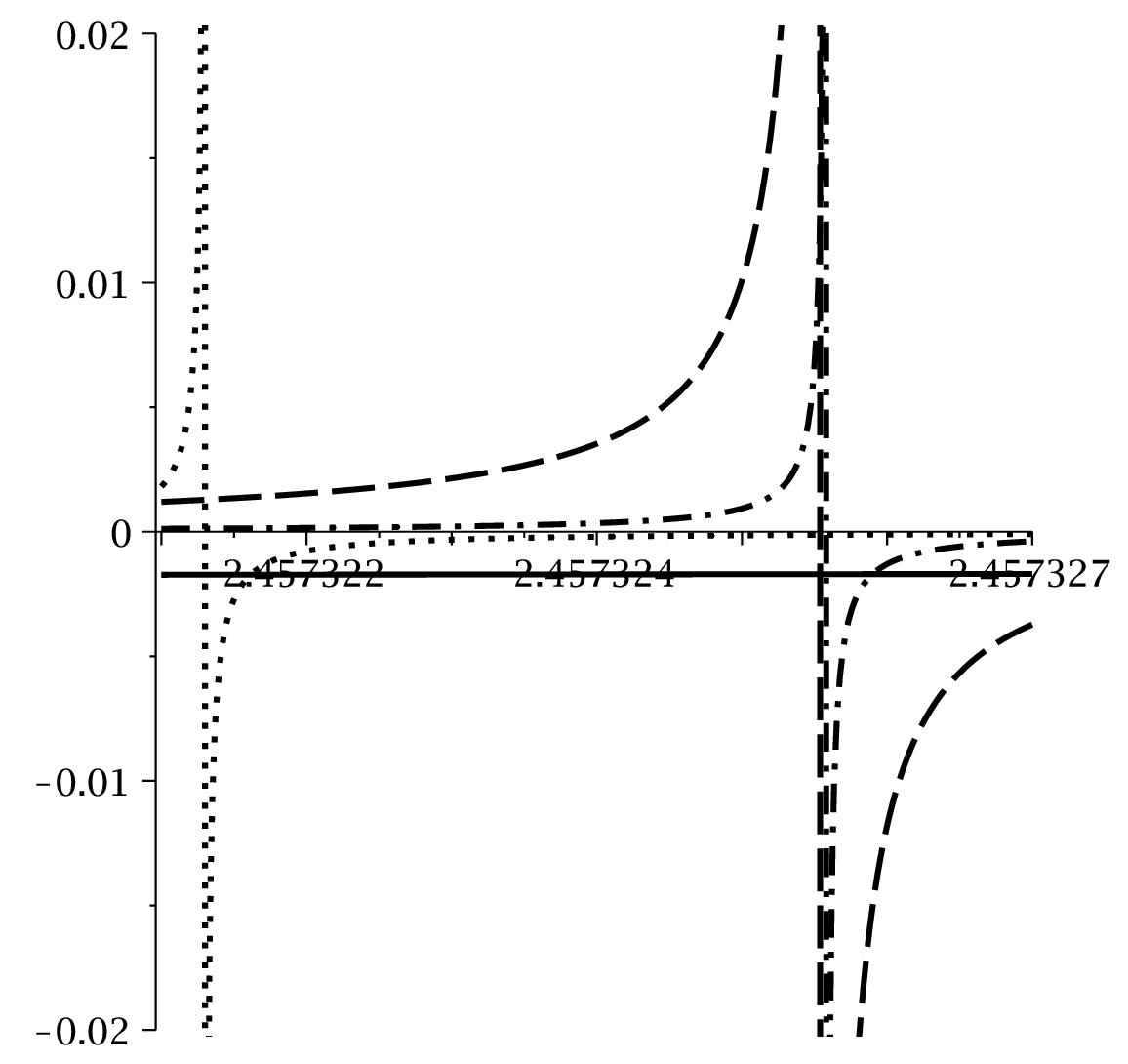} & \epsfxsize=5.5cm %
\epsffile{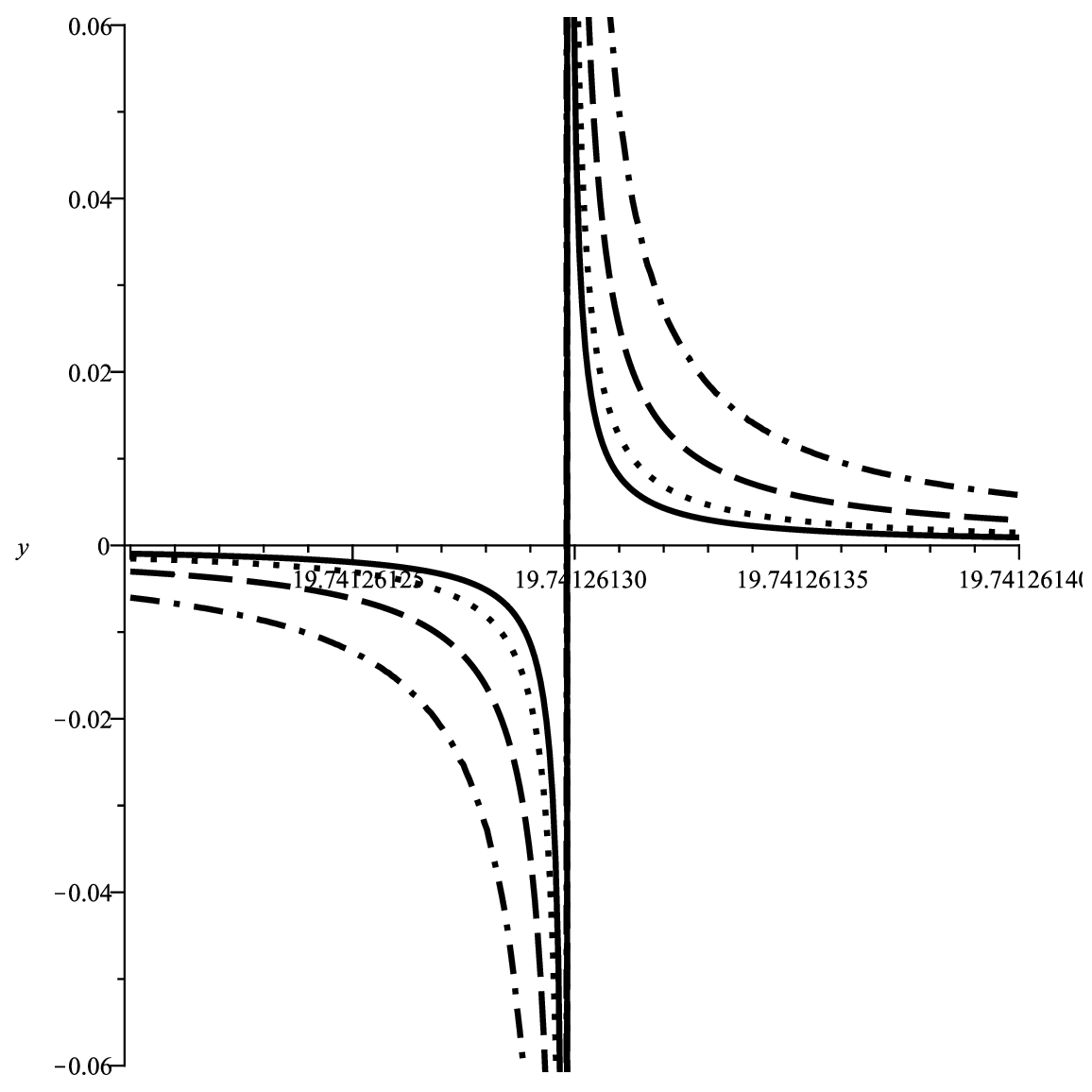}%
\end{array}
$%
\caption{For different scales: $C_{Q}$ and $T$ (only right up panel) versus $%
r_{+} $ for $q=1$, $\Lambda =-1$, $c=c_{1}=c_{2}=2$, $c_{3}=c_{4}=0.2$, $m=3$%
, $\protect\alpha=0.5$, $d=6$ and $\protect\kappa=1$; $\protect\beta=0.1$
(continues line), $\protect\beta=1$ (dotted line), $\protect\beta=10$
(dashed line) and $\protect\beta=100$ (dashes-dotted line).}
\label{Fig3}
\end{figure}

\begin{figure}[tbp]
$%
\begin{array}{ccc}
\epsfxsize=5.5cm \epsffile{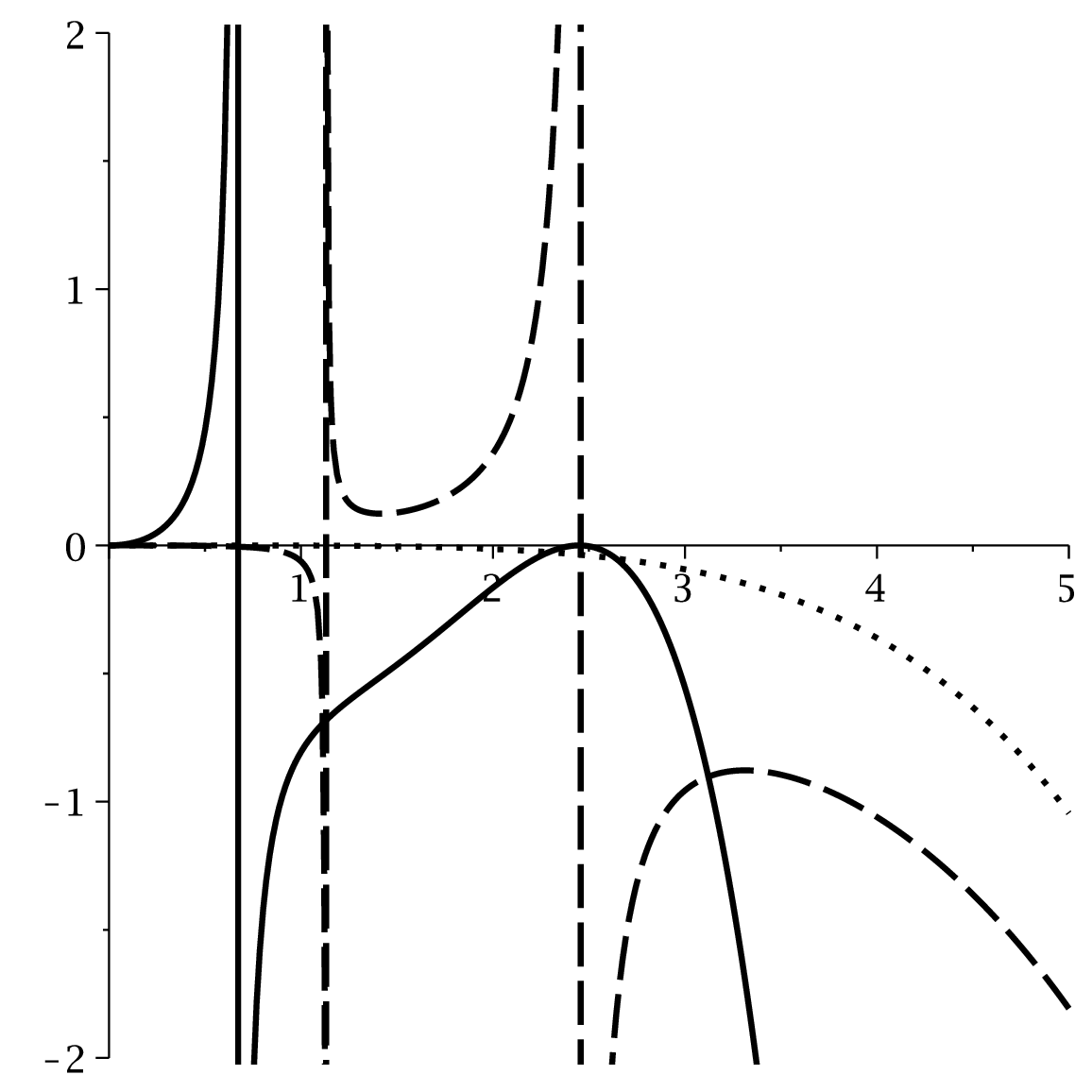} & \epsfxsize=5.5cm %
\epsffile{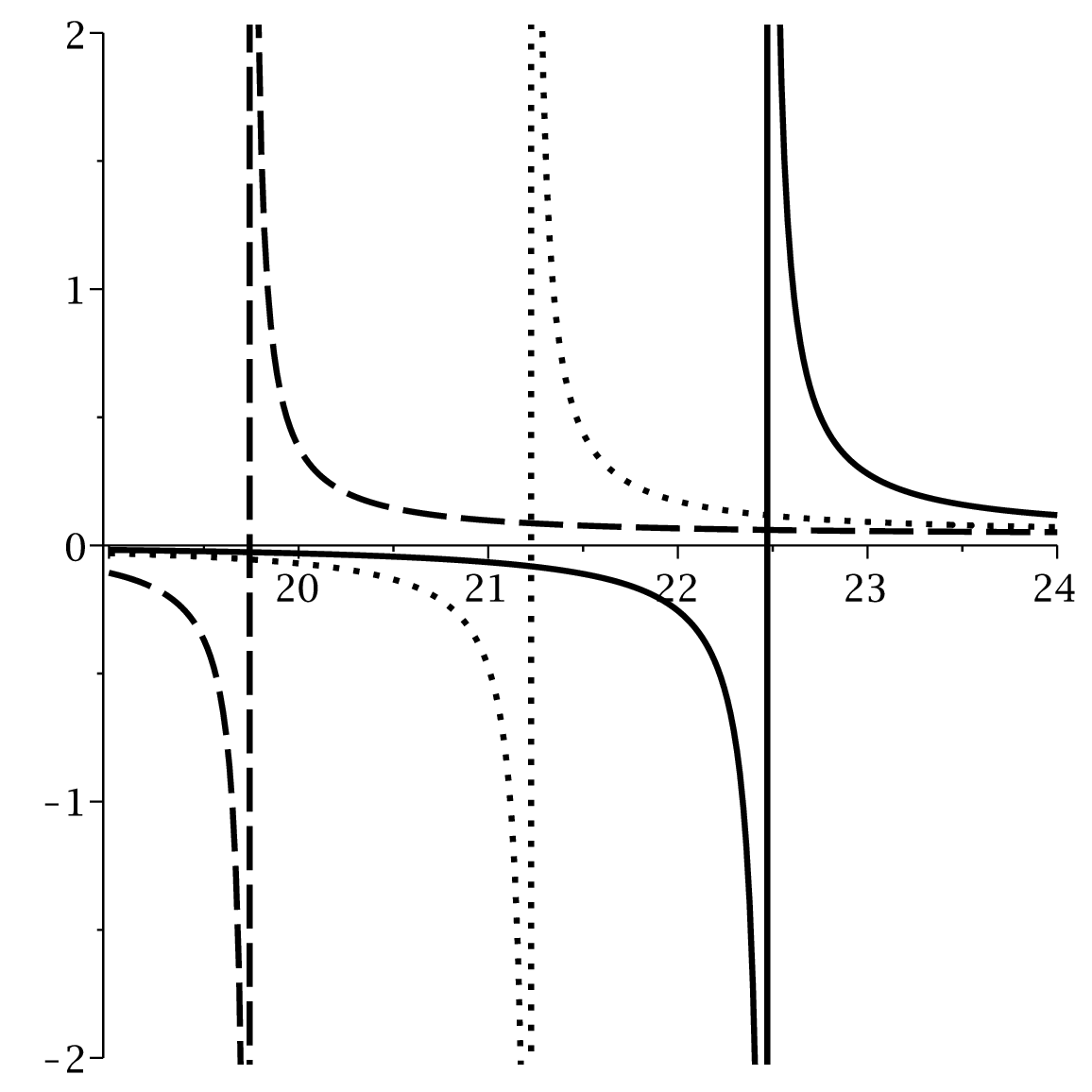} & \epsfxsize=5.5cm \epsffile{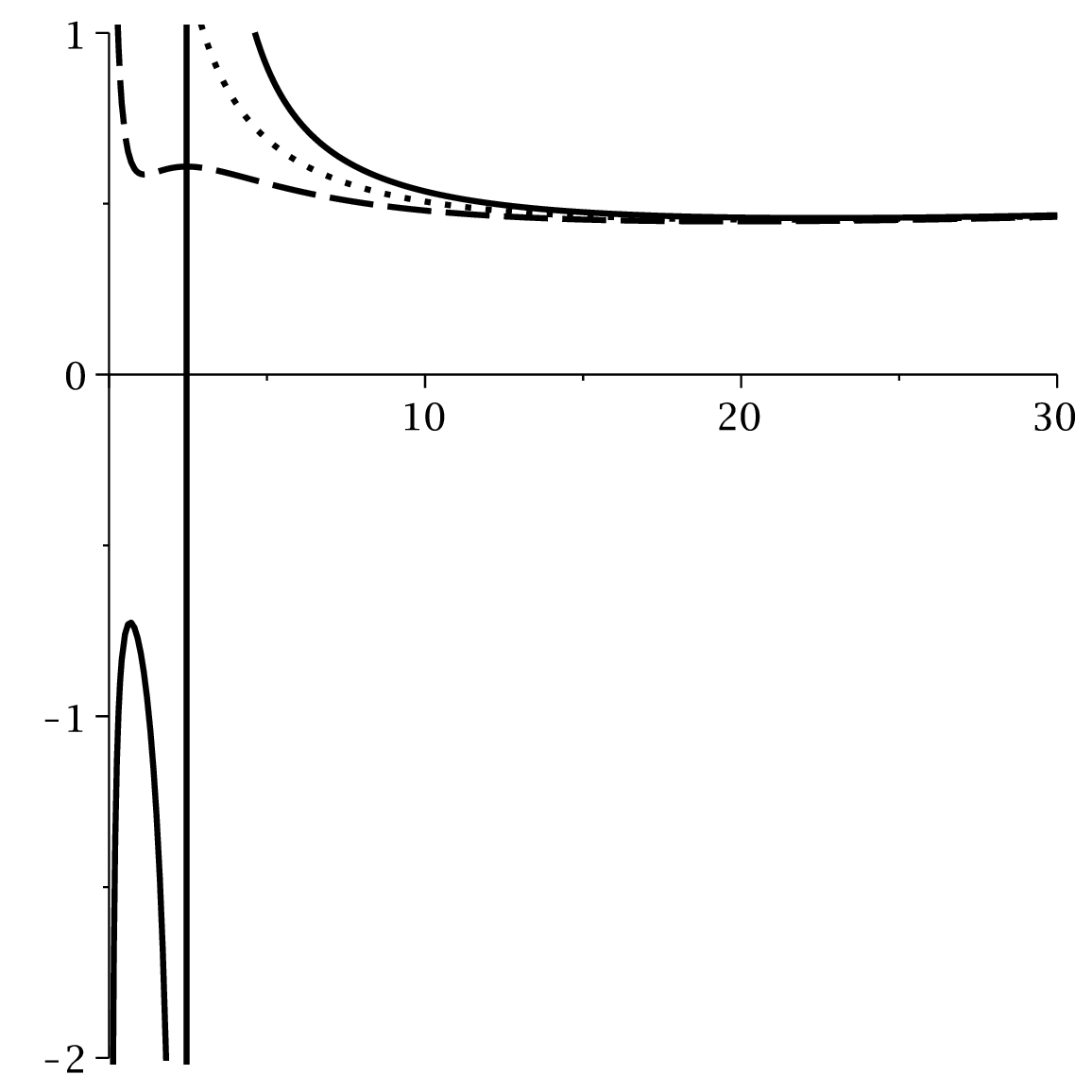}%
\end{array}
$%
\caption{For different scales: $C_{Q}$ (left and middle panels) and $T$
(right panel) versus $r_{+} $ for $q=1$, $\Lambda =-1$, $c=c_{1}=c_{2}=2$, $%
c_{3}=c_{4}=0.2$, $m=3$, $\protect\beta=0.5$, $d=6$ and $\protect\alpha=0.5$%
; $\protect\kappa=-1$ (continues line), $\protect\kappa=0$ (dotted line) and
$\protect\kappa=1$ (dashed line).}
\label{Fig4}
\end{figure}

\begin{figure}[tbp]
$%
\begin{array}{ccc}
\epsfxsize=5.5cm \epsffile{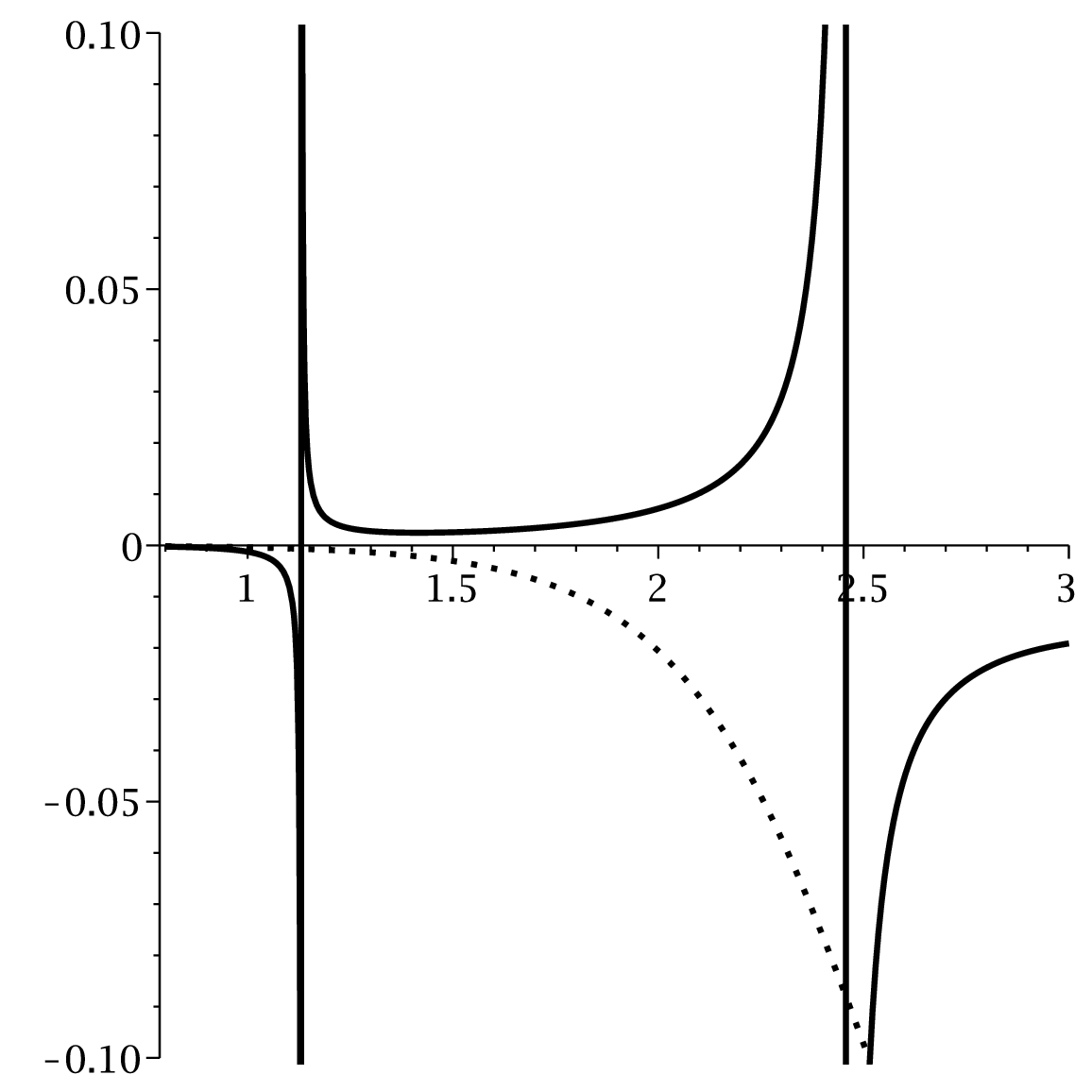} & \epsfxsize=5.5cm %
\epsffile{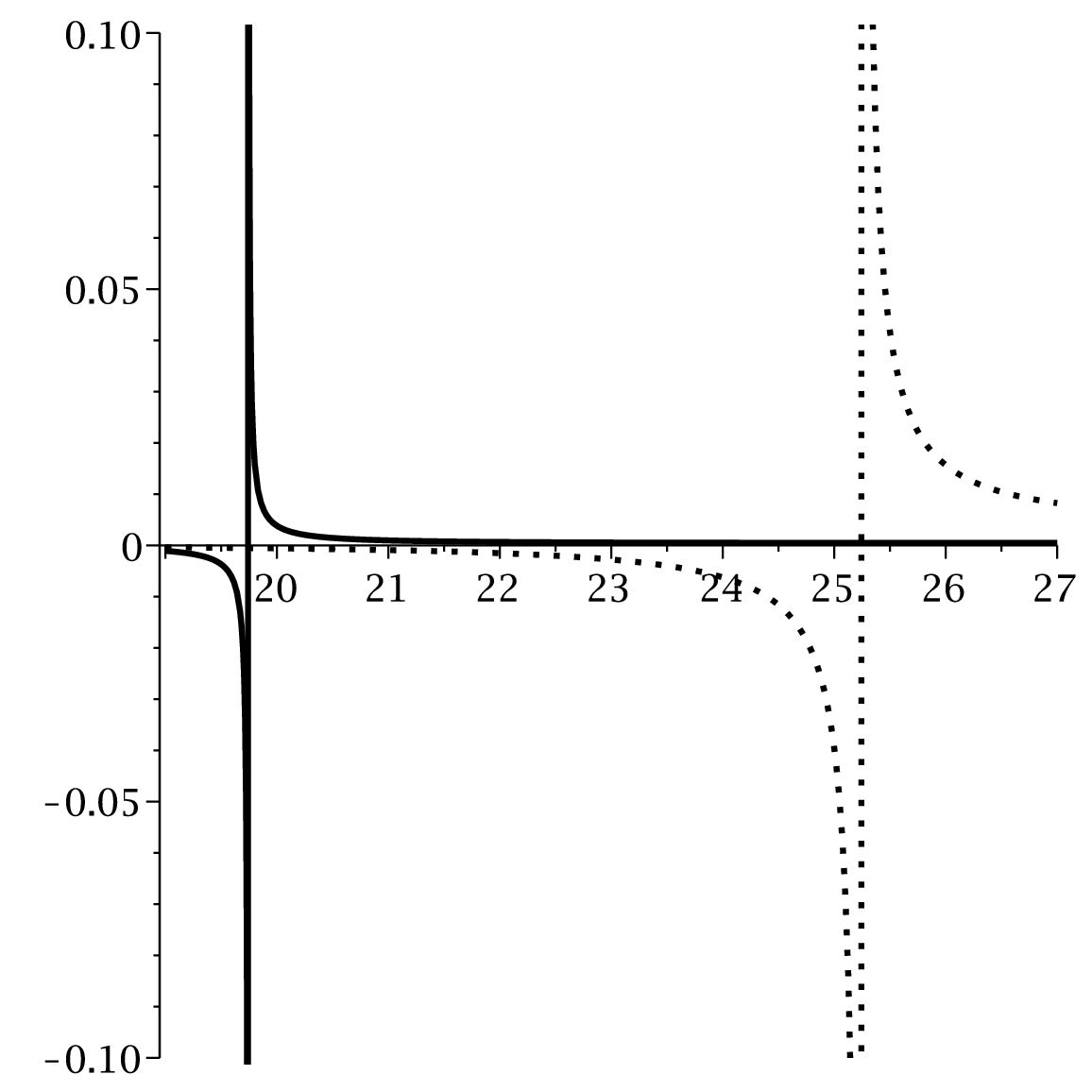} & \epsfxsize=5.5cm \epsffile{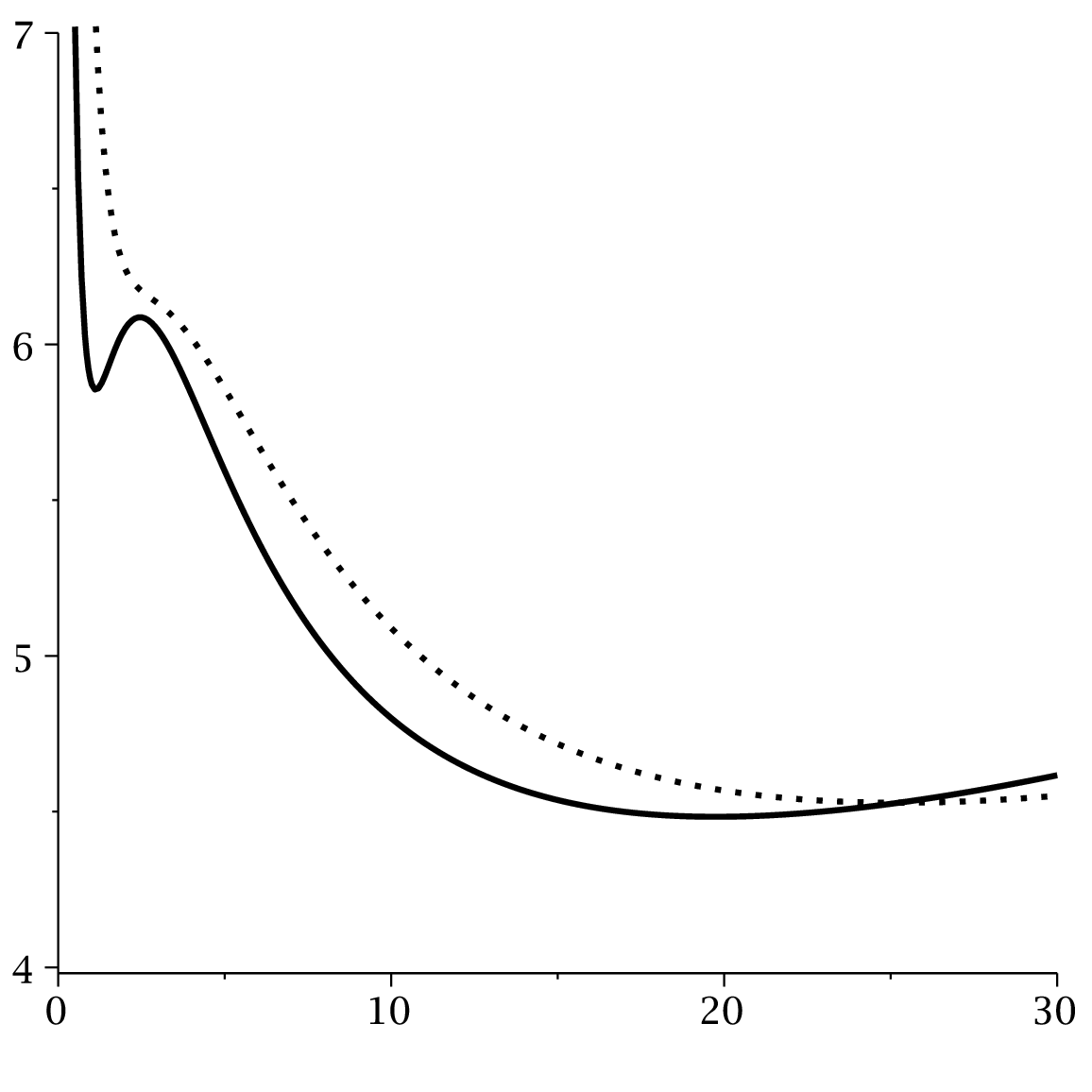}%
\end{array}
$%
\caption{For different scales: $C_{Q}$ (left and middle panels) and $T$
(right panel) versus $r_{+} $ for $q=1$, $\Lambda =-1$, $c=c_{1}=c_{2}=2$, $%
c_{3}=c_{4}=0.2$, $m=3$, $\protect\beta=0.5$, $\protect\kappa=1$ and $%
\protect\alpha=0.5$; $d=6$ (continues line) and $d=7$ (dotted line).}
\label{Fig5}
\end{figure}

\begin{figure}[tbp]
$%
\begin{array}{cc}
\epsfxsize=6cm \epsffile{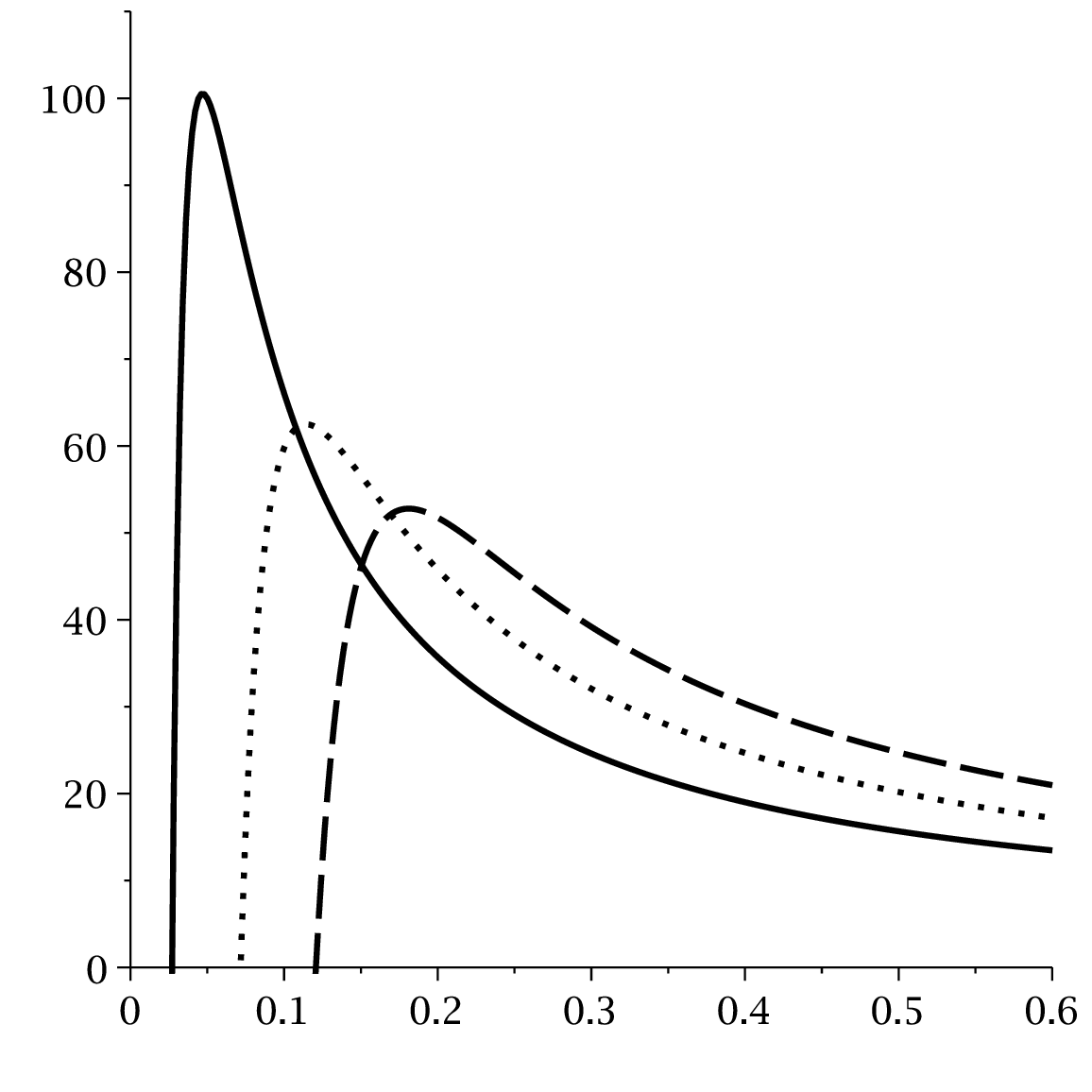} & \epsfxsize=6cm %
\epsffile{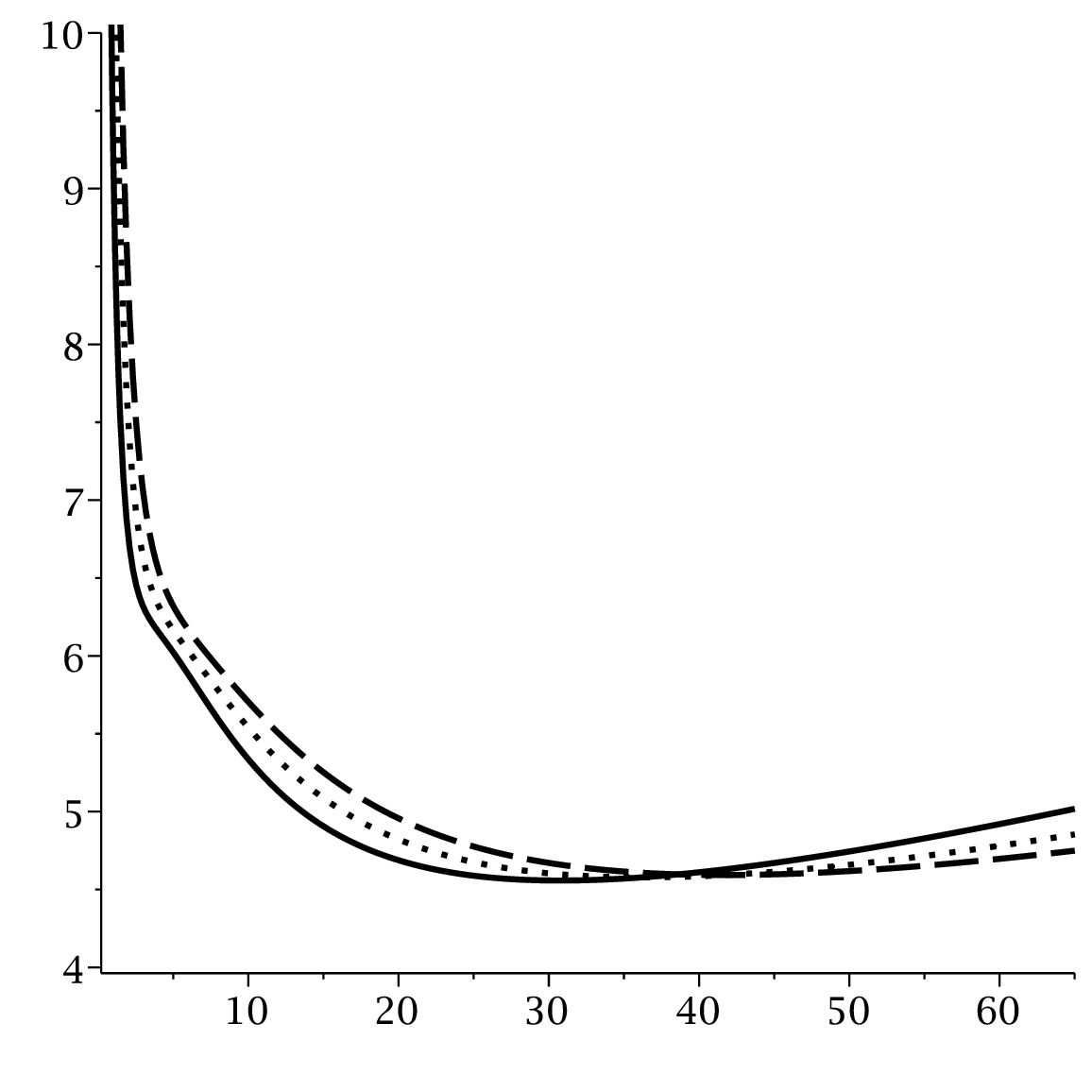} \\
\epsfxsize=6cm \epsffile{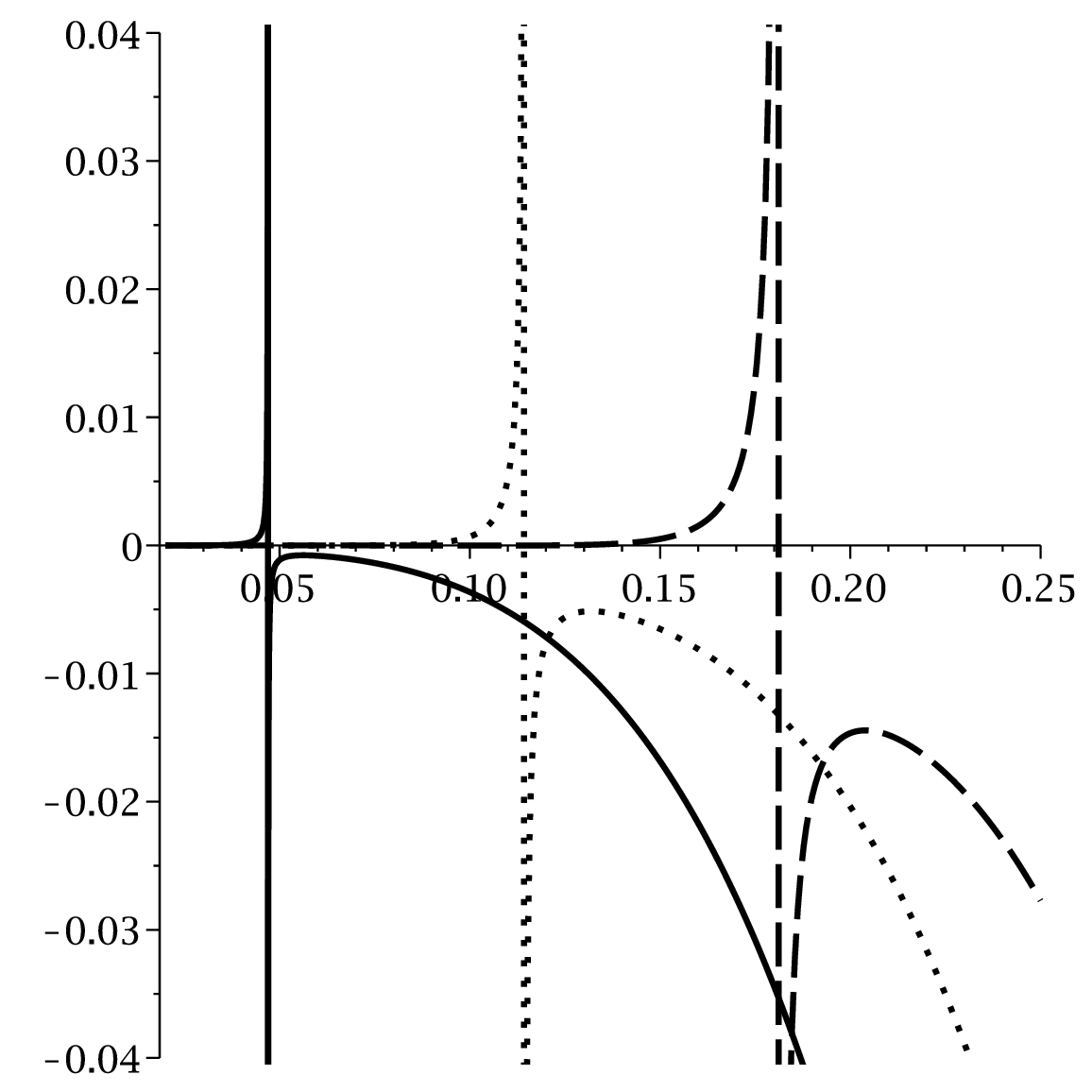} & \epsfxsize=6cm %
\epsffile{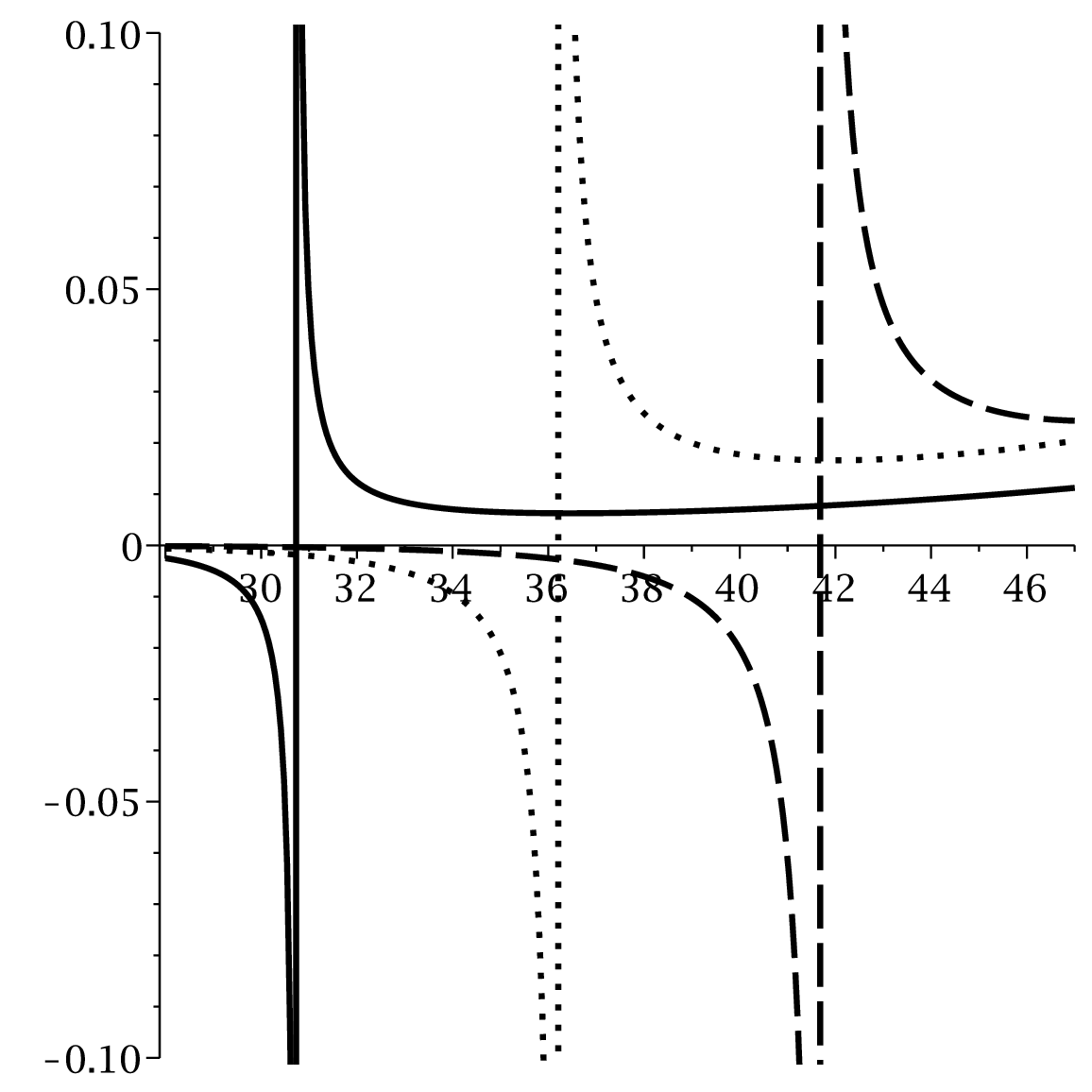}%
\end{array}
$%
\caption{For different scales: $C_{Q}$ (down panels) and $T$ (up panels)
versus $r_{+} $ for $q=1$, $\Lambda =-1$, $c=c_{1}=c_{2}=2$, $%
c_{3}=c_{4}=0.2 $, $m=3$, $\protect\beta=0.5$, $\protect\kappa=1$ and $%
\protect\alpha=0.5$; $d=8$ (continues line), $d=9$ (dotted line) and $d=10$
(dashed).}
\label{Fig6}
\end{figure}

Eqs. (\ref{TotalTT}) and (\ref{CQ}) show that investigation of heat capacity
in analytical form is not easy, and therefore, we plot $T$ and $C_{Q}$ to
study their behaviors (see Figs. \ref{Fig1} - \ref{Fig6}).

It is evident that the temperature and its specific properties are functions
of the massive, GB and BI parameters as well as topological structure of the
black holes and dimensionality. In other words, variation of one parameter
while the other parameters are specifically fixed, leads to changes in
number (and place) of the extremum and root that temperature can obtain (
see Figs. \ref{Fig1} and \ref{Fig2} up panels, \ref{Fig4} and \ref{Fig5}
right panels ). In some cases, the variation of one parameter for different
domains may results into different (and in some cases opposite) behavior for
the temperature. These modifications lead to change in number and places of
the extrema and roots of the temperature. ( see Figs. \ref{Fig3} up right
panel and \ref{Fig6} up panels ). In case of the topological effect, by
adjusting parameters properly, one can find a divergency for the case of $%
k=-1$. This singular point of temperature comes from the GB modification of
the temperature and does not appear in the Einstein gravity ( see Eq. \ref%
{TotalTT} ).

The heat capacity is constructed by employing temperature and its
derivation. The derivation of the temperature resides in denominator of the
heat capacity ( see Eq. \ref{CQ} ). Therefore, one can conclude that roots
and extrema of the temperature are places in which heat capacity acquires
roots and divergencies, respectively. The root and divergencies of the heat
capacity are interpreted as physical limitation point and phase transitions
of the black holes, respectively. Now, considering what was mentioned in
last paragraph regarding the effects of variation of different parameters on
number of root and extremum and their corresponding places, one can conclude
that consequently, the number of divergencies and root are functions of
different parameters.

The thermal stability conditions are indicated by the number of roots and
divergencies. In other words, the thermal stability conditions are dictated
according to the type of extremum, its number and roots. In root, usually a
transition from non-physical solutions to physical ones takes place. In
place of minimum in temperature, a phase transition from larger unstable to
smaller stable black holes happens whereas in maximum of the temperature the
opposite (phase transition of smaller unstable to larger stable black holes)
takes place.

According to what was stated, depending on number of the extremum, their
types and roots, the physical and stability conditions for these black holes
vary. By adjusting different parameters properly, one can see that these
black holes may have different phase transitions and limitation point; a)
only one phase transition of smaller unstable to larger stable ( see Figs. %
\ref{Fig1} and \ref{Fig2} down panels, \ref{Fig4} and \ref{Fig5} two left
panels), b) one phase transition of smaller unstable to larger stable and
another phase transition of larger unstable to smaller stable ( see Figs. %
\ref{Fig1} and \ref{Fig2} down panels ), c) one limitation point from
non-physical to physical and two phase transitions of smaller unstable to
larger stable and larger unstable to smaller stable ( see Fig. \ref{Fig6}
two left panels ) and d) one limitation point for non-physical to physical
and two phase transitions of smaller unstable to larger stable and two phase
transitions of larger unstable to smaller stable ( see Fig. \ref{Fig3}).


\section{$P-V$ criticality of black hole solutions in GB-BI-massive gravity}

In this section, we study the phase transition points of black holes in
GB-BI-massive gravity through the use of $P-V$ criticality and related phase
diagrams in spherically symmetric spacetime ($\kappa=1$). To do so, we
consider following relationship between thermodynamical pressure and
cosmological constant
\begin{equation}
P=-\frac{\Lambda }{8\pi }.  \label{P}
\end{equation}

From thermodynamical point of view, one can point it out that conjugating
thermodynamical variable corresponding to pressure would be thermodynamical
volume. Therefore, in order to calculate the thermodynamical volume of the
solutions, one should use
\begin{equation}
V=\left( \frac{\partial H}{\partial P}\right) _{S,Q}.  \label{V}
\end{equation}

Considering cosmological constant as thermodynamical pressure leads to an
interpretation of mass not only as internal energy but as Enthalpy of
thermodynamical system. This interpretation leads to following relation for
the Gibbs free energy of the system
\begin{equation}
G=H-TS=M-TS,  \label{G}
\end{equation}

There are several methods for studying the critical behavior of the system
near critical points. Among them one can name $P-V$, $T-V$ and $G-T$
diagrams. The similarity that was observed in studying $P-V$ diagrams of
black holes and van der Waals liquid/gas system, brought a new insight to
black holes thermodynamics. $T-V$ diagrams enable one to investigate single
state region of different phases for thermodynamical systems. In case of
black holes, these single state of different regions are denoted as small
and large black holes, which phase transitions take place between them. The
formation of the swallow-tail for pressures smaller than critical pressure,
makes the $G-T$ diagrams one of the easiest ways for detecting a phase
transition. According to what was mentioned, we will study critical behavior
of these black holes by plotting $P-V$, $T-V$ and $G-T$ diagrams.

In order to find critical points, we use the inflection points that are
obtainable through the use of isotherm $P-V$ diagrams. We use following
relations for obtaining critical values

\begin{equation}
\left( \frac{\partial P}{\partial r_{+}}\right)_{T} =\left( \frac{\partial
^{2}P}{\partial r_{+}^{2}}\right)_{T} =0.  \label{infel}
\end{equation}

Using Eqs. (\ref{TotalTT}), (\ref{TotalS}), (\ref{Mm}), (\ref{P}) and (\ref%
{G}), one can find following relations for pressure and Gibbs free energy

\begin{eqnarray}
P &=&\frac{d_{2}\left( 2\kappa \alpha ^{\prime }+r_{+}^{2}\right) T}{%
4r_{+}^{3}}-\frac{m^{2}cd_{2}\left( d_{3}d_{4}c^{2}\left(
d_{5}cc_{4}+c_{3}r_{+}\right) +r_{+}^{2}\left( d_{3}cc_{2}+c_{1}r_{+}\right)
\right) }{16\pi r_{+}^{4}}  \notag \\
&&+\frac{\beta ^{2}\left( \sqrt{1+\eta _{+}}-1\right) }{4\pi }-\frac{%
d_{2}\kappa \left( d_{5}\kappa \alpha ^{\prime }+d_{3}r_{+}^{2}\right) }{%
16\pi r_{+}^{4}},  \label{PP} \\
&&  \notag \\
\notag \\
G &=&\frac{d_{2}}{16\pi }\left[ \kappa r_{+}^{d_{3}}\left( 1+\frac{\kappa
\alpha ^{\prime }}{r_{+}^{2}}\right) +\frac{m^{2}cr_{+}^{d_{5}}\left(
d_{2}d_{3}c^{2}\left( d_{4}cc_{4}+c_{3}r_{+}\right) +r_{+}^{2}\left(
d_{2}c_{2}c+c_{1}r_{+}\right) \right) }{d_{2}}\right.  \notag \\
&&\left. +\frac{4r_{+}^{d_{1}}}{d_{1}d_{2}}\left( \beta ^{2}\left( 1-\sqrt{%
1+\eta _{+}}\right) +4\pi P\right) +\frac{2d_{2}q^{2}}{d_{1}r_{+}^{d_{3}}}%
\mathcal{H}_{+}\right] -\frac{r_{+}^{d_{2}}\left( 1+\frac{2d_{2}\kappa
\alpha ^{\prime }}{d_{4}r_{+}^{2}}\right) }{16\pi \left( 2\kappa \alpha
^{\prime }+r_{+}^{2}\right) }\left[ \frac{d_{3}\kappa \left( r_{+}^{2}+\frac{%
d_{5}\kappa \alpha ^{\prime }}{d_{3}}\right) }{r_{+}}\right.  \notag \\
&&\left. +\frac{m^{2}c\left( d_{3}d_{4}c^{2}\left(
d_{5}cc_{4}+c_{3}r_{+}\right) +r_{+}^{2}\left( d_{3}c_{2}c+c_{1}r_{+}\right)
\right) }{r_{+}}+\frac{4r_{+}^{3}}{d_{2}}\left( \beta ^{2}\left( 1-\sqrt{%
1+\eta _{+}}\right) +4\pi P\right) \right] ,  \label{GG}
\end{eqnarray}

where $\alpha ^{\prime }=d_{3}d_{4}\alpha $.

Now, by employing Eqs. (\ref{infel}) and (\ref{PP}), one can find following
relation for calculating critical horizon radius
\begin{eqnarray}
&&\sqrt{2}r_{+}^{2d_{2}}\beta \sqrt{1+\eta _{+}}\left\{ 2\beta
^{2}d_{3}\left( \kappa +m^{2}c_{2}c^{2}\right)
r_{+}^{2d_{-2}}-12m^{2}cr_{+}^{2d_{-3/2}}\beta ^{2}\left( c_{1}\kappa \alpha
^{\prime }-\frac{d_{3}d_{4}c_{3}c^{2}}{2}\right) +12d_{2}d_{3}q^{2}r_{+}^{4}%
\mathcal{D}\right.  \notag \\
&&\left. -24r_{+}^{2d_{-1}}\beta ^{2}\left( \kappa ^{2}\alpha ^{\prime }-%
\frac{d_{3}d_{4}d_{5}}{2}m^{2}c_{4}c^{4}+\frac{d_{3}}{2}m^{2}c_{2}c^{2}%
\kappa \alpha ^{\prime }\right) +24d_{5}\kappa \alpha ^{\prime
}r_{+}^{2d}\beta ^{2}\left( \kappa ^{2}\alpha ^{\prime
}+d_{3}d_{4}m^{2}c_{4}c^{4}\right) \right\}  \notag \\
&&-6\sqrt{2}d_{2}d_{3}q^{2}\beta \left[ \frac{2d_{5/2}\beta ^{2}}{3}%
r_{+}^{2d_{-3}}+4d_{7/2}\kappa \alpha ^{\prime }r_{+}^{2d_{-2}}\beta
^{2}+d_{2}d_{3}q^{2}r_{+}^{8}\left( d\left( \kappa \alpha ^{\prime }+\frac{%
r_{+}^{2}}{6}\right) -\frac{r_{+}^{2}}{2}-5\kappa \alpha ^{\prime }\right) %
\right] =0  \label{cr}
\end{eqnarray}%
where $\mathcal{D}$ is
\begin{eqnarray}
\mathcal{D} &=&d_{3}d_{4}d_{5}m^{2}c_{4}c^{4}\left( \kappa \alpha ^{\prime }+%
\frac{r_{+}^{2}}{2}\right) +\frac{d_{3}d_{4}m^{2}c_{3}c^{3}r_{+}^{3}}{4}-%
\frac{d_{3}m^{2}c_{2}c^{2}r_{+}^{2}}{2}\left( \kappa \alpha ^{\prime }-\frac{%
r_{+}^{2}}{6}\right) -\frac{\kappa \alpha ^{\prime }m^{2}c_{1}cr_{+}^{3}}{2}
\notag \\
&&+\kappa \left( \frac{d_{3}r_{+}^{4}}{12}-\kappa \alpha ^{\prime }\left(
r_{+}^{2}+d_{5}\kappa \alpha ^{\prime }\right) \right) .
\end{eqnarray}

\begin{center}
\begin{tabular}{ccccc}
\hline\hline
$m$ & $r_{c}$ & $T_{c}$ & $P_{c}$ & $\frac{P_{c}r_{c}}{T_{c}}$ \\
\hline\hline
$0.000000$ & $2.020913551$ & $0.1396519429$ & $0.02602417127$ & $%
0.3765976991 $ \\ \hline
$0.500000$ & $2.378161673$ & $0.4423844412$ & $0.0555448739$ & $0.2985970525$
\\ \hline
$1.000000$ & $2.518902071$ & $1.3647986130$ & $0.1515266147$ & $0.2796608232$
\\ \hline
$1.500000$ & $2.555959475$ & $2.9044194570$ & $0.3125563944$ & $0.2750571980$
\\ \hline
$2.000000$ & $2.570059221$ & $5.0602917940$ & $0.5381806700$ & $0.2733352639$
\\ \hline
\end{tabular}
\\[0pt]
\vspace{0.1cm} Table ($1$): $q=1$, $\alpha =0.5$, $\beta =0.5$, $%
c=c_{1}=c_{2}=2$, $c_{3}=0.2$, $c_{4}=-0.2$ and $d=6$. \vspace{0.5cm}

\begin{tabular}{ccccc}
\hline\hline
$\beta $ & $r_{c}$ & $T_{c}$ & $P_{c}$ & $\frac{P_{c}r_{c}}{T_{c}}$ \\
\hline\hline
$10^{-9}$ & $2.308160166$ & $0.4455199824$ & $0.0568662372$ & $0.2946139089$
\\ \hline
$10^{-5}$ & $2.308163661$ & $0.4455195772$ & $0.0568660973$ & $0.2946138981$
\\ \hline
$0.100000$ & $2.352654221$ & $0.4429849335$ & $0.0558166578$ & $0.2964373857$
\\ \hline
$5.000000$ & $2.379768359$ & $0.4423515057$ & $0.0555299142$ & $0.2987405516$
\\ \hline
$50.00000$ & $2.379784726$ & $0.4423511724$ & $0.0555296666$ & $0.2987414994$
\\ \hline
$500.0000$ & $2.379784890$ & $0.4423511689$ & $0.0555364292$ & $0.2987779041$
\\ \hline
$5000.000$ & $2.379784892$ & $0.4423511693$ & $0.0552976969$ & $0.2974935588$
\\ \hline
\end{tabular}
\\[0pt]
\vspace{0.1cm} Table ($2$): $q=1$, $m=0.5$, $\alpha =0.5$, $c=c_{1}=c_{2}=2$%
, $c_{3}=0.2$, $c_{4}=-0.2$ and $d=6$. \vspace{0.5cm}

\begin{tabular}{ccccc}
\hline\hline
$\alpha $ & $r_{c}$ & $T_{c}$ & $P_{c}$ & $\frac{P_{c}r_{c}}{T_{c}}$ \\
\hline\hline
$0.000000$ & $1.655138693$ & $0.7701822534$ & $0.1684536184$ & $0.3620105508$
\\ \hline
$0.800000$ & $2.745465406$ & $0.3752283006$ & $0.0381288304$ & $0.2789805157$
\\ \hline
$1.400000$ & $3.430032297$ & $0.3026420039$ & $0.0223913368$ & $0.2537751123$
\\ \hline
$2.000000$ & $4.066801555$ & $0.2625706266$ & $0.0152864154$ & $0.2367622723$
\\ \hline
$7.000000$ & $%
\begin{array}{c}
0.4604432507 \\
8.5661098700%
\end{array}%
$ & $%
\begin{array}{c}
0.04557012100 \\
0.16263006060%
\end{array}%
$ & $%
\begin{array}{c}
-0.619424375 \\
0.0032210312%
\end{array}%
$ & $%
\begin{array}{c}
-6.258701240 \\
0.1696593344%
\end{array}%
$ \\ \hline
$8.000000$ & $%
\begin{array}{c}
0.6532944257 \\
9.3813191800%
\end{array}%
$ & $%
\begin{array}{c}
0.05173846074 \\
0.15530428400%
\end{array}%
$ & $%
\begin{array}{c}
-0.319158907 \\
0.0026797563%
\end{array}%
$ & $%
\begin{array}{c}
-4.029975610 \\
0.1618735099%
\end{array}%
$ \\ \hline
$9.000000$ & $%
\begin{array}{c}
0.7978966829 \\
10.181225130%
\end{array}%
$ & $%
\begin{array}{c}
0.05408684451 \\
0.14928940340%
\end{array}%
$ & $%
\begin{array}{c}
-0.219549753 \\
0.0022720106%
\end{array}%
$ & $%
\begin{array}{c}
-3.238828615 \\
0.1549463726%
\end{array}%
$ \\ \hline
\end{tabular}
\\[0pt]
\vspace{0.1cm} Table ($3$): $q=1$, $m=0.5$, $\beta =0.5$, $c=c_{1}=c_{2}=2$,
$c_{3}=0.2$, $c_{4}=-0.2$ and $d=6$. \vspace{0.5cm}
\end{center}

\begin{figure}[tbp]
$%
\begin{array}{ccc}
\epsfxsize=5cm \epsffile{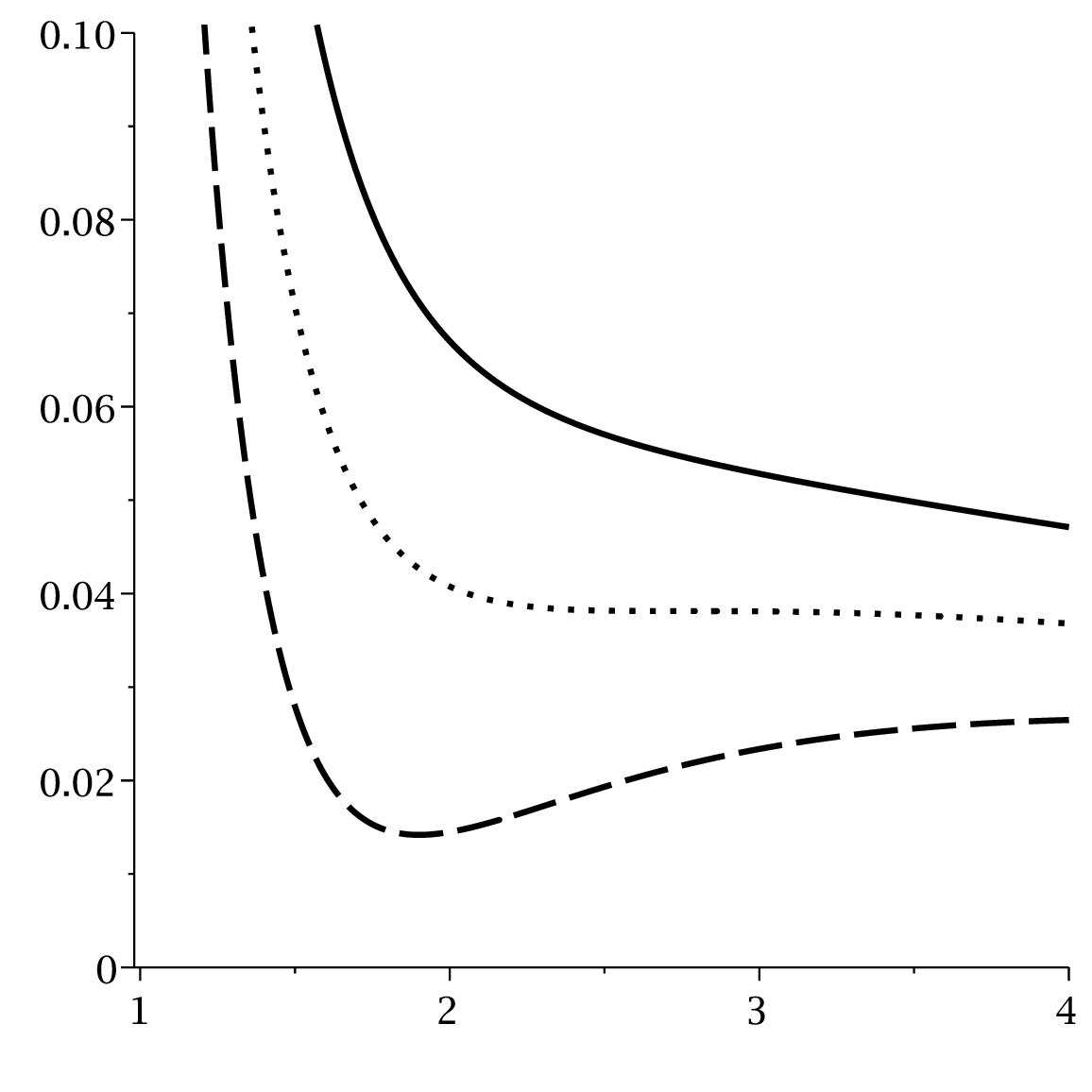} & \epsfxsize=5cm %
\epsffile{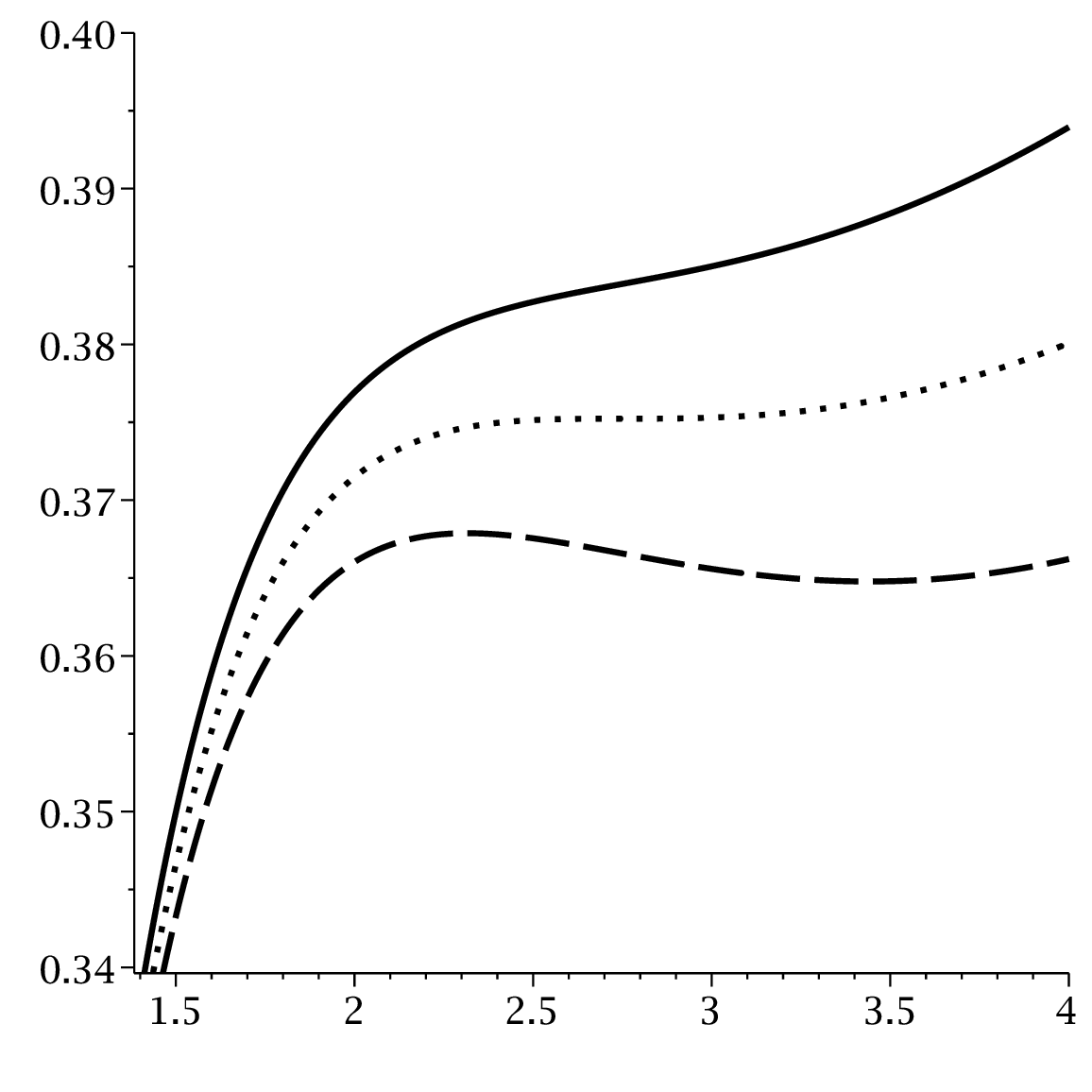} & \epsfxsize=5cm \epsffile{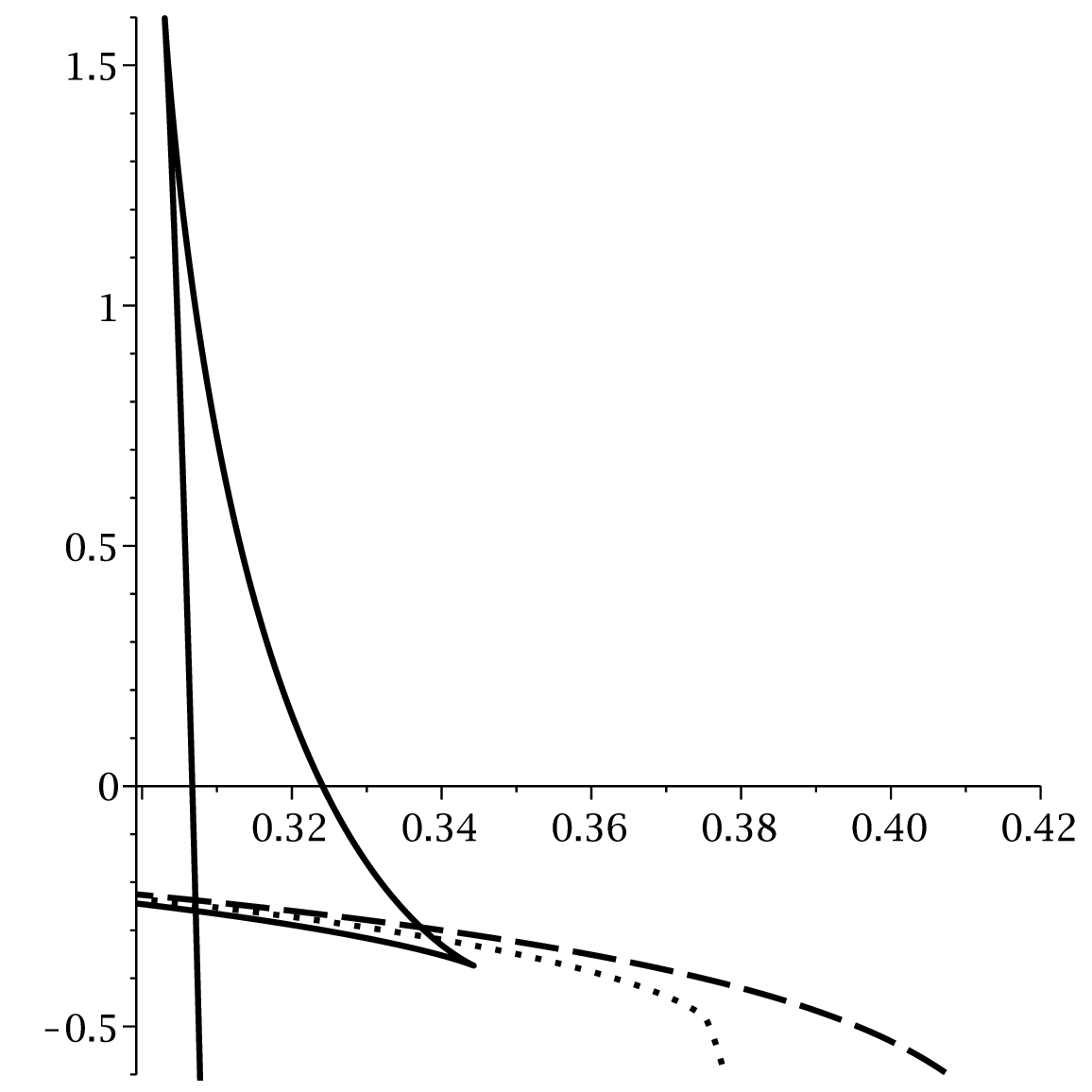}%
\end{array}
$%
\caption{ $P-r_{+}$ (left), $T-r_{+}$ (middle) and $G-T$ (right) diagrams
for $\protect\beta=0.5$, $q=1$, $m=0.5$, $\protect\alpha=0.8$, $%
c=c_{1}=c_{2}=2$, $c_{3}=0.2$, $c_{4}=-0.2$ and $d=6$. \newline
$P-r_{+}$ diagram, from up to bottom $T=1.1T_{c}$, $T=T_{c}$ and $T=0.9T_{c}$
, respectively. \newline
$T-r_{+}$ diagram, from up to bottom $P=1.1P_{c}$, $P=P_{c}$ and $P=0.9P_{c}$
, respectively. \newline
$G-T$ diagram for $P=0.5P_{c}$ (continuous line), $P=P_{c}$ (dotted line)
and $P=1.5P_{c}$ (dashed line). }
\label{Fig7}
\end{figure}

\begin{figure}[tbp]
$%
\begin{array}{ccc}
\epsfxsize=5cm \epsffile{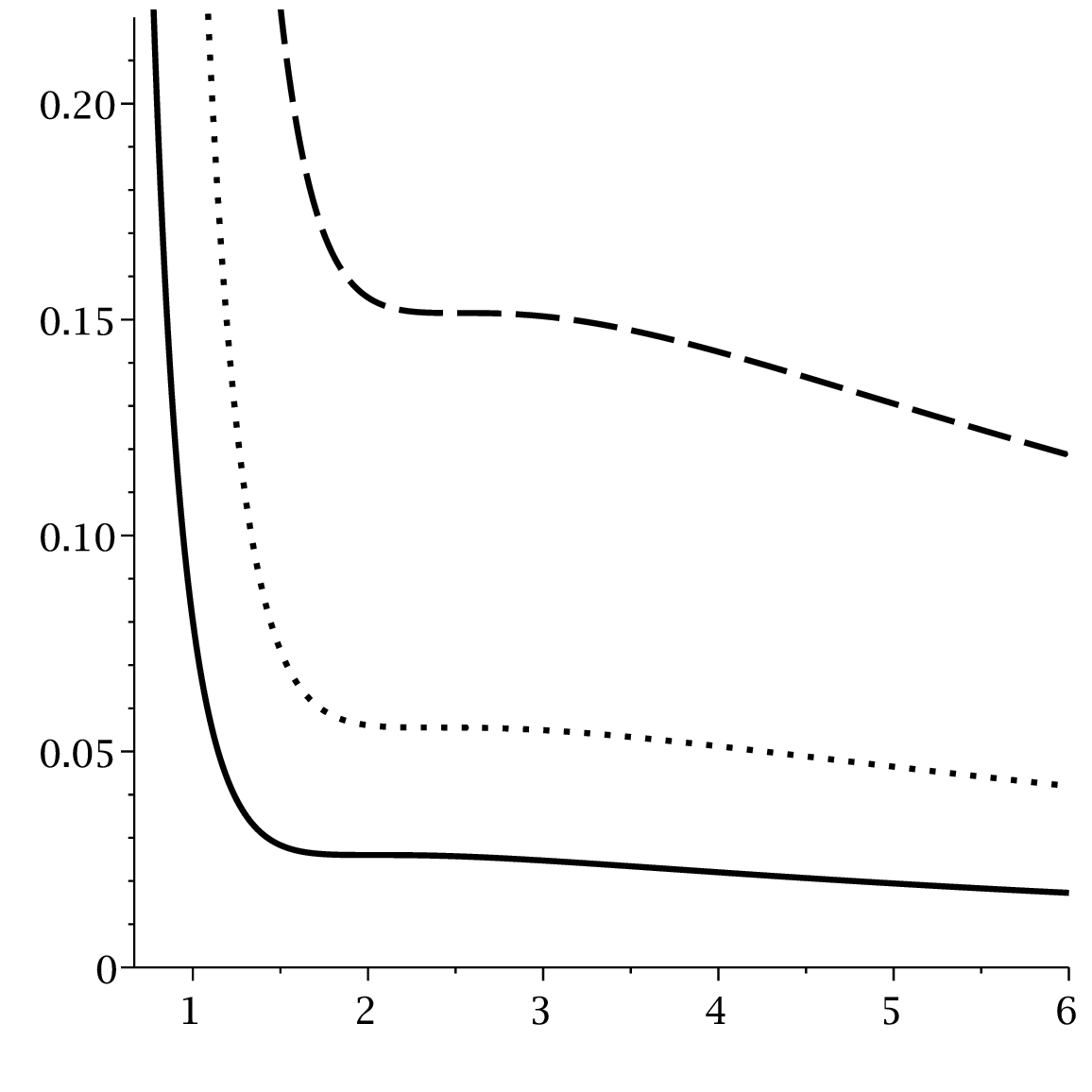} & \epsfxsize=5cm %
\epsffile{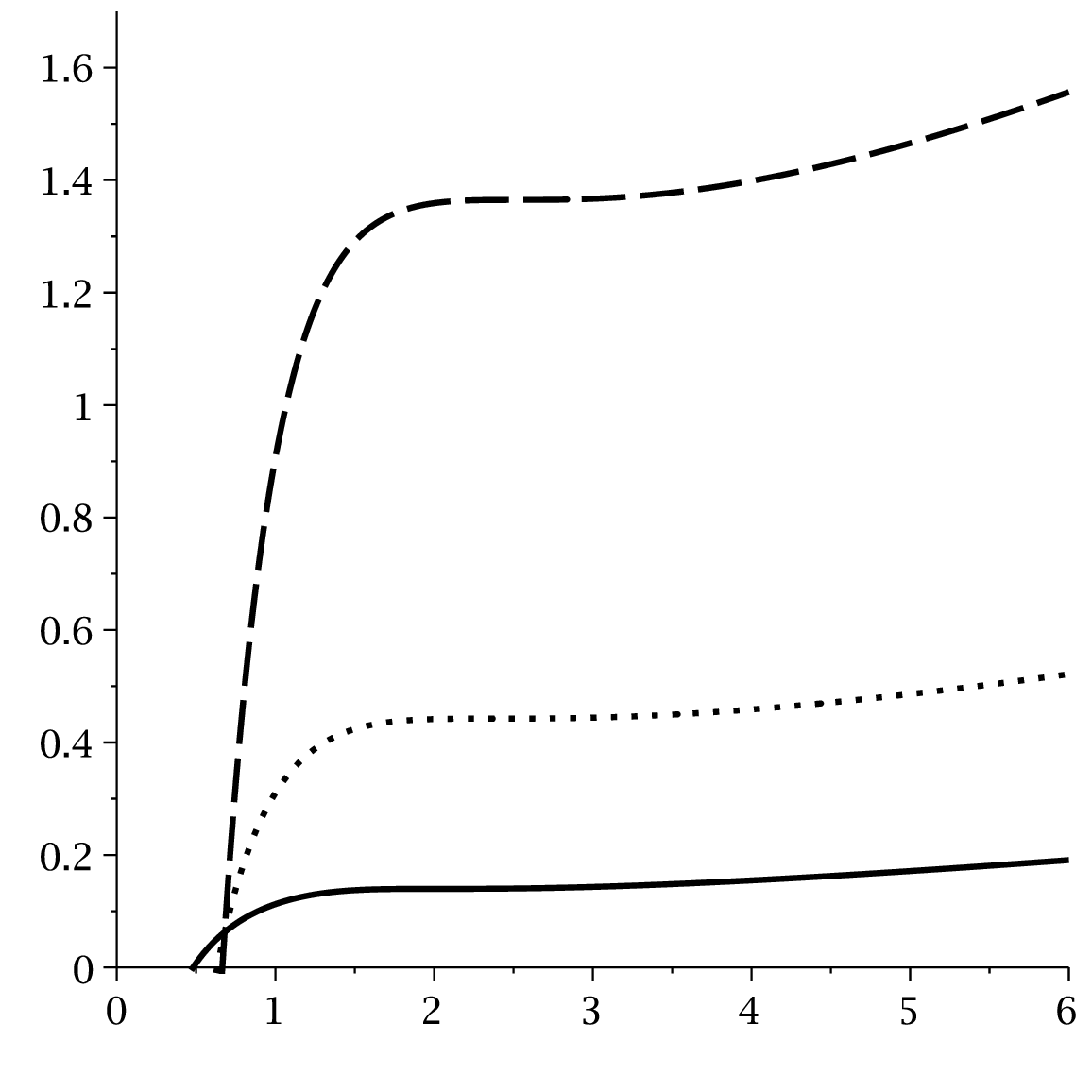} & \epsfxsize=5cm \epsffile{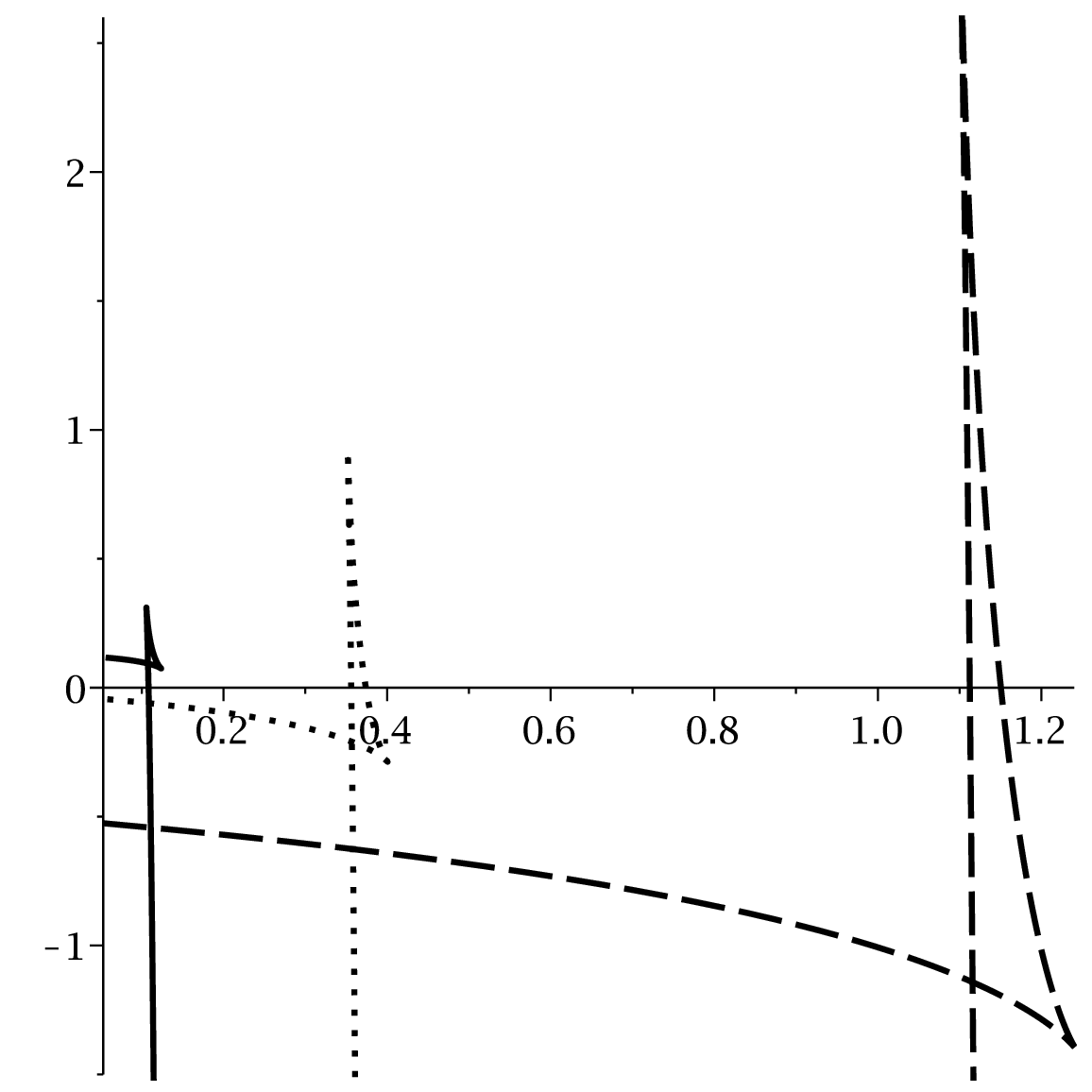}%
\end{array}
$%
\caption{ $P-r_{+}$ (left), $T-r_{+}$ (middle) and $G-T$ (right) diagrams
for $\protect\beta=0.5$, $q=1$, $\protect\alpha=0.8$, $c=c_{1}=c_{2}=2$, $%
c_{3}=0.2$, $c_{4}=-0.2$, $d=6$, $m=0$ (continuous line), $m=0.5$ (dotted
line) and $m=1$ (dashed line). \newline
$P-r_{+}$ diagram for $T=T_{c}$, $T-r_{+}$ diagram for $P=P_{c}$ and $G-T$
diagram for $P=0.5P_{c}$. }
\label{Fig8}
\end{figure}

\begin{figure}[tbp]
$%
\begin{array}{ccc}
\epsfxsize=5cm \epsffile{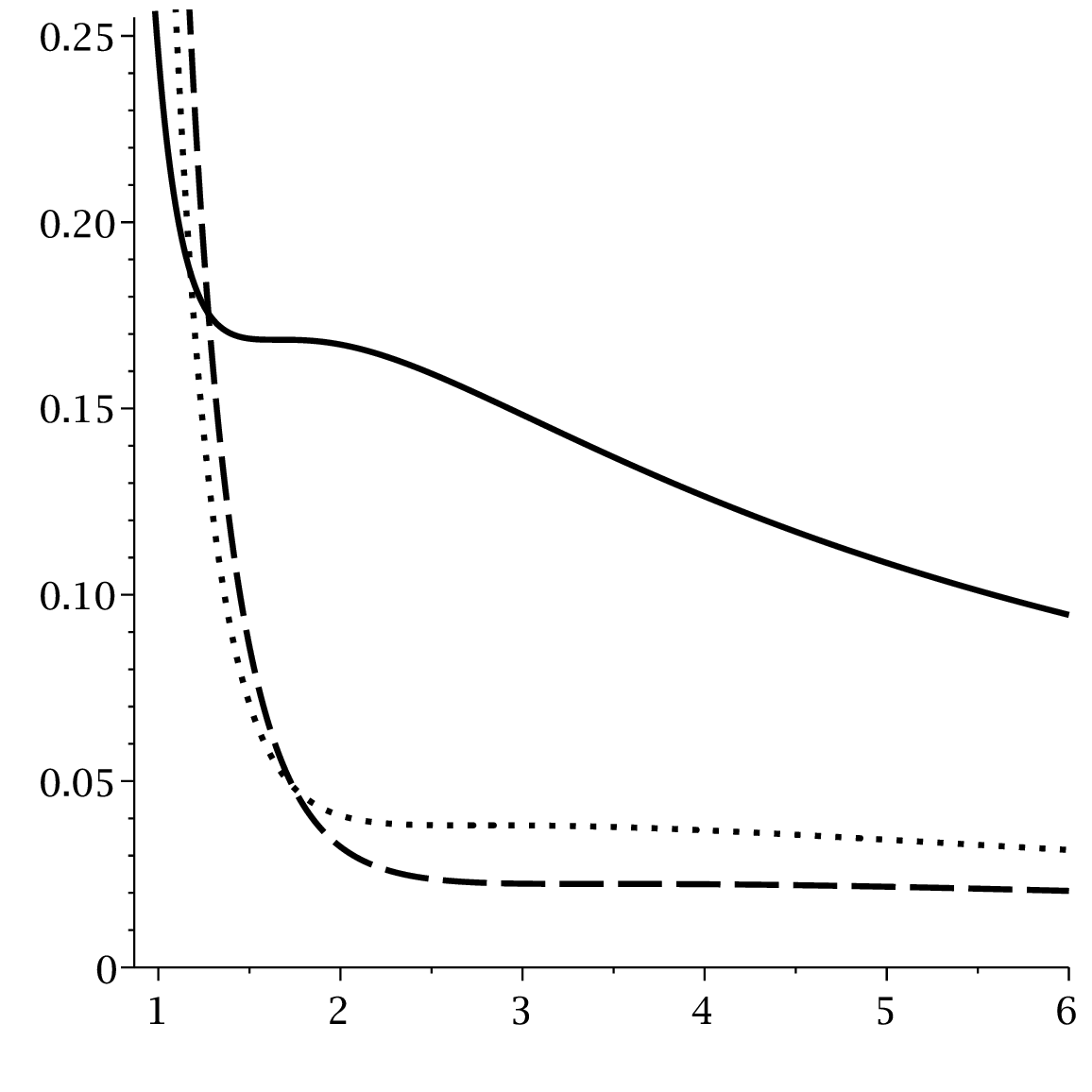} & \epsfxsize=5cm %
\epsffile{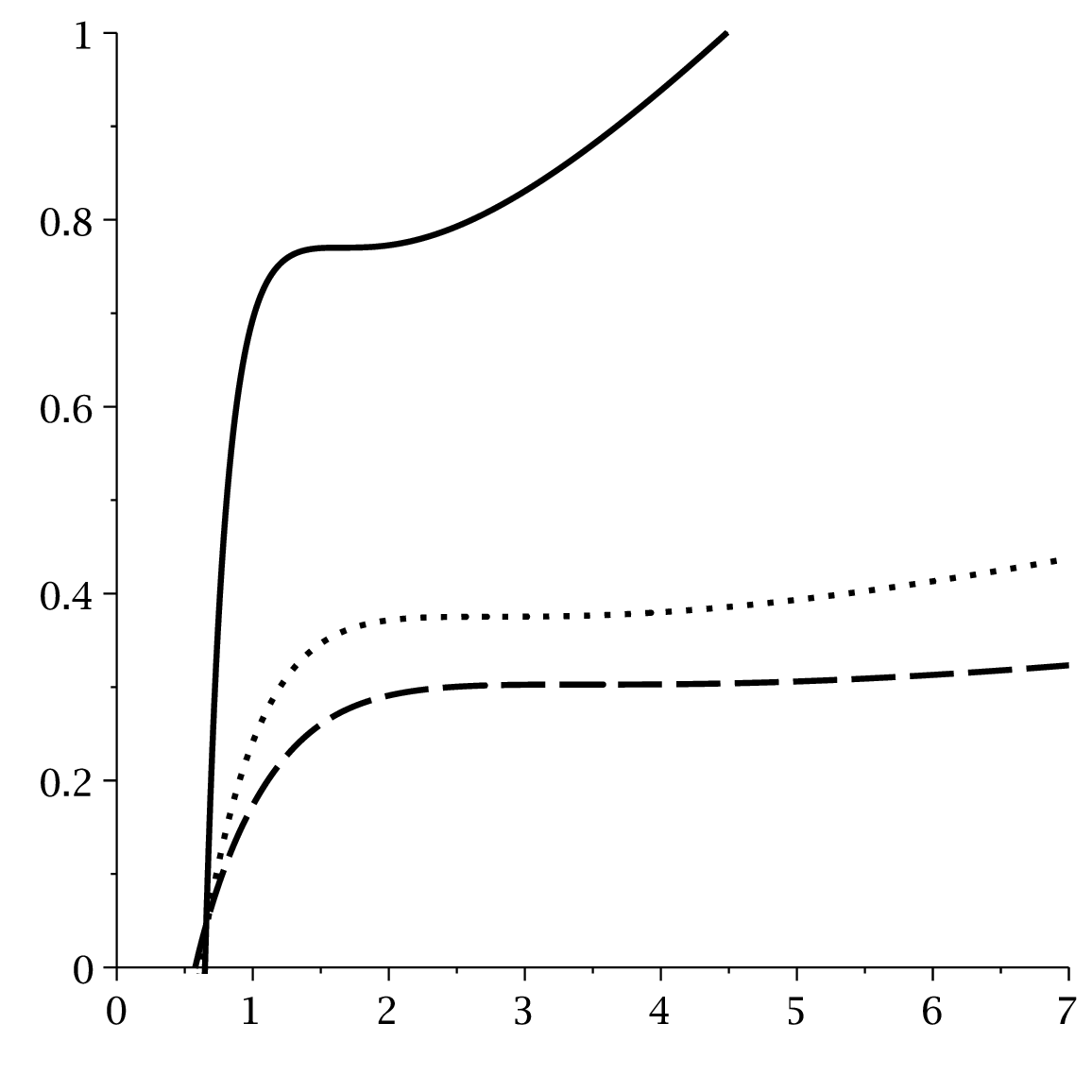} & \epsfxsize=5cm \epsffile{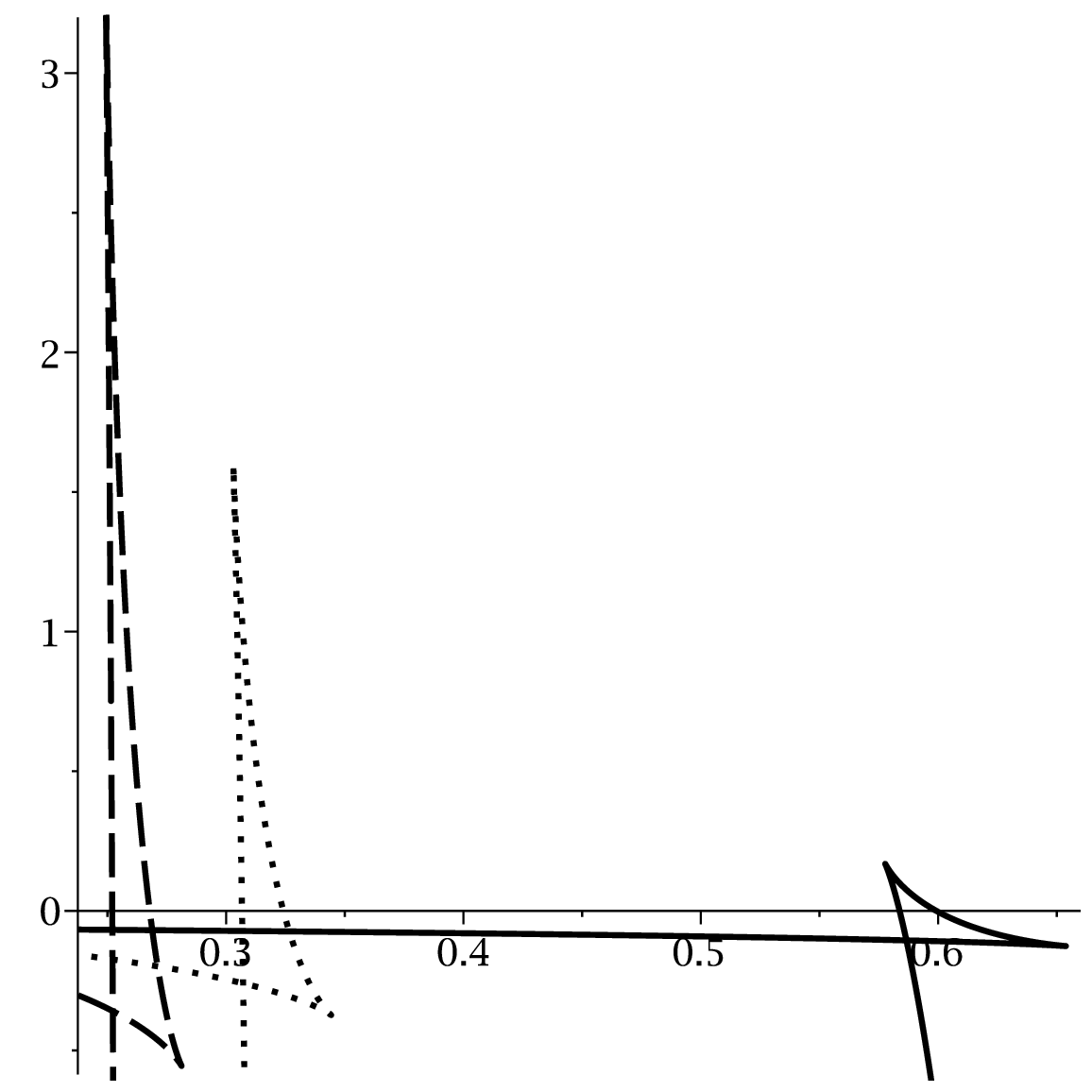}%
\end{array}
$%
\caption{ $P-r_{+}$ (left), $T-r_{+}$ (middle) and $G-T$ (right) diagrams
for $\protect\beta=0.5$, $q=1$, $m=0.5$, $c=c_{1}=c_{2}=2$, $c_{3}=0.2$, $%
c_{4}=-0.2$, $d=6$, $\protect\alpha=0$ (continuous line), $\protect\alpha%
=0.8 $ (dotted line) and $\protect\alpha=1.4$ (dashed line). \newline
$P-r_{+}$ diagram for $T=T_{c}$, $T-r_{+}$ diagram for $P=P_{c}$ and $G-T$
diagram for $P=0.5P_{c}$. }
\label{Fig9}
\end{figure}

\begin{figure}[tbp]
$%
\begin{array}{ccc}
\epsfxsize=5cm \epsffile{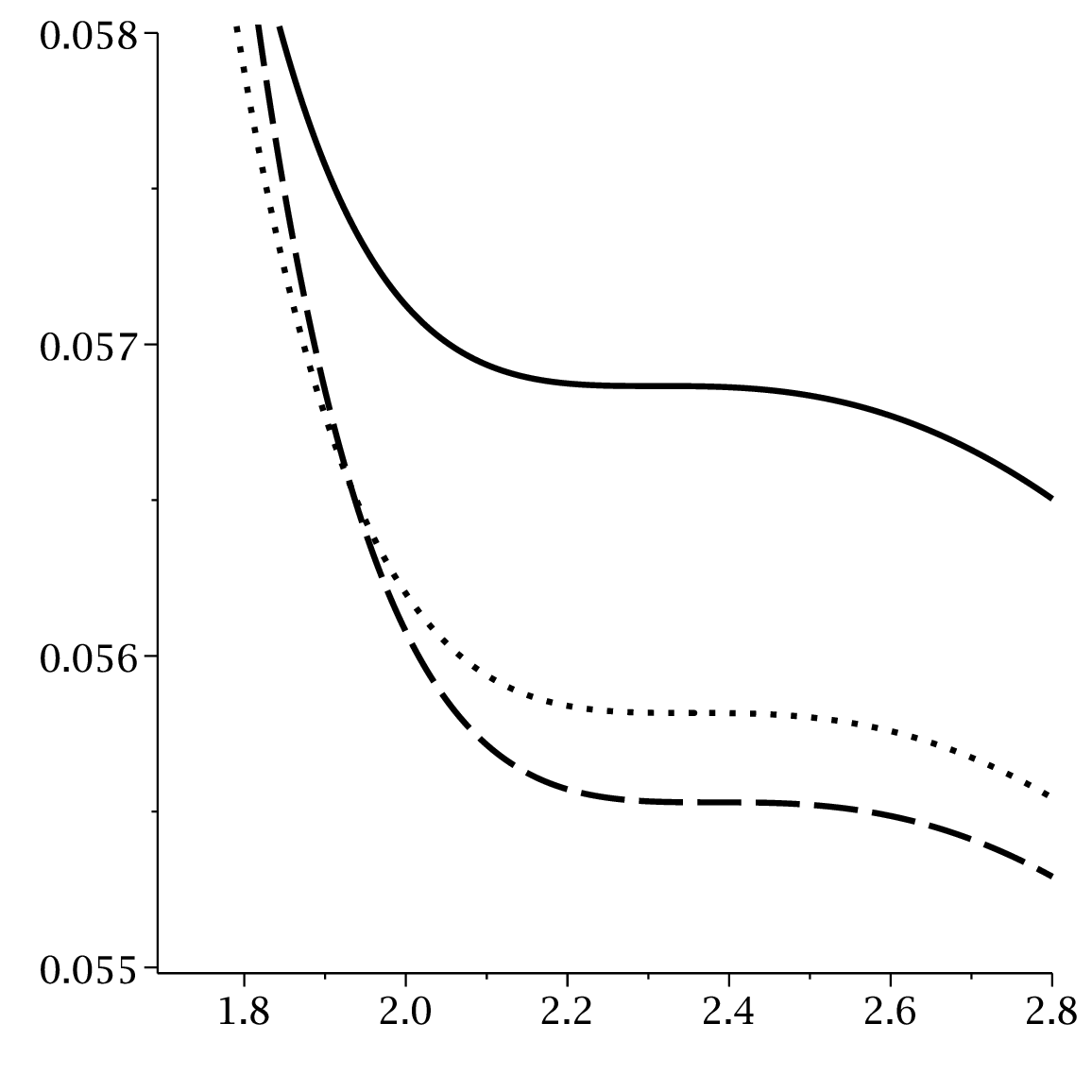} & \epsfxsize=5cm %
\epsffile{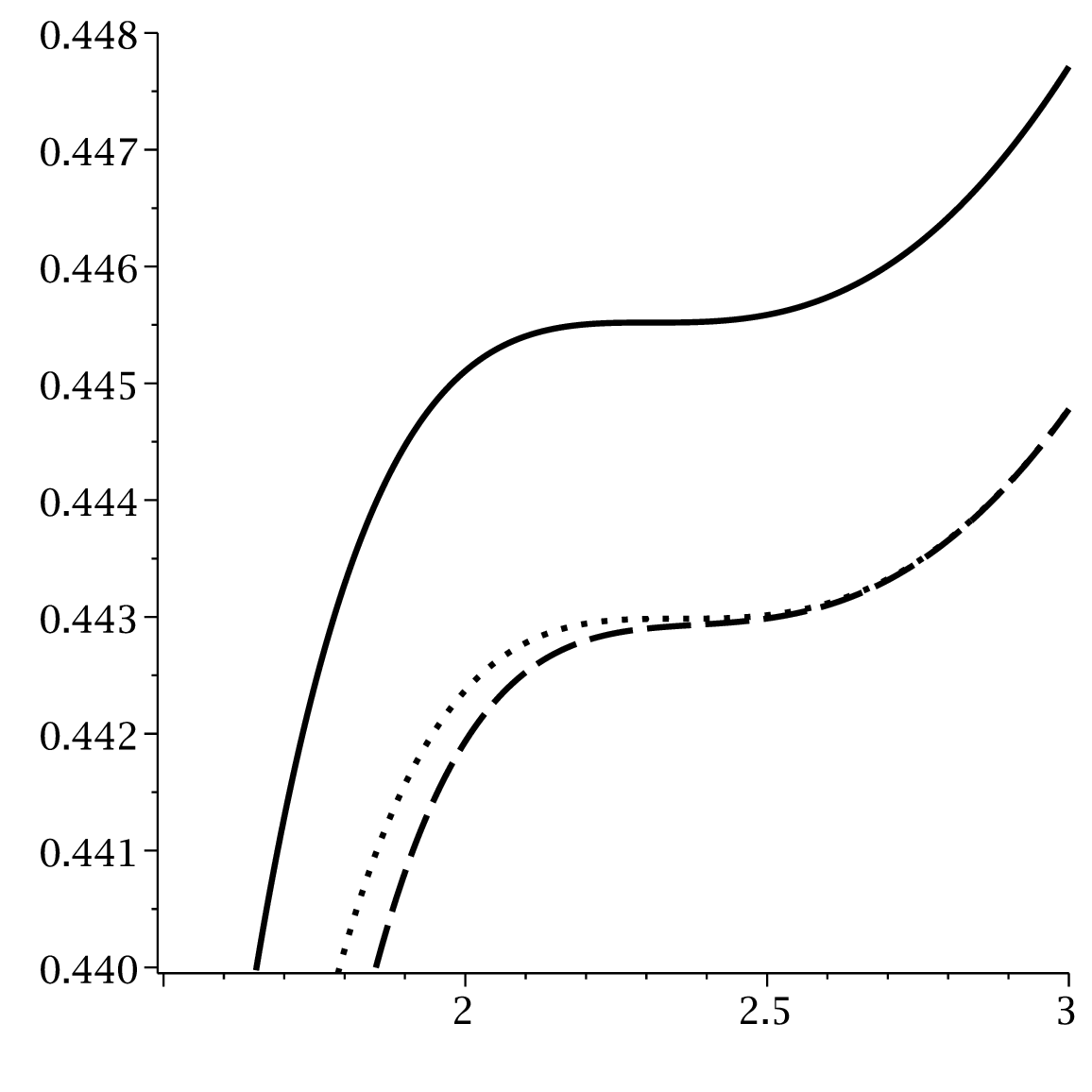} & \epsfxsize=5cm \epsffile{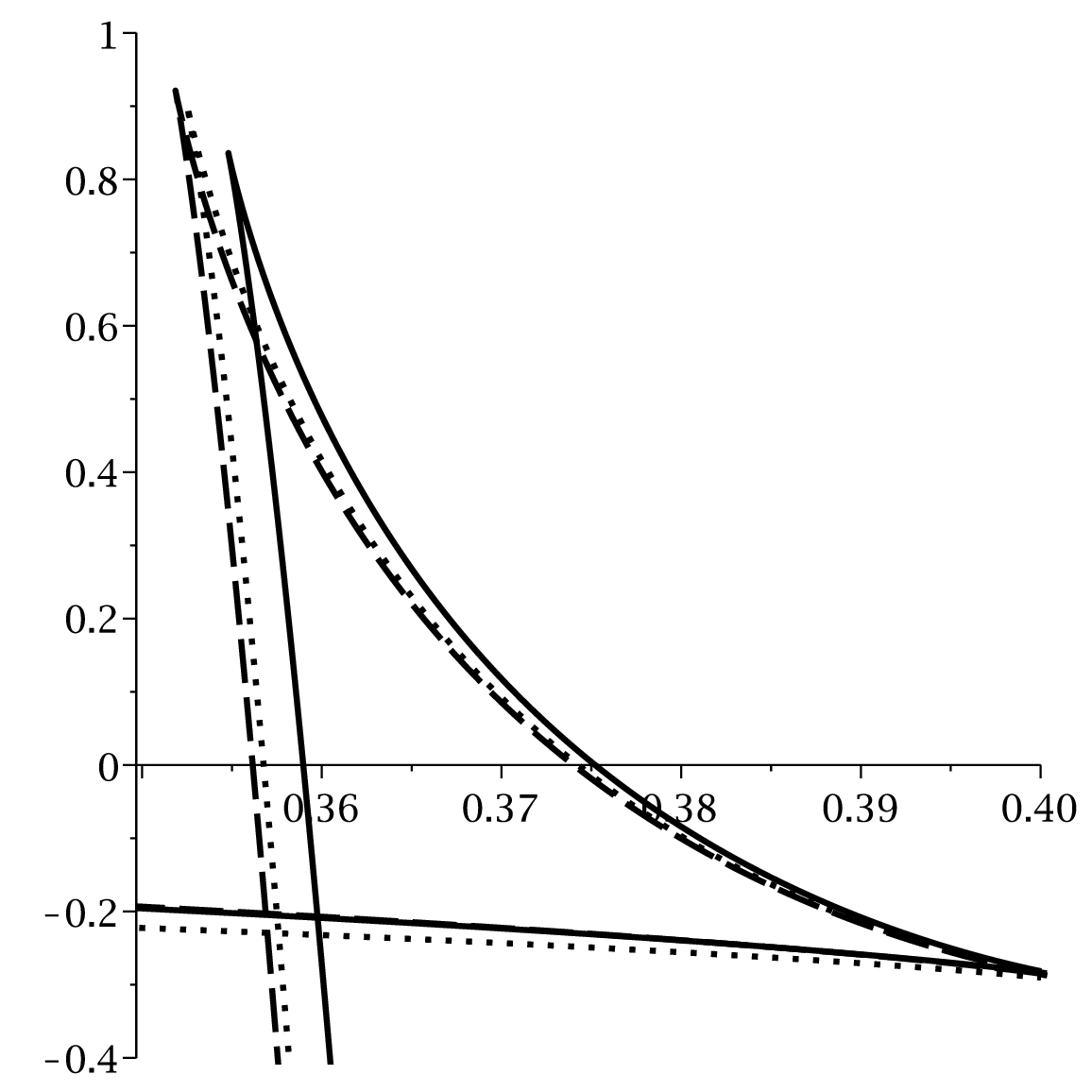}%
\end{array}
$%
\caption{ $P-r_{+}$ (left), $T-r_{+}$ (middle) and $G-T$ (right) diagrams
for $\protect\alpha=0.5$, $q=1$, $m=0.5$, $c=c_{1}=c_{2}=2$, $c_{3}=0.2$, $%
c_{4}=-0.2$, $d=6$, $\protect\beta=10^{-9}$ (continuous line), $\protect\beta%
=0.1$ (dotted line) and $\protect\beta=5$ (dashed line). \newline
$P-r_{+}$ diagram for $T=T_{c}$, $T-r_{+}$ diagram for $P=P_{c}$ and $G-T$
diagram for $P=0.5P_{c}$. }
\label{Fig10}
\end{figure}

\begin{figure}[tbp]
$%
\begin{array}{ccc}
\epsfxsize=5cm \epsffile{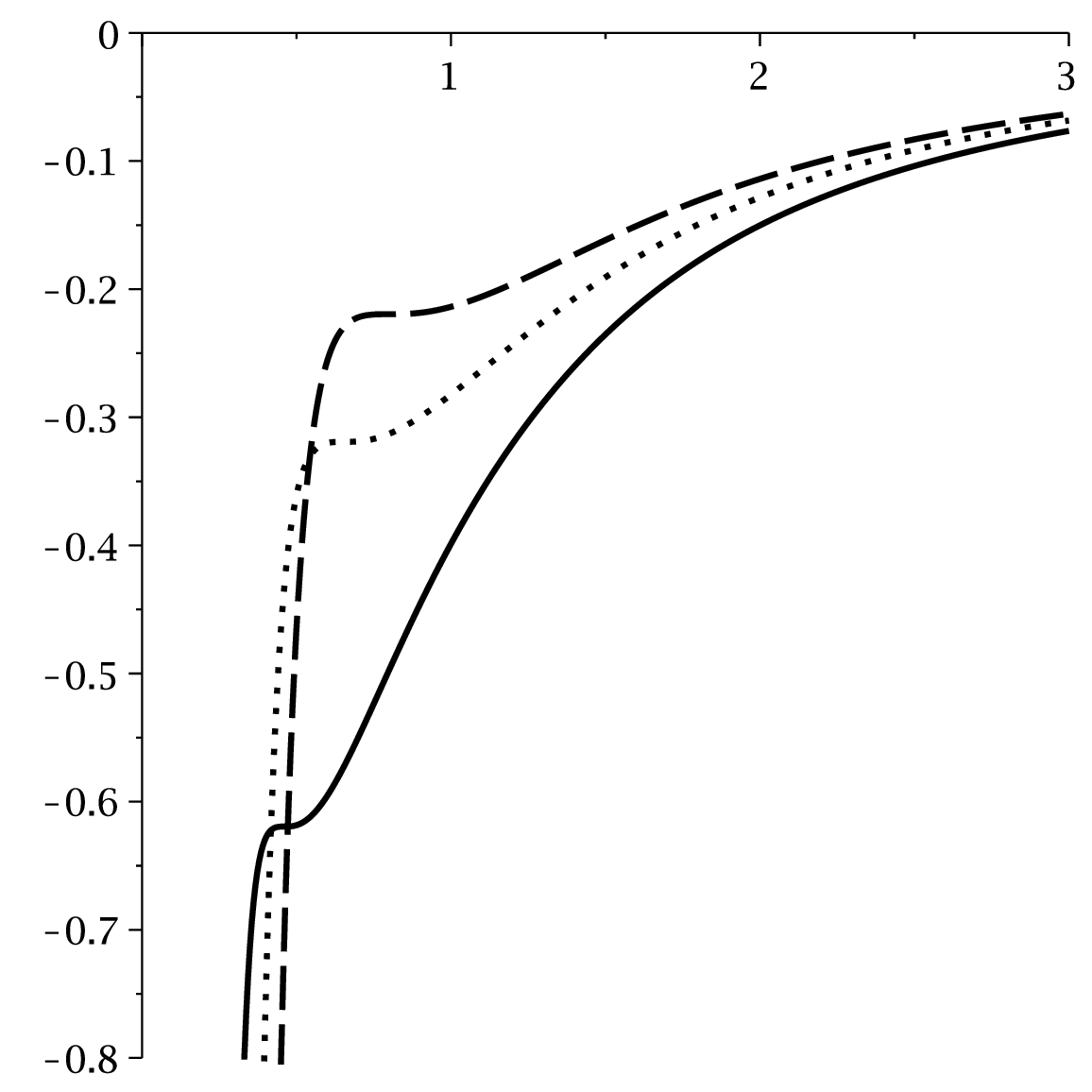} & \epsfxsize=5cm %
\epsffile{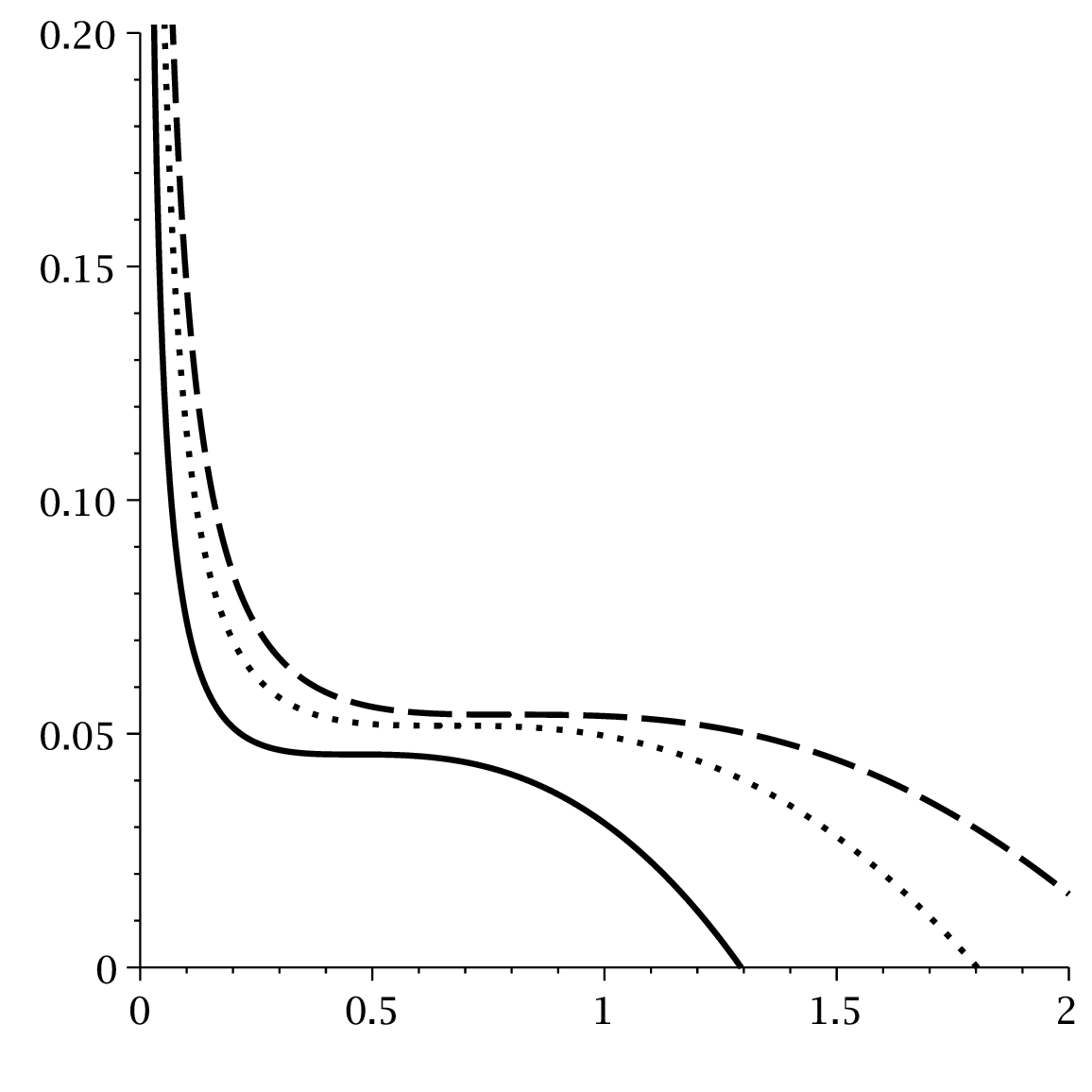} & \epsfxsize=5cm \epsffile{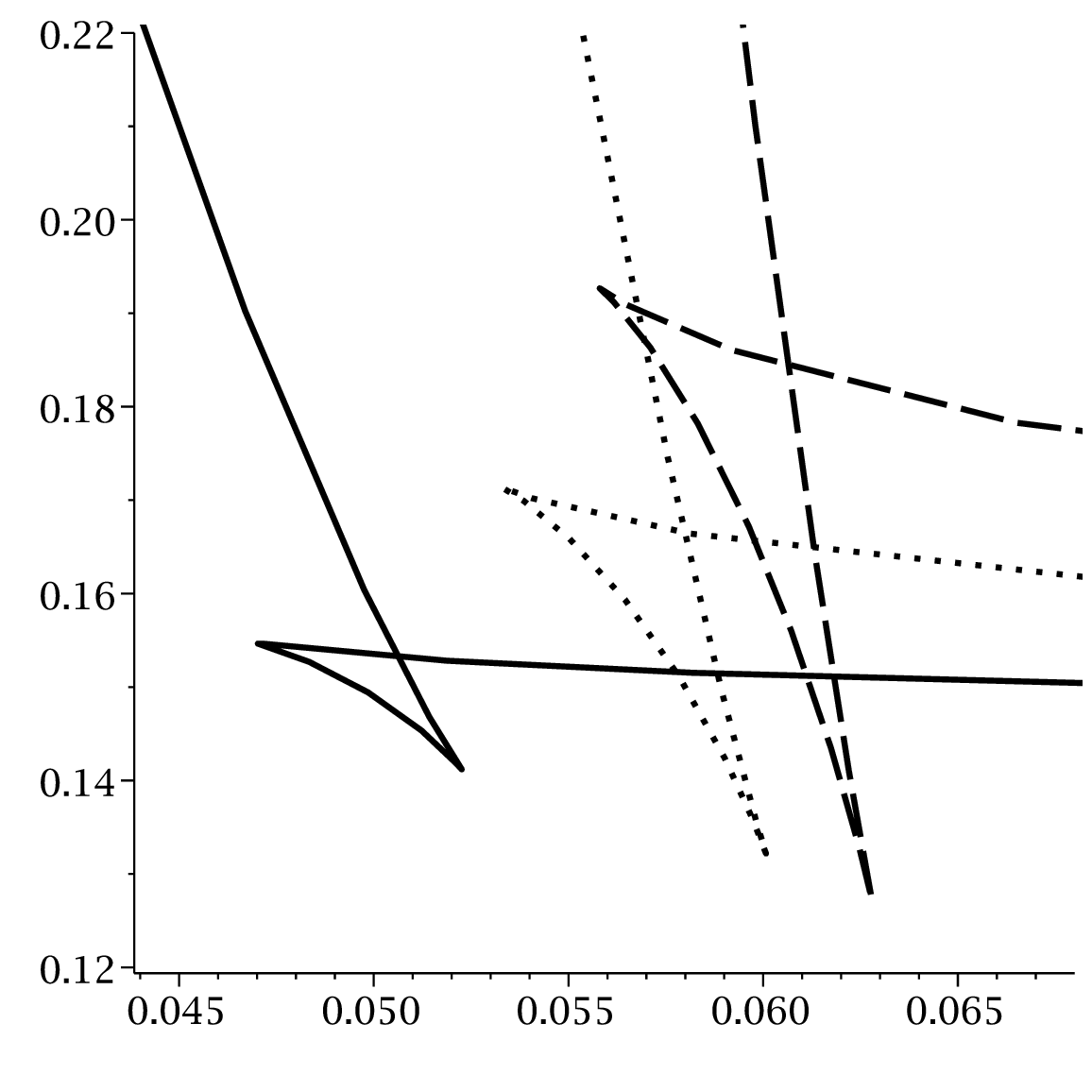}%
\end{array}
$%
\caption{ $P-r_{+}$ (left), $T-r_{+}$ (middle) and $G-T$ (right) diagrams
for $\protect\beta=0.5$, $q=1$, $m=0.5$, $c=c_{1}=c_{2}=2$, $c_{3}=0.2$, $%
c_{4}=-0.2$, $d=6$, $\protect\alpha=7$ (continuous line), $\protect\alpha=8$
(dotted line) and $\protect\alpha=9$ (dashed line). \newline
$P-r_{+}$ diagram for $T=T_{c}$, $T-r_{+}$ diagram for $P=P_{c}$ and $G-T$
diagram for $P=0.5P_{c}$, \emph{For smaller roots.} }
\label{Fig11}
\end{figure}

\begin{figure}[tbp]
$%
\begin{array}{ccc}
\epsfxsize=5cm \epsffile{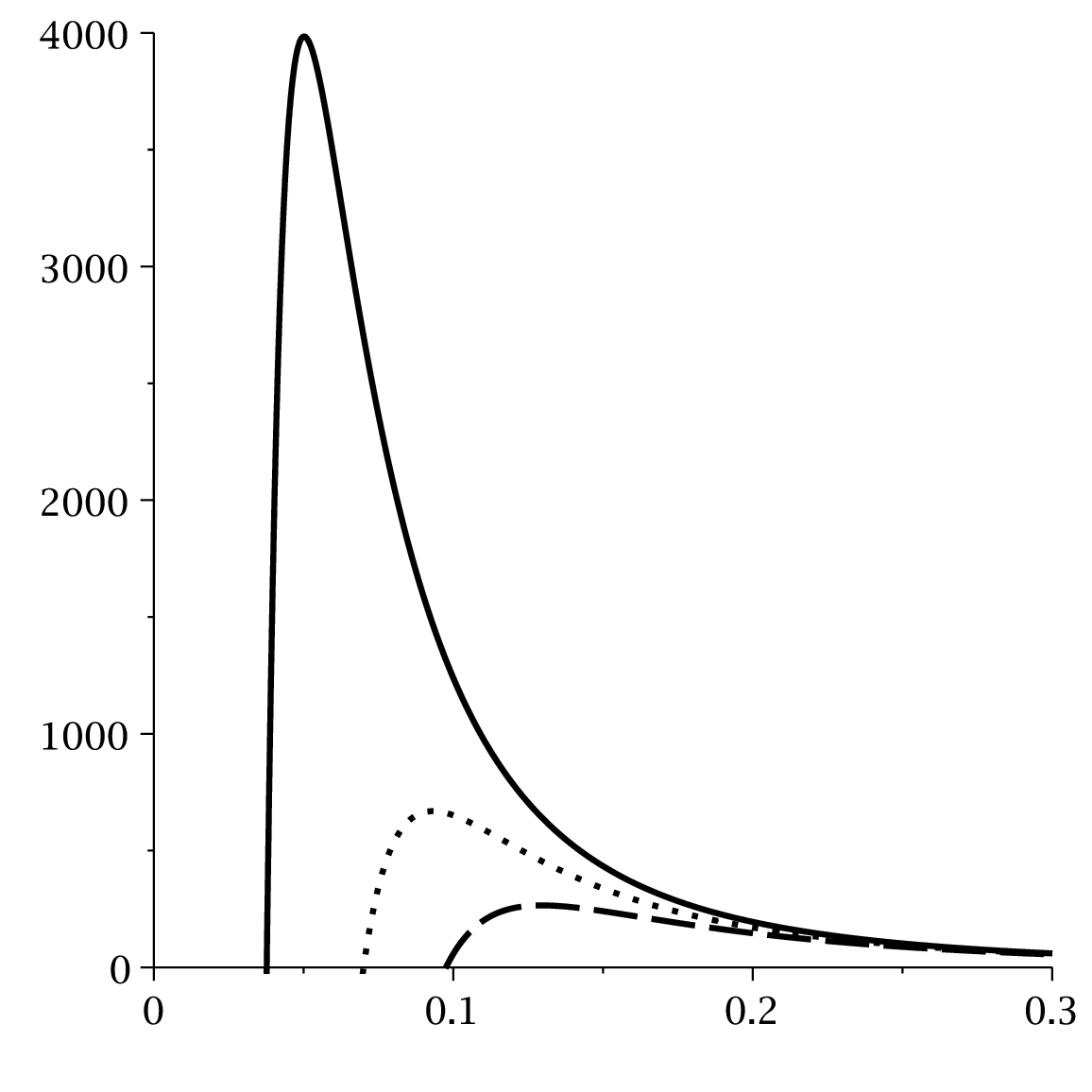} & \epsfxsize=5cm %
\epsffile{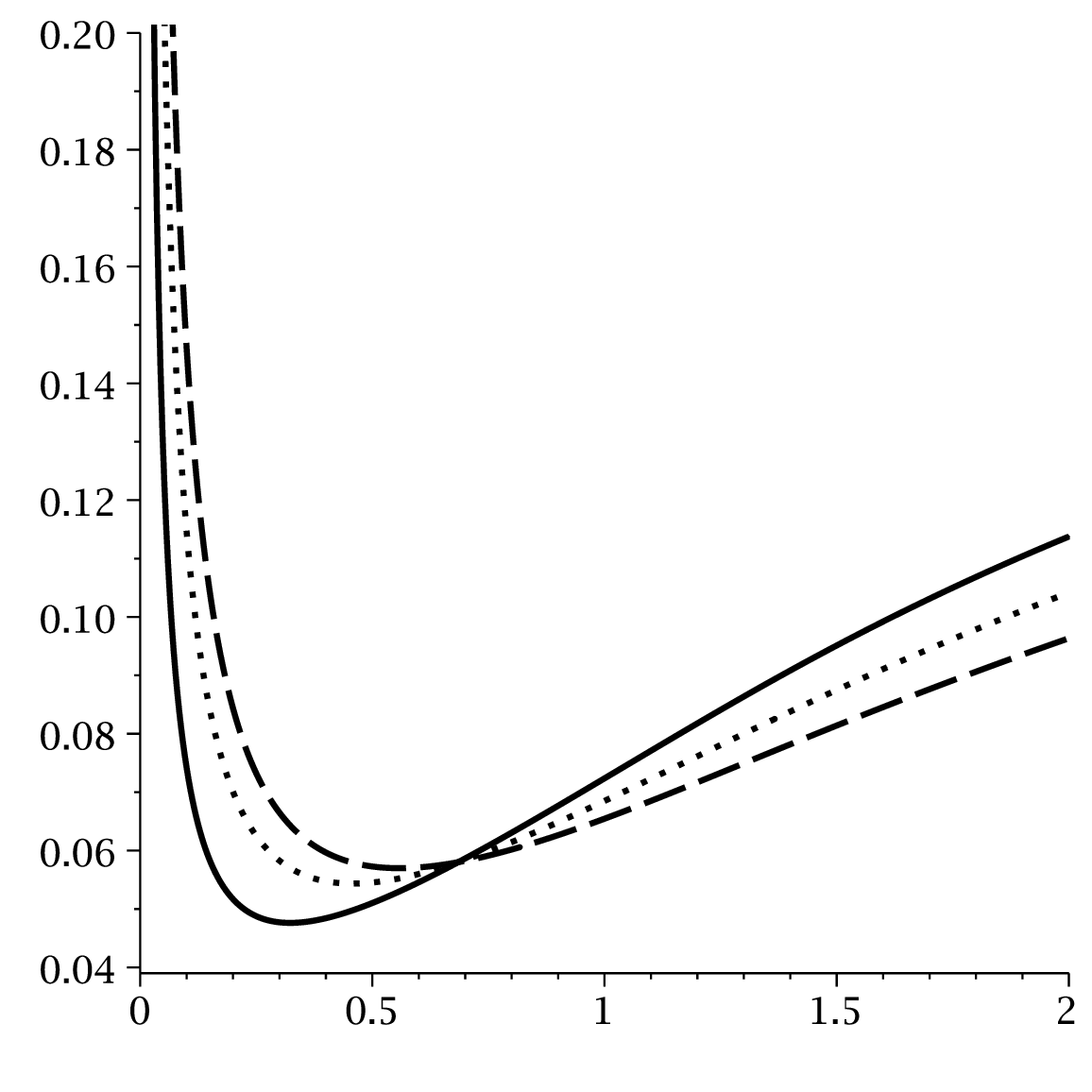} & \epsfxsize=5cm \epsffile{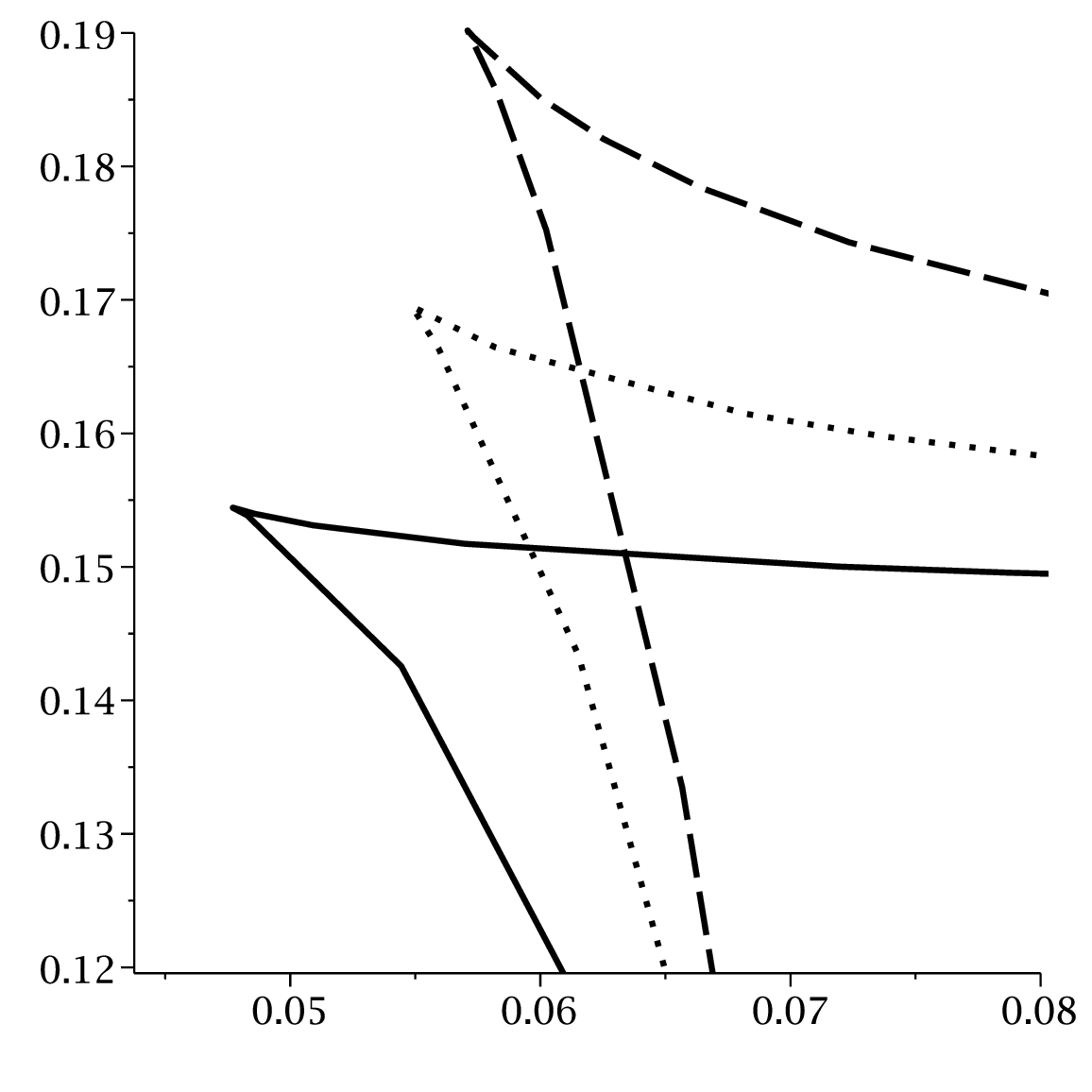}%
\end{array}
$%
\caption{ $P-r_{+}$ (left), $T-r_{+}$ (middle) and $G-T$ (right) diagrams
for $\protect\beta=0.5$, $q=1$, $m=0.5$, $c=c_{1}=c_{2}=2$, $c_{3}=0.2$, $%
c_{4}=-0.2$, $d=5$, $\protect\alpha=7$ (continuous line), $\protect\alpha=8$
(dotted line) and $\protect\alpha=9$ (dashed line). \newline
$P-r_{+}$ diagram for $T=T_{c}$, $T-r_{+}$ diagram for $P=P_{c}$ and $G-T$
diagram for $P=0.5P_{c}$, \emph{For larger roots.} }
\label{Fig12}
\end{figure}

\begin{figure}[tbp]
$%
\begin{array}{ccc}
\epsfxsize=5cm \epsffile{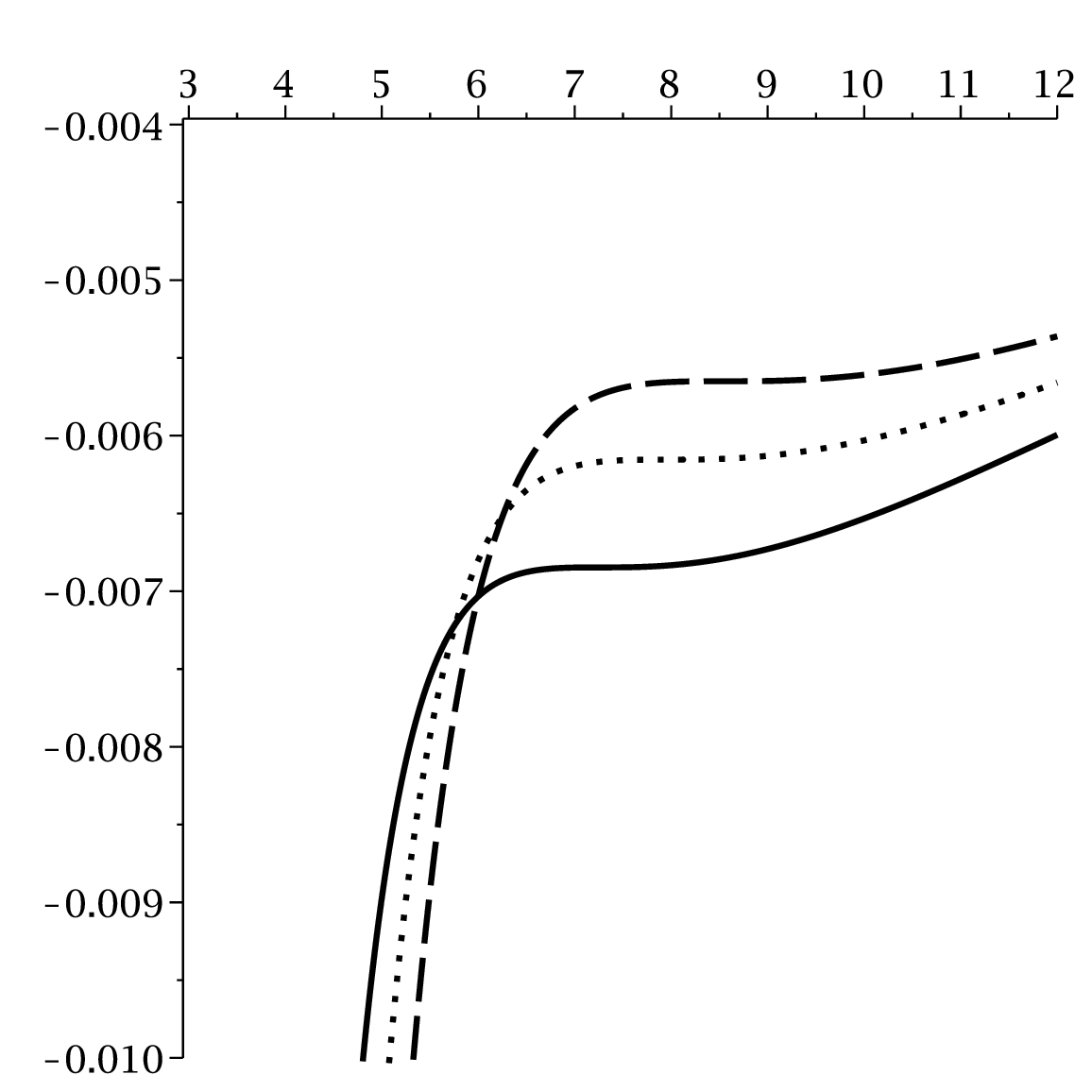} & \epsfxsize=5cm %
\epsffile{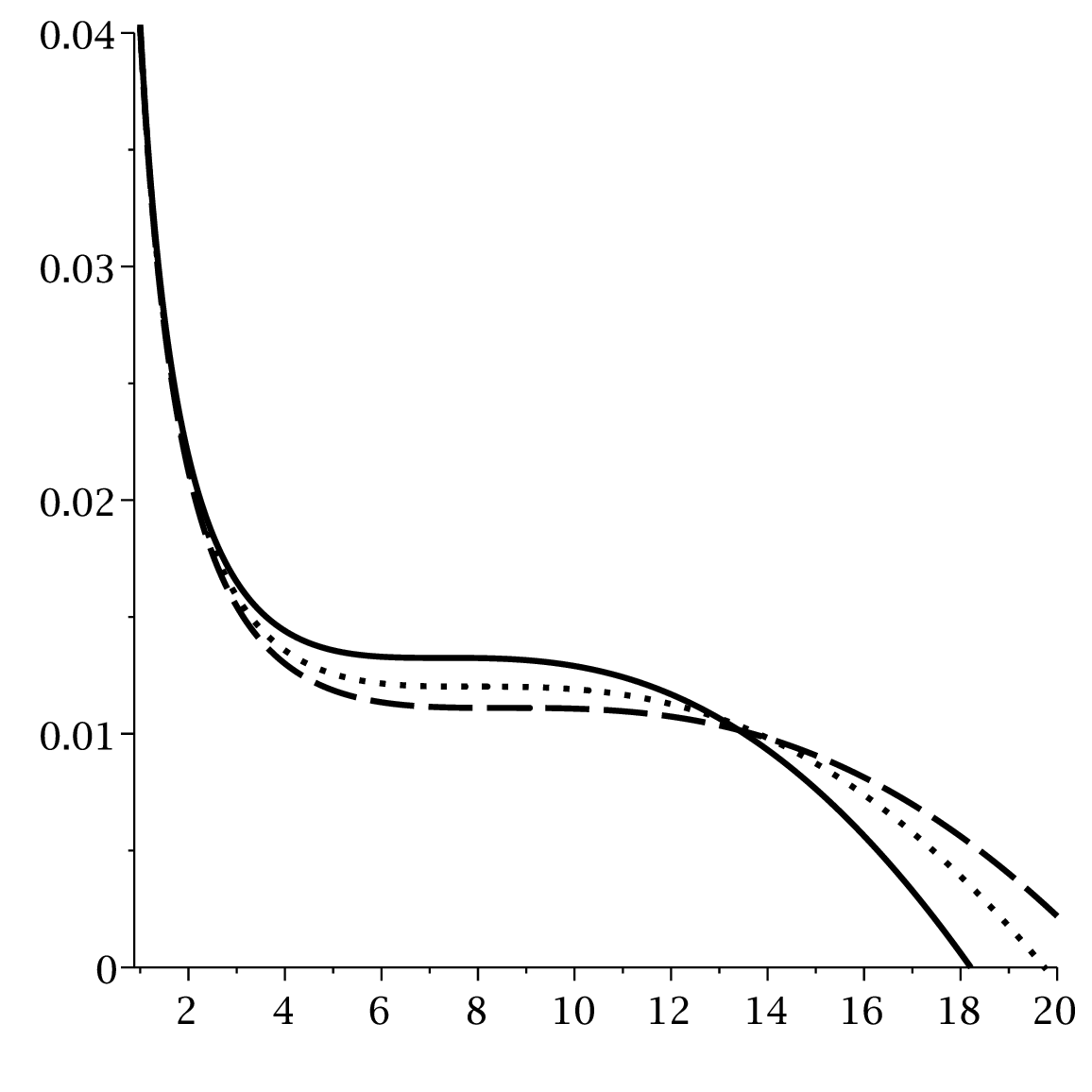} & \epsfxsize=5cm \epsffile{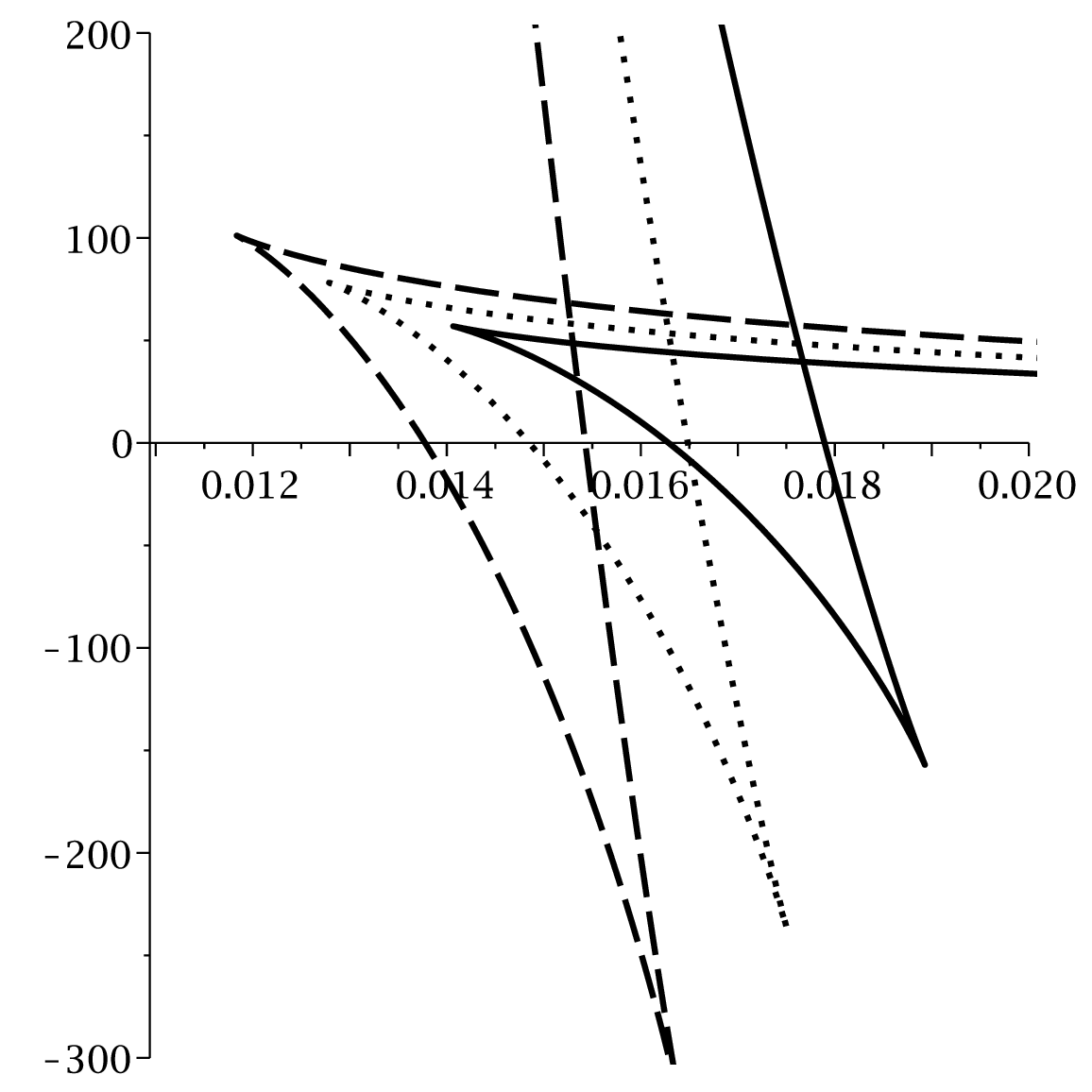}%
\end{array}
$%
\caption{ $P-r_{+}$ (left), $T-r_{+}$ (middle) and $G-T$ (right) diagrams
for $\protect\beta=0.5$, $q=1$, $m=0.5$, $c=c_{1}=c_{2}=2$, $c_{3}=0.2$, $%
c_{4}=-0.2$, $d=5$, $\protect\alpha=400$ (continuous line), $\protect\alpha%
=500$ (dotted line) and $\protect\alpha=600$ (dashed line). \newline
$P-r_{+}$ diagram for $T=T_{c}$, $T-r_{+}$ diagram for $P=P_{c}$ and $G-T$
diagram for $P=0.5P_{c}$. }
\label{Fig13}
\end{figure}

It is evident that due to existence of the swallow-tail in $G-T$ diagrams
for $P<P_{c}$, system enjoys the existence of second order phase transition
for specific values of different parameters (Fig. \ref{Fig7} ). The critical
horizon radius is an increasing function of $m$, $\beta$ and $\alpha$.
Whereas the critical temperature and pressure are increasing functions of
massive parameter (Fig. \ref{Fig8} left and middle panels) and decreasing
functions of GB (Fig. \ref{Fig9} right and middle panels) and nonlinearity
(Fig. \ref{Fig10} left and middle panels) parameters. In addition, the ratio
$\frac{P_{c}r_{c}}{T_{c}}$ is a decreasing function of the GB, massive and
nonlinearity parameters.

On the other hand, the size of swallow-tail is an increasing function of the
massive (Fig. \ref{Fig8} right panel) and GB (Fig. \ref{Fig9} right panel)
parameters and is not a highly sensitive function of the $\beta$ (Fig. \ref%
{Fig10} right panel). The length of subcritical isobars which is
representing the phase transition region, is an increasing function of GB
parameter (Fig. \ref{Fig9} middle panel) and a decreasing function of the
massive parameter (Fig. \ref{Fig8} middle panel). It is worthwhile to
mention, similar to the case of size of swallow-tail, the length of
subcritical isobar is not changed considerably by variation of nonlinearity
parameter (Fig. \ref{Fig10} middle panel).

It should be pointed out that for large values of nonlinearity parameter,
the nonlinear strength will be weak and the behavior of the system will be
Reissner-Nordstr\"{o}m like. On the contrary, for sufficiently small values
of nonlinearly parameter, the black holes present a Schwarzschild like
behavior and the effects of nonlinearity increases drastically.
Interestingly, in case of these black holes, due to their structures, the
critical behaviors are not highly modified for variation of the nonlinearity
parameter comparing to massive and GB parameters (see tables $1-3$ for more
details).

Critical temperature and pressure were highly sensitive to variation of the
massive parameter. Whereas in case of the critical horizon radius, it has
considerable modification due to variation of GB parameter.

Interestingly, for small values of GB parameter, there exists a second order
phase transition and the usual van der Waals like liquid/gas phase
transition (Fig. \ref{Fig7}). On the other hand, for specific range of GB
parameter, black holes enjoy two critical horizon radii which are increasing
functions of this parameter. Critical temperature and pressure for smaller
critical horizon radius are increasing functions of GB parameter whereas in
case of larger horizon radius, they are decreasing functions of it (see
table $1$).

In case of smaller critical horizon radius a negative pressure is obtained.
Plotted phase diagrams represent inverse van der Waals like diagrams. In
other words, the existence of swallow-tail, subcritical isobars and
inflection point are observed in $G-T$, $T-r_{+}$ and $P-r_{+}$,
respectively. But the behavior of diagrams before and after critical point
is opposite of van der Waals like diagrams (Fig. \ref{Fig11}). On the other
hand, all the critical parameters in case of larger critical horizon radius
are positive. But here, instead of swallow-tail, a cusp is observed which
represents the existence of the first order phase transition (Fig. \ref%
{Fig12}).

Next, for large values of $\alpha$, one critical horizon radius with
corresponding positive and negative critical temperature and pressure,
respectively, is found. The plotted phase diagrams are similar to the ones
that were observed for smaller critical point in previous case. In other
words, around critical points opposite behavior to the case of van der Waals
like black holes is observed (Fig. \ref{Fig13}).

Clearly, the type of phase transition is different for the three cases that
were observed in plotted phase diagrams. The cases of two horizon radii
represents a boundary case. Here, the second order phase transition is
vanished, a first order phase transition and another type of phase
transition is observed. Increasing GB parameter leads to vanishing the first
order phase transition and the existence of other mentioned phase transition.

The GB parameter is a free parameter which represents the power of higher
derivative gravity. It is evident that depending on the gravitational power,
the type and number of the phase transition may vary. The phase structure of
these black holes with this specific configuration (GB-BI-massive) goes
under three modifications. These modifications and their corresponding
properties are determined by the value of GB parameter. In other words,
observed critical behaviors are functions of power of gravity.

The van der Waals like behavior of the $4$-dimensional black holes in the
presence of Born-Infeld nonlinear electromagnetic field was first
investigated in Ref. \cite{bornphase}. The generalization to higher
dimensions and the effects of the dimensionality on critical behavior were
investigated in Ref. \cite{bornphasehigher}. Furthermore, the effects of the
Lovelock gravity coupled with Born-Infeld nonlinear electromagnetic field on
critical behavior of the black holes have been explored \cite%
{bornphaseLovelock}. In these works, it was shown that existence
of the critical behavior, depends on gravities and matter fields
under consideration and it is possible to obtain limits for the
presence/absence of van der Waals like behavior. In this paper, we
have considered one more generalization: massive gravitons. The
mentioned generalization introduced new critical behaviors. It was
shown that depending on the choices of different parameters, these
black holes could have three distinctive critical behaviors which
were highlighted in tables and plotted diagrams. The usual van der
Waals like phase transition was related to only one of these three
behaviors while the other ones were not observed before. These
three groups of behaviors have their own characteristic properties
such inverse van der Waals like behavior or existence of valid
critical values for set of parameters and absence of valid ones
for the others. Depending on the critical point being located in
one of these three categories, the critical behavior of the system
around it was different. These new phenomena in critical behavior
of the system are rooted in specific set up the we have considered
in this paper. In addition to these new phenomena, it was also
shown that the values of the critical points were sensitive to
variation of the graviton's mass. This shows that critical nature
of the system could be modified depending on values that the mass
of the graviton could acquire.

\section{Closing Remarks}

In this paper, we have generalized Einstein-BI gravity by considering GB and
massive gravities. It was shown that obtained black hole solutions in this
case can enjoy the existence of multiple horizons. Considering the
configurations of the horizon, different phenomenologies could be described.
Conserved and thermodynamic quantities, that were calculated for these black
holes, satisfied the first law of thermodynamics.

Next, we studied thermal stability of these black holes and their
corresponding phase transition in the context of canonical ensemble. We also
investigated the effects of different parameters on the behavior of the
temperature. We found that considering the contributions of different
parameters, the behavior of the temperature could be highly modified and
resulted into different number of phase transitions, hence, different
stability conditions.

It was seen that in case of divergencies for the heat capacity, GB
and massive parameters have opposing effects, although the
existence of phase transitions were observed for specific values
of $\alpha$. On the other hand, strong nonlinearity parameter
modified the behavior of the temperature completely. A region of
non-physical and phase transitions related to root and four
divergencies were observed.

It is evident that matter field (nonlinear electrodynamics), gravitational
field (GB gravity) and massive gravity contribute highly to thermodynamical
structure of these black holes. Each of these factors add an effective
parameter to the phase transitions and thermal stability structure of these
black holes which enable one to modify/control the effects of other
parameters.

It was also pointed out that dimensionality modifies the stability and
thermodynamical behavior of the system. The cases of $d<7$ and $d>7$ had
different behaviors regarding stability conditions. These conditions were
originated from number of roots and divergencies in the heat capacity. The
case of $d=7$ was also different from the other dimensions.

Here, we should point out that in cases of the maximums in temperature, a
phase transition from larger unstable to smaller stable exists. Whereas, for
the minimums of temperature smaller unstable to larger stable phase
transition takes place which shows that temperature itself presents an
independent picture for studying phase transitions and stability conditions.

Next, we have studied the critical behavior of the GB-BI-massive black holes
through $P-V$, $T-V$ and $G-T$ diagrams. We employed the analogy and linear
proportionality between cosmological constant and thermodynamical pressure
in extended phase space. It was shown that variation of $m$, $\alpha$ and $%
\beta$ affect the critical values, phase transition region and size of the
swallow-tail. The variation of the GB parameter leads to interesting second
order, first order and another type of phase transition. The case of the
second order phase transition was related to existence of usual van der
Waals like behavior. Whereas, first order phase transition was due to
existence of cusp. The other type of phase transition was for the cases in
which around the phase transition point, the behavior of the system was
opposite (and also symmetric) to the van der Waals manner. Therefore, three
different behaviors were observed around critical points for these black
holes. Considering these three behaviors around the critical points, one can
conclude that in strong higher derivative gravity the phase structure of
these black holes will be drastically different and the behavior of the
system will be even opposite of the van der Waals behavior. To our
knowledge, this is a unique behavior which is observed only for these black
holes.

Considering obtained types of phase transition, it will be worthwhile to
study these phase transitions in the context of superconductors. Also, it
will be interesting to study the structure of new phase transition that was
observed in this paper in more details. In other words, one may think about
the possible relation between obtained negative critical pressure with the
nature of dark energy.

\begin{acknowledgements}
We would like to thank anonymous referees for their enlightening
comments. We thank Shiraz University Research Council. This work
has been supported financially by the Research Institute for
Astronomy and Astrophysics of Maragha, Iran.
\end{acknowledgements}

\end{document}